\documentclass[10pt]{article}
\usepackage{amsmath}
\usepackage{amsthm}
\usepackage{graphicx}
\usepackage{amsfonts}
\usepackage{amssymb}
\usepackage{enumerate}
\usepackage[margin=0.85in]{geometry}
\usepackage{comment}
\usepackage{xcolor}
\usepackage{hyperref}
\usepackage{cleveref}
\usepackage{upgreek}
\usepackage{todonotes}

\usepackage[english]{babel}
\usepackage[latin1]{inputenc}
\usepackage[numbers]{natbib}

\usepackage[normalem]{ulem}
\usepackage{bbm}

\DeclareOldFontCommand{\bf}{\normalfont\bfseries}{\mathbf}
\DeclareOldFontCommand{\cal}{\normalfont\bfseries}{\mathcal}

\newtheorem{theorem}{Theorem}[section]
\newtheorem{lemma}[theorem]{Lemma}
\newtheorem{proposition}[theorem]{Proposition}
\newtheorem{corollary}[theorem]{Corollary}

\theoremstyle{definition}

\newtheorem{definition}[theorem]{Definition}
\newtheorem{remark}[theorem]{Remark}
\newtheorem{condition}[theorem]{Condition}
\crefname{condition}{condition}{conditions}
\numberwithin{equation}{section}

\newcommand{\diff}{\,\mathrm{d}}

\newcommand{\ceil}[1]{\lceil{#1}\rceil}
\renewcommand{\d}{\mathrm{d}}
\newcommand{\e}{\mathrm{e}}
\newcommand{\wstar}{W^{*}}
\newcommand{\etai}{\eta_i}
\newcommand{\opt}{\widetilde{\pi}}
\newcommand{\paran}[1]{\left({#1}
\right)}
\newcommand{\abs}[1]{\lvert#1\rvert}
\newcommand{\sumj}{\sum\limits_{j \neq i}^n}
\newcommand{\sumall}{\sum\limits_{j=1}^n}

\newcommand\cA{\mathcal A}
\newcommand\cB{\mathcal B}

\newcommand\cF{\mathcal F}

\newcommand\cI{\mathcal I}
\newcommand\cL{\mathcal L}

\newcommand\EE{\mathbb E}

\newcommand\PP{\mathbb P}

\def \E{\mathbb{E}}
\def \F{\mathbb{F}}
\def \G{\mathbb{G}}
\def \H{\mathbb{H}}

\def \L{\mathbb{L}}

\def \N{\mathbb{N}}

\def \P{\mathbb{P}}
\def \Q{\mathbb{Q}}
\def \R{\mathbb{R}}
\def \S{\mathbb{S}}

\def\Ac{\mathcal{A}}
\def\Bc{\mathcal{B}}

\def\Ec{\mathcal{E}}
\def\Fc{\mathcal{F}}
\def\Gc{\mathcal{G}}

\def\Ic{\mathcal{I}}
\def\Jc{\mathcal{J}}

\def\Tc{\mathcal{T}}

\def\Vc{\mathcal{V}}

\def\Yc{\mathcal{Y}}
\def\Zc{\mathcal{Z}}
\def \QQ{\mathbb{Q}}

\DeclareMathOperator*{\esssup}{ess\,sup}

\pdfoutput=1
\usepackage{color}
\begin{document}


\title{Optimal Investment in a Large Population of Competitive and Heterogeneous Agents
\footnote{Both authors gratefully acknowledge partial support from the NSF grants DMS-2005832 and CAREER DMS-2143861. We thank the anonymous referees and the editor for their contructive feedback as well as Stefanie Hesse for fruitful discussions.}}
\author{ Ludovic Tangpi\footnote{Princeton University, ORFE, ludovic.tangpi@princeton.edu}
$\qquad \qquad$  Xuchen Zhou \footnote{Princeton University, ORFE, xuchen.zhou@princeton.edu}}

\maketitle

\paragraph{\textbf{Abstract}} 
This paper studies a stochastic utility maximization game under relative performance concerns in finite agent and infinite agent settings, where a continuum of agents interact through a graphon (see definition below). 
We consider an incomplete market model in which agents have CARA utilities, and we obtain characterizations of Nash equilibria in both the finite agent and graphon paradigms. 
Under modest assumptions on the denseness of the interaction graph among the agents, we establish convergence results for the Nash equilibria and optimal utilities of the finite player problem to the infinite player problem. 
This result is achieved as an application of a general backward propagation of chaos type result for systems of interacting forward-backward stochastic differential equations, where the interaction is \emph{heterogeneous} and through the control processes, and the generator is of quadratic growth. 
In addition, characterizing the graphon game gives rise to a novel form of infinite dimensional forward-backward stochastic differential equation of Mckean-Vlasov type, for which we provide well--posedness results. An interesting consequence of our result is the computation of the competition indifference capital, i.e., the capital making an investor indifferent between whether or not to compete.


\section{Introduction}

We consider agents investing in a common riskless bond and a vector of stocks of their own choosing. Each agent aims to maximize their utility as a function of their terminal wealth, benchmarked by the industry average. In addition, agents' utilities are of constant absolute risk aversion type. 
This problem was first investigated by \citeauthor*{espinosa_touzi_2013} \cite{espinosa_touzi_2013,EspinosaThesis} in the setting of a complete market with a finite number of agents. 
In these works, the benchmark for a particular agent was taken to be an empirical average of the other palyers' terminal wealth, multiplied by a constant factor between $0$ and $1$ representing the sensitivity of this particular agent to their peers' performance. 
Such a utility maximization problem under \emph{relative performance concerns} has since been explored extensively using various techniques, see for instance \cite{frei2014splitting,Frei-Reis11,lacker2019mean,guanxing2020mean,Lacker-Soret20,hu2022n} and references therein for a small sample of works on the question.
We also refer to \cite{dos2022forward,anthropelos2022competition,dos2019forward} for more recent articles studying relative performance concerns through the lens of the forward criteria of \citeauthor*{MarekZari008} \cite{MarekZari008}.
As in \cite{espinosa_touzi_2013}, \cite{frei2014splitting,Frei-Reis11,guanxing2020mean} approach the problem from a purely probabilistic perspective through characterizing the game using systems of (forward) backward stochastic differential equation (F)BSDEs. 
In particular, \citeauthor*{guanxing2020mean} \cite{guanxing2020mean} explore an extension of the game in an incomplete market framework in the following sense: all agents invest in the same vector of stocks (with dimension $d$) and their strategies can take values in $\R^d$. In addition to having a common Brownian motion $W^*$ representing the market uncertainty or ``common noise'' in the price dynamics of all stocks, they allow each stock to be driven by a separate Brownian motion representing the ``idiosyncratic noise''. 
These Brownian motions are i.i.d and independent of $W^*$.
In this setting, the characterizing system of BSDEs has the particular feature that it is quadratic in the control variable.
As first observed in the work of \citeauthor*{Frei-Reis11} \cite{Frei-Reis11}, such systems are not always globally solvable, making the analysis of the problem in full generality particularly challenging.
For instance, \citeauthor*{Frei-Reis11} \cite{Frei-Reis11,frei2014splitting} provide specific counter examples when equilibria do not exist even in the case of a complete market where stocks are driven by idiosyncratic noise only. 
\citeauthor*{frei2014splitting} \cite{frei2014splitting} showed that multidimensional quadratic BSDEs are in general only locally solvable (i.e.\ the solutions exist only on small time intervals), and he provided equilibria for the original game of \citeauthor*{espinosa_touzi_2013} \cite{espinosa_touzi_2013} using the existence of local solutions to the characterizing BSDEs. \citeauthor*{guanxing2020mean} \cite{guanxing2020mean} established existence and uniqueness of the characterizing BSDEs in an incomplete market framework where all investors invest in two stocks only and the strategies are \emph{unconstrained}.

Recent developments in mean field games provide new avenues to approach the above described utility maximization game by considering the infinite population case.
In fact, standard mean field games heuristics of \citeauthor*{MR2295621} \cite{MR2295621}, \citeauthor*{huang2007invariance} \cite{huang2007invariance} and \citeauthor*{MR3752669} \cite{MR3752669} suggest that in a \emph{homogeneous} game, that is when agents are \emph{symmetric} and \emph{identical}, the infinite population analogue of the game can be solved by considering a \emph{single} representative player whose best response is obtained as solution of a (one--dimensional) McKean--Vlasov BSDE.
More precisely, one thus expects to bypass the subtleties coming with studying multi--dimensional quadratic BSDEs by analyzing a one--dimensional McKean--Vlasov quadratic BSDE.
Despite also having quadratic growth, the latter equation seems much easier to analyze (both analytically and numerically) than the former, making the mean field game paradigm particularly attractive for this game.
The mean field setting was first considered by
\citeauthor*{lacker2019mean} \cite{lacker2019mean} and \citeauthor*{Lacker-Soret20} \cite{Lacker-Soret20} in the Markovian setting with deterministic, constant coefficients, and equilibria was derived using Hamilton--Jacobi--Bellman SPDE methods.
More recently, probabilistic techniques were proposed by \citeauthor*{guanxing2020mean} \cite{guanxing2020mean} and \citeauthor*{dos2019forward} \cite{dos2019forward}.

In the present paper we rigorously study the link between the finite and infinite population games.
Our main modeling assumptions can be summarized as follows: we consider an incomplete market in which
\begin{itemize}
    \item agents are allowed to invest in different vectors of stocks with \emph{random coefficients}, driven by idiosyncratic noise and common noise,
    \item agents' strategies are \emph{constrained} to be in a closed, convex set,
    \item agents benchmark their performance by a \emph{weighted average} of other agents' terminal wealth.
\end{itemize}
Let us elaborate on the latter and probably less studied model feature mentioned above. 
For a particular agent $i$, instead of having a single factor $\lambda_i$ representing their sensitivity to a plain average of the other agents' terminal wealth, we allow this agent to have different sensitivity factors $(\lambda_{ij})_{j \neq i}$ towards each agent. This assumption is a lot more realistic in the sense that funds usually aim to out-perform a small, specific group of competitors, and are usually completely indifferent of the performance of other funds that are, for example, on a much smaller or larger scale, utilize completely different strategies, or operate in a widely different market sector.
This leads to a \emph{heterogeneous} game set on a (random) graph and deviates from the standard symmetric agent interaction assumption, which is arguably the main limitation of the mean field game formulation.
Following the seminal works of \citeauthor*{Lovasz-Szegedy06} \cite{Lovasz,Lovasz-Szegedy06} on the convergence of graphs to the so--called \emph{graphons} (see precised definition and discussion below), the natural infinite population analogue of the game we consider in the present heterogeneous setting is a utility maximization graphon game.
It is worth pointing out that in addition to the methodology, the main modeling difference between the present work and \cite{guanxing2020mean} resides in the heterogeneous interactions among agents and the consideration of constrained strategies here.
These give rise to infinite dimensional Mckean--Vlasov type (F)BSDEs with quadratic generators, making the analyses more demanding and requiring new techniques.

Similar to mean field games, graphon games provide an alternative to study large scale network games that in general suffers less from the curse of dimensionality. However, unlike mean field games, agents in a graphon game are no longer anonymous, as mentioned by \citeauthor*{carmona2022stochastic} \cite{carmona2022stochastic}. The benefit associated with a graphon interaction is that agents are now aware of who their neighbors are, and are allowed to possess different preference metrics towards different neighbors. As a result, when deriving the optimal strategy for a specific agent, one needs to take into account (an \textit{aggregation} of) a continuum of infinitely many other agents, where the aggregation is established through the graphon. The analysis of graphon games has gained traction in recent years, mostly in the engineering community.
We refer for instance to \cite{8619367,caines2019graphon,gao2021lqg,parise2019graphon}.
 \citeauthor*{parise2019graphon} \cite{parise2019graphon} was the first work to analyze equilibria for static graphon games. In \cite{8619367,caines2019graphon}, \citeauthor*{8619367} studied decentralized control for graphon mean field games, and established an $\epsilon$-Nash theory that relates the equilibria for an infinite population game to that of a finite population game. \citeauthor*{gao2021lqg} \cite{gao2021lqg} explored linear quadratic Gaussian mean field games. 
Outside of the engineering community \citeauthor*{carmona2022stochastic} \cite{carmona2022stochastic} studied various static graphon games. \citeauthor*{aurell2021stochastic}  \cite{aurell2021stochastic}  studied stochastic graphon games in a linear--quadratic setting. 
We also refer to the recent works of \citeauthor*{Lacker-Soret22} \cite{Lacker-Soret22} and \citeauthor*{Bayr-Wu-Zhang22} \cite{Bayr-Wu-Zhang22} for recent results, on generic games.
The main contributions of the present work can be summarized as follows:
\begin{itemize}
    \item We derive explicit characterization properties of the Nash equilibria in the finite and the graphon utility maximization games.
    \item We show that if the sensitivity matrix in the finite-agent game stems from a graphon and follows a Bernouilli distribution, then the heterogeneous finite-agent game converges to the graphon game in the sense that every sequence of Nash equilibria converges (up to a subsequence) to a graphon equilibrium along with the associated value functions.
    \item We prove solvability of the graphon utility maximization game.
\end{itemize}
For the characterization properties, we adopt an extension of a well--known methodology proposed by \citeauthor{hu2005utility} \cite{hu2005utility}. The convergence and existence results are more involved. Convergence is obtained as a byproduct of a general \emph{backward propogation of chaos} type result which appears to be of independent interest: 
Consider a general system of weakly interacting FBSDEs, in which the interaction is given through a random graph (appropriately) stemming from a graphon.
We prove strong convergence of the interacting particle system to a limit consisting of infinitely many coupled particles. Backward propagation of chaos type results and their link to the mean field game convergence problem was first developed in recent works by \citeauthor*{lauriere2022backward} \cite{lauriere2022backward}, \citeauthor*{Luo-Tangpi21} \cite{Luo-Tangpi21} and \citeauthor{possamai2021non} \cite{possamai2021non}. 
Also note that in these works, generators are Lipschitz--continuous. Our work contributes to the theory by extending it to systems in heterogeneous interactions through the control processes, and where the generators are of quadratic growth. 
A case of FBSDEs with heterogeneous interactions was posted on ArXiv a week before the present work by \citeauthor*{Bayr-Wu-Zhang22} \cite{Bayr-Wu-Zhang22}.
See also \citeauthor*{bayraktar2020graphon} \cite{bayraktar2020graphon} \cite{bayraktar2022graphon} for results along the same lines for \emph{forward} particle systems. 
The results and methods of the present work further allows us to introduce and compute the so--called competition--indifference capital, which is the capital allowing to make the investor indifferent between being concerned by their peer's performance or not.

The paper is organized as follows. In \Cref{sec:setting}, we first introduce the probabilistic setting and the market model, followed by the finite--agent model and the graphon model, and lastly the main result of this paper, namely the convergence of the finite-agent Nash equilibirum to the graphon Nash equilibirum. 
The BSDE characterizations of the finite-agent game and the graphon game are presented in \Cref{sec:charac.bsde} and \Cref{sec:charac.graph} respectively. 
\Cref{sec:proofs} is dedicated to the proofs of the existence results. 
In \Cref{sec:gct} we prove the main results, which are propagation of chaos for heterogeneous particle systems.
\Cref{sec:wellposedness} establishes existence and uniqueness of general graphon FBSDEs of Mckean-Vlasov type allowing to derive well--posedness of the graphon game.

\section{Probabilistic setting and main results}
\label{sec:setting}
Let us now present the probabilistic setting underpinning this work.
In this section we will also describe the market model as well as the finite and infinite population games under consideration.
At the end of the section we present the main results    of the article.

\subsection{The market model} \label{setup: n-agent}
We fix a finite time horizon $T > 0$ and integers $n,d \in \N$.
Let $(W^i)_{i \ge1 }$ be a sequence of independent $d$-dimensional Brownian motions supported on the probability space $(\Omega, \mathcal{F}, \PP)$. 
In addition, this probability space supports another independent one-dimensional Brownian motion $W^*$ and independent $\mathbb{R}$--valued random variables $(\xi^i)_{i\ge1}$.
We denote by $\F^n := (\cF^n_t)_{t \in [0,T]}$ the $\P$--completion of the natural filtration of $\{(W^i)_{i =1,\dots,n}, W^*,\xi^1,\dots, \xi^n\}$. 
Let us define spaces and norms that will be used throughout the paper.
Fix a generic finite--dimensional normed vector space $(E,\|\cdot\|_E)$, let $\G$ be a filtration, and $\Gc$ a sub-$\sigma$-algebra of $\Fc$ in the probability space $(\Omega,\Fc,\P)$. 

\medskip
$\bullet$ For any $p\in[1,\infty]$, $\L^p(E,\Gc)$ is the space of $E$-valued, $\Gc$-measurable random variables $R$ such that 
\[
    \|R\|_{\L^p(E,\Gc)}:=\Big(\E\big[\|R\|_E^p\big]\Big)^{\frac1p}<\infty,\; \text{when}\; p<\infty,\; \|R\|_{\L^\infty(E,\Gc)}:=\inf\big\{\ell\geq0:\|R\|_E\leq \ell,\; \P\text{\rm --a.s.}\big\}<\infty.
\]

\medskip
$\bullet$ For any $p\in[1,\infty)$, $\H^p(E,\G)$ is the space of $E$-valued, $\G$-predictable processes $Z$ such that 
\begin{equation*}
    \|Z\|_{\H^p(E,\G)}^p:=\EE\bigg[\bigg(\int_0^T\|Z_s\|_E^2\mathrm{d}s\bigg)^{p/2}\bigg]<\infty.
\end{equation*}

\medskip
$\bullet$ 
$\cL^2(E,\G)$ is the space of $E$-valued, $\G$-predictable processes $Z$ such that 
\begin{equation*}
    \int_0^T\|Z_s\|_E^2\mathrm{d}s<\infty\quad \P\text{--a.s}.
\end{equation*}
$\bullet$ For any $p\in[1,\infty]$, $\S^p(E,\G)$ is the space of $E$-valued, continuous, $\G$-adapted processes $Y$ such that 
\begin{equation*}
    \|Y\|_{\S^p(E,\G)}:=\bigg(\EE\bigg[\sup_{t\in[0,T]}\|Y_t\|_E^p\bigg]\bigg)^{\frac1p}<\infty,\;  \text{when}\; p<\infty,\; \|Y\|_{\S^\infty(E,\G)}:=\bigg\|\sup_{t\in[0,T]}\|Y_t\|_E\bigg\|_{\L^\infty(E,\Gc_T)}<\infty.
\end{equation*}
When the probability measure in the definition of these norms is different, say another probability measure $\Q$ on $(\Omega,\Fc)$, we will specify this by writing  $\L^p(E,\Gc,\Q)$, $\H^p(E,\G,\Q)$, and $\S^p(E,\G,\Q)$.


\medskip

The financial market consists of $n$ agents trading in a common risk-less bond with interest rate $r=0$ and $n\times d$ stocks. In particular, each agent trades in a $d$-dimensional vector of stocks $S^i$
{\color{black}with price evolution following the dynamics
\begin{align*}
    \d S_t^{i} = \mathrm{diag}(S^i_t)\big(\mu^{i}_t \d t + \sigma_t^i\d W^{i}_t  + \sigma^{*i}_t d \wstar_t\big) \quad i=1,\dots,n,
\end{align*}
}
where we denote by $diag(x)$ the square matrix with entries $x \in \R^d$ on the diagonal and $0$ everywhere else.
The coefficients {\color{black}$\mu^i$}, $\sigma^i$ and $\sigma^{*i}$ are predictable stochastic processes {\color{black}assumed to be  bounded.
Let $\Sigma_t^i:= (\sigma_t^i, \sigma_t^{*i})$.
We assume throughout that for all $i\in \{1,\dots,N\}$, the matrix $\Sigma^i(\Sigma^i)^\top$ is uniformly elliptic, that is, $KI_d\ge \Sigma^i(\Sigma^i)^\top\ge \varepsilon I_d$ $\P$--a.s. for some constants $K>\varepsilon>0$.
Let us introduce the process $\theta^i$ given by 
$$\theta_t^i := (\Sigma_t^i)^\top(\Sigma_t^i(\Sigma_t^i)^\top)^{-1}\mu_t^i.$$ 
}

\subsection{The \texorpdfstring{$n$}--agent game}
\label{sec: n-agent}

A portfolio strategy is an $\F^n$-predictable, $\R^d$-valued process $(\pi_t)_{t \in [0,T]}$, with each component representing the amount invested in the corresponding stock at time $t$. 
Let $X_t^{i, \pi}$ denote the wealth of agent $i$ at time $t$ when starting with the initial position $\xi^i$ and employing the trading strategy $\pi$, which we assume to be self-financing. 
Then $X_t^{i,\pi}$ satisfies 
\begin{align*}
    {\color{black}\d X_t^{i, \pi} =   \pi_t \cdot \left(\Sigma_t^{i} \theta_t^i \d t + \sigma_t^i\d W_t^i  +  \sigma_t^{*i} \d W_t^{*} \right),\quad X^{i,\pi}_0 = \xi^i.}
\end{align*}
Each agent aims at maximizing their own utility\footnote{In the non--competitivive case, utility maximization has a very long history, and solvability and characterization issues are settled. See for instance \cite{hu2005utility,optimierung,Kra-Sch,Cvi-Sch-Wang,Delbaen-expo-uti,rouge2000pricing,schachermayer2004utility,tahar2010merton} for a very incomplete list of references.} from the terminal wealth, and in this work we assume the utility function to be exponential.
In addition, each agent is concerned with the relative performance of their peers; see e.g. \cite{espinosa_touzi_2013,Frei-Reis11,EspinosaThesis} for early works on the problems.
Thus, the terminal wealths are benchmarked by a \emph{weighted average} of the other agents' terminal values\footnote{Throughout this work we use $\sum_{j \neq i} x^j$ as a shorthand notation for $\sum_{j \in \{1,\dots, n\}\setminus \{i\}} x^j$.}: $ \frac{1}{n-1} \sum_{j \neq i} \frac{\lambda_{ij}}{\beta_n} X_T^{j,\pi}$.
The main modeling novelty considered in the present work is the addition of the term $\frac{\lambda_{ij}}{\beta_n}$ which measures agent $i$'s sensitivity to agent $j$'s wealth. 
{\color{black}The point is that each agent will try to perform better than the average of the other agents in the market, but they are not concerned with the performance of \emph{all} agents.
Think for instance of hedge funds.
They will typically compete with ``similar'' hedge funds, for instance those raising capital from same investors.
Thus, $\lambda_{ij} = 1$ if agent $i$ is concerned with agent $j$'s performance and $\lambda_{ij}=0$ if not.}
Denote for simplicity
$$\lambda_{ij}^n:=\frac{1}{n-1}\frac{\lambda_{ij}}{\beta_n}\quad \text{with}\quad \lambda^n_{ii} := 0.$$
The terminal utility of agent $i$ takes the form
\begin{align}
\label{eqn:utility-n-agent}
    U_i\Big(X_T^{i,\pi^i},\sum\limits_{j \neq i}\lambda_{ij}^n X_T^{j,\pi^j} \Big) := -\exp\bigg\{-\frac{1}{\etai}\bigg(X_T^{i,\pi^i} - {\color{black}\rho}\sum\limits_{j \neq i} \lambda_{ij}^n X_T^{j,\pi^j}\bigg)\bigg\},
\end{align}
where $\eta^i \in (0,1)$ measures the risk preference level for agent $i$ {\color{black}and $\rho$ models the interaction weight.
Since we are interested in competition, we fix $\rho\in (0,1]$ throughout the article, see e.g. \citeauthor*{hu2022n} \cite{hu2022n}.}
Let $\cA_i$ denote the set of admissible strategies for agent $i$ (which we will define shortly). To avoid bulky notations, we will use the abbreviated $\pi^i$ for the rest of this section with the understanding that the strategy depends on the size $n$ of the game.
The optimization problem for agent $i$ thus takes the form
\begin{align}\begin{split}
\label{eq: n-agent obj}
 V_0^{i,n}
    &:=V_0^{i,n}((\pi^j)_{j\neq i})  \\
    &:= \sup_{\pi \in \cA_i} \E \bigg[ -\exp\bigg\{-\frac{1}{\etai}\bigg(X_T^{i,\pi} - \rho\sum\limits_{j \neq i} \lambda_{ij}^n X_T^{j,\pi^j}\bigg)\bigg\} \bigg].
    \end{split}
\end{align}

\begin{definition}[Admissibility]
\label{defn: admissible strategy}
Let $A_i$ be a closed convex subset of $\R^d$ that we will call constraint set.
A strategy $\pi^i$ for player $i$
is admissible if $\pi^i \in \H^2(A_i,\F^n )$ and for every $j \in \{1,\dots,n\}$, there is $p>2$ such that
 the family
{\color{black}
\[\big\{\e^{\frac{p}{\eta^i}\frac{\rho\lambda_{ij}}{\beta_n}X_{\tau}^{i,\pi^i} };\; \text{with $\tau$ a $\F^n$--stopping time on $[0,T]$}  \big\}\]
}
is uniformly integrable.
In this case we will say that $\pi^i \in \cA_i$.
\end{definition}
 As usual we will be interested in Nash equilibria, whose definition we recall:

\begin{definition}
    A vector $(\opt^{1}, \opt^{2}, \dots, \opt^{n})$ of admissible strategies in $\cA_1 \times \cA_2 \times \dots \times \cA_n$  is a Nash equilibrium if for every $i = 1,\dots, n$, 
    the strategy $\opt^{i}$ is a solution to the portfolio optimization problem given in \Cref{eq: n-agent obj} with value $V_0^i(\opt^{1},\dots, \opt^{i-1}, \opt^{i+1},\dots,,\opt^{n})$.
    That is, for each $i$, 
    \begin{equation*}
        V^{i,n}_0((\tilde\pi^j)_{j\neq i}) =  \E \bigg[ -\exp\bigg\{-\frac{1}{\etai}\bigg(X_T^{i,\tilde\pi^i} - \rho\sum\limits_{j \neq i} \lambda_{ij}^n X_T^{j,\tilde\pi^j}\bigg)\bigg\} \bigg].
    \end{equation*}
\end{definition}
In this work we will assume that $(\lambda_{ij})_{1\le i,j\le n}$ are realizations of i.i.d. random variables, which are independent of the randomness source $(W^i,W^*,\xi^i)_{i\in \{1,\dots,n\}}$.
In particular, $(\lambda_{ij})_{1\le i,j\le n}$ is defined on a different probability space $(\mathfrak{D}, \mathfrak{F}, \mathfrak{P})$ and results will be proved for almost every realization of the graph.
Therefore, we are actually working on the product space $(\Omega\times\mathfrak{D}, \mathcal{F}\otimes\mathfrak{F}, \P\otimes\mathfrak{P})$.
We will often use $\P$ to simplify the exposition.
The interaction parameters $(\lambda_{ij})_{1\le i,j\le n}$ give rise to an \emph{undirected random graph}. Notice at this point already that our setting will include Erd\"os-Renyi graphs and the traditional complete graph.

Let us conclude this subsection by introducing some more notation that will be used in the paper.
Given a vector $\boldsymbol{y} = (y^1,\dots, y^n)$, we put
\begin{equation*}
    \overline{y}^i := \sum_{j \neq i}\lambda_{ij}^ny^j,
\end{equation*}
the weighted average of the vector $\boldsymbol{y}$ (taking out $y^i$). Let $X_t^{\pi^i}$ be a short hand notation for $X_t^{i,\pi^i}$ and given a Nash equilibrium, $(\opt^{1}, \opt^{2}, \dots, \opt^{n})$, denote $\overline{X}_t^i := \sum_{j \neq i}\lambda_{ij}^nX_t^{\tilde{\pi}^j}$ the weighted average of the portfolio values for agents $j \neq i$ when they all use the Nash equilibrium strategy $\opt^j$. These notation will be used in the statement of the main results.

\subsection{The graphon game}\label{sec: graphon game}
\label{setup:graphon} 
Let $I =[0,1]$ denote the unit interval. Intuitively, in the context of an infinite-player graphon game, we will label by $u \in I$ a given agent amid a continuum. The following probabilistic setup models the infinite population game.

Let $\cB_I$ be the Borel $\sigma$-field of $I$, and $\mu_I$ be the Lebesgue measure on $I$. Given a probability space $(I, \cI, \mu)$ extending the usual Lebesgue measure space $(I, \cB_I, \mu_I)$, and the sample space $(\Omega, \cF, \P)$, consider a rich Fubini extension $(I \times \Omega, \cI \boxtimes \cF, \mu \boxtimes \P)$ of the product space $(I \times \Omega, \cI \otimes \cF, \mu \otimes \P)$. Unfamiliar readers can consult \citeauthor*{sun2006exact} \cite{sun2006exact} for a self--contained presentation of the theory of rich Fubini extensions. 
Let $C([0,T];\R^d)$ denote the space of continuous functions from $[0,T]$ to $\R^d$. By \cite{sun2006exact},
we can construct $\cI \boxtimes \cF$-measurable processes $(W,\xi) : I \times \Omega \rightarrow C([0,T],\R^d) \times \R$ with \emph{essentially pairwise independent} (e.p.i.)\footnote{Here, following \cite[Definition 2.7]{sun2006exact}, essentially pairwise independent means that for $\mu$-almost every $u \in I$ and $\mu$-almost every $v \in I$, the processes $(W^u,\xi^u)$ and $(W^v,\xi^v)$ are independent.}, and identically distributed random variables $(W^u, \xi^u)_{u \in I}$, such that  for each $u \in I$, the process $W^u = (W_t^u)_{0 \leq t \leq T}$ is a $d$-dimensional Brownian motion supported on the probability space $(\Omega, \cF, \P)$, and $\xi^u$ represents the starting wealth of agent $u$. Suppose that in addition to $(W^u)_{u \in I}$, the probability space $(\Omega, \cF, \P)$ supports the independent one-dimensional Brownian motion $W^*$. 

\begin{remark} \label{rmk: fubini property}
    By \cite[Lemma 2.3]{sun2006exact}, we have the usual Fubini property on the rich product space $(I \times \Omega, \cI \boxtimes \cF, \mu \boxtimes \P)$, i.e, we are free to exchange order of integrations.
    That is, given a measurable and integrable function $f$ on $(I \times \Omega, \cI \boxtimes \cF, \mu \boxtimes \P)$ we can write
    \begin{equation*}
        \int_{I\times \Omega}f(u,\omega)\mu\boxtimes \P(\d \omega,\d u) = \int_I\E[f(u)]\mu(\d u) = \E\Big[\int_If(u)\mu(\d u)\Big].
    \end{equation*}
    This will be used often in the proof without further mention of \cite[Lemma 2.3]{sun2006exact}.
    Moreover, we will write $$\mu(\d u) \equiv \d u$$ to lighten the notation.
\end{remark}
Let $\F^u$ denote the completion of the filtration generated by $(W^u, W^*,\xi^u)$, and let $\F$ denote the completion of filtration generated by $((W^u)_{u \in I}, W^*, (\xi^u)_{u\in I})$. 
As above, we assume to be given a continuum of stocks $S^u$ with dynamics 
\begin{equation*}
    \d S^u_t = \mathrm{diag}(S^u_t)(\mu^u_t\d t + \sigma^u_t\d W^u_t + \sigma^{*u}_t\d W^*_t),\quad u\in I
\end{equation*}
so that the wealth process for agent $u$ when employing strategy $\pi$ follows the dynamics
{\color{black}
\begin{align}
\label{eqn:mf-portfolio}
   \d X_t^u &= \pi_t \cdot \left(\Sigma^u_t\theta^u_t \d t  + \sigma_t^u\d W_t^u  +   \sigma_t^{*u} \d W_t^* 
   \right), \quad X^u_0 = \xi^u
\end{align}}
{\color{black}
where $\Sigma$, $\theta$, $\sigma$ and $\sigma^*$ are $\mathcal{B}([0,T])\otimes\cI \boxtimes \cF$-measurable stochastic processes on $[0,T]\times I \times \Omega$, bounded uniformly in $u\in I$, with
\begin{equation*}
    \Sigma^u_t := (\sigma^u_t, \sigma^{*u}_t)\quad \text{and}\quad \theta^u_t:= {\Sigma^u_t}^\top\big(\Sigma_t^u{\Sigma^u_t}^\top\big)^{-1}\mu^u_t,
 \end{equation*} 
 with $\Sigma_t^u{\Sigma^u_t}^\top$ assumed to be uniformly elliptic, and where for almost every $u\in I$, $\sigma^u,\sigma^{*u}$ and $\mu^u$ are $\F^u$--predictable.
} 
We finally assume that $(\sigma^u)_{u \in I}$, $(\sigma^{*u})_{u \in I}$ and $(\mu^u)_{u \in I}$ are e.p.i. and identically distributed.

\begin{definition}
    A strategy profile is a family $(\pi^u)_{u \in I}$ of $\F^u$-progressive processes taking values in $\R^d$ and such that $(u, t, \omega)\mapsto\pi^u$ is $\cI\otimes \Bc([0,T])\otimes \Fc$--measurable.
\end{definition}

Let the mapping  $\eta: I \rightarrow (0,1)$ be $\cI$-measurable and bounded away from zero uniformly in $u$. 
Assume that the agent $u$ is an exponential utility maximizer with risk aversion parameter $\eta^u$
and is additionally concerned with the performance of their peers.
The \emph{interaction} among the continuum of agents will be modeled by a graphon, which is a symmetric and measurable function
\begin{equation*}
    G:I\times I\to I.
 \end{equation*} 
Throughout the paper, we fix a graphon $G$.
The utility function for a representative agent $u$ is similar in form to that of \eqref{eqn:utility-n-agent}. 
In particular, let $\F^* := (\cF_t^*)_{t \in [0,T]}$ denote the $\P$--completion of the filtration generated by $W^*$.
Given $u \in I$, consider the utility maximization problem 
\begin{align}\begin{split}
    \label{eq:graphon-obj}
    V_0^{u,G} &= V_0^{u,G}\left( (\pi^v )_{v \neq u}\right) \\
    & := \sup_{\pi^u \in \cA^{G}} \E\left[-\exp\left(-\frac{1}{\eta_u}\left(X_T^{u,\pi^u} - \E\Big[ \rho\int_I X_T^{v, {\pi}^v} G(u,v) \d v \big|\cF_T^* \Big] \right)\right) \right].
\end{split}\end{align}
The set of admissible strategies $\Ac^G$ in the infinite population game is defined as:
\begin{definition}\label{defn:graphon adm strat}
    Let $u \in I$ and let $A_u$ be a closed convex subset of $\R^d$. 
    A strategy profile $(\pi^u)_{u \in I}$ is admissible if 
    $\mu$--almost every $u \in I$, it holds $\pi^u\in \H^2(A^u,\F^u)$ and $\int_I\|\pi^u\|_{\H^2(A^u, \F^u)}\d u<\infty$. 
\end{definition}
Taking inspiration from the theory of mean field games, see e.g. \citeauthor*{MR3752669} \cite{MR3753660,MR3752669} or \citeauthor*{MR2295621} \cite{MR2295621}, we are interested in \emph{graphon Nash equilibria} defined as follow:
\begin{definition}
    A family of admissible strategy profiles $(\widetilde \pi^u)_{u\in I}$ is called a graphon Nash equilibrium if for $\mu$--almost every $u$ the strategy $\widetilde \pi^u$ is optimal for \eqref{eq:graphon-obj} with $(\pi^v)_{v\neq u}$ replaced by $(\widetilde \pi^v)_{v\neq u}$.
    That is, 
    \begin{equation*}
        V_0^{u,G}\left( (\widetilde\pi^v)_{v \neq u}\right)  :=  \E\left[-\exp\left(-\frac{1}{\eta^u}\left(X_T^{u,\widetilde\pi^u} - \E\Big[ \rho\int_I X_T^{v, {\widetilde\pi}^v} G(u,v) \d v \big|\cF_T^* \Big] \right)\right) \right].
    \end{equation*}
\end{definition}

\subsection{Main results}
\label{sec:main.results}
Let us now present the main results of this work.
These are essentially the existence of graphon games, the convergence of the finite population game to the graphon game and a new notion of competition--indifference capital.

\subsubsection{Existence of the graphon game}
We will begin with the solvability of the graphon utility maximization problem.
Existing results on well--posedness of graphon games largely focus on linear quadratic games or static games, see e.g. \citeauthor*{aurell2021stochastic} \cite{aurell2021stochastic} and \citeauthor*{carmona2022stochastic} \cite{carmona2022stochastic}; we also refer to the more recent works by \citeauthor*{Lacker-Soret22} \cite{Lacker-Soret22} and \citeauthor*{Bayr-Wu-Zhang22} \cite{Bayr-Wu-Zhang22} for more general settings.
Moreover, the case of games with common noise has remained untouched.
The existence result given here  relies on general solvability of graphon BSDEs and FBSDEs discussed in the final section of the paper.

{\color{black}
\begin{theorem}
\label{thm:game.existence}
    Assume that $\xi^u \in L^2(\mu\otimes \P)$.
    Then the following hold:
    \begin{itemize}
        \item[(i)] If $A^u=\R^d$ for all $u \in I$, and $\rho$ satisfies $\rho<\e^{-(2\|\theta\|_\infty + \frac12)T}(2\|\Sigma\|_\infty\vee\|\theta\|_\infty)^{-1}$,
        then the graphon game admits a graphon Nash equilibrium.
        \item[(ii)] If $\sigma^{*u}=0$ for all $u\in I$, then the graphon game admits a graphon Nash equilibirum.
    \end{itemize}
\end{theorem}}
In the existence \Cref{thm:game.existence} above, we consider two cases.
The first one is the common noise case.
Here, we make the simplifying assumption that the strategies are unconstrained.
This is a standard assumption in the literature.
We additionally require the competition parameter $\delta$ to be sufficiently small.
The second case $(ii)$ is the non--common noise case.
Here, existence is obtained in full generality.

\subsubsection{Convergence}
The second result states that as $n \rightarrow \infty$, the $n$-agent problem \emph{converges in the strong sense} to the graphon problem, of course given some link between the sensitivity parameters $(\lambda_{ij})_{1 \leq i,j \leq n}$ of the $n$-agent problem and its counterpart $G(u,v)$ in the graphon problem.
Essentially, we will assume below that $(\lambda_{ij})_{1\leq i,j\le n}$ forms the adjacency matrix of a (random) graph converging in an appropriate sense to the graph represented by the graphon $G$. 
See for instance \citeauthor*{Lovasz} \cite{Lovasz} and \citeauthor*{Lovasz-Szegedy06} \cite{Lovasz-Szegedy06} for extensive accounts on convergence of graphs an link between graphs and graphons.  Here we remind the readers the definitions of cut metric, which will be used in the main results.
\begin{definition}
The cut norm for a graphon $G$ is defined by 
\[\|G\|_{\Box} := \sup_{E,E' \in \Bc_I}\Big|\int_{E \times E'} G(u,v) \d u \d v \Big|, \]
and the corresponding cut metric for two graphons $G_1$ and $G_2$ is defined by $d_{\Box}(G_1,G_2):=\|G_1 - G_2 \|_{\Box}$.
\end{definition}
Although $\|\cdot\|_\Box$ is not exactly a norm, we can make it one by identifying graphons which agree almost everywhere.
We will also consider the usual $\L^2$ norm on graphons, which is defined as
\begin{equation*}
    \|G\|_2:=\bigg( \int_{I\times I}|G(u,v)|^2\diff u\diff v\bigg)^{1/2}.
\end{equation*}
Thus, we make the following assumptions:
\begin{condition}\label{cdn:G}
\begin{enumerate}
  \item[(1)]  There is a sequence $(\beta_n)_{n\ge1}$ in $\R_+$ such that $\lim_{n \rightarrow \infty} n\beta_n^2 =\infty$;
  \item[(2)] there exists 
  a sequence of graphons $(G_n)_{n\ge1}$ such that:
   \begin{enumerate}
    \item[(2a)] the graphons $G_n$ are step functions, i.e. they satisfy
       \begin{align*}
           &G_n(u,v) = G_n\Big(\frac{\ceil{nu}}{n},\frac{\ceil{nv}}{n}\Big) \quad \text{for}\quad (u,v) \in I \times I,\quad \text{and for every } n\in \N, 
       \end{align*}
       and it holds
       \[n\|G_n -G\|_{2} \xrightarrow[n\to\infty]{} 0, 
       \] 
       \item[(2b)]   $\lambda_{ij} = \lambda_{ji} = \mathrm{Bernoulli}(\beta_n G_n(\frac{i}{n},\frac{j}{n}))$ independently for $1 \leq i,j \leq n$, and independently of \\ $(\xi^u, \sigma^u, \theta^u, \sigma^{u*}, \eta^u)_{u\in I}$, $ W^{*}$, and $(W^u)_{u \in I}$.
   \end{enumerate}
\end{enumerate}
\end{condition}
The graphons $G_n$ introduced above are called step graphons, given that they are piecewise constant.
The conditions $(1)$ and $(2)$ are the important modeling conditions. 
By \cite[Theorem 11.22]{Lovasz}, $(2b)$ says that the graph on which the finite population game is written converges (in the cut metric) to a infinite population graph.
 $(2b)$ implicitly implies that $\beta_nG_n(\frac{i}{n}, \frac{j}{n}) \in [0, 1]$, and means that the finite population graph is a simple graph with weights $\{0,1\}$ depending on the outcome of a "coin toss".
 The parameter $\beta_n$ can be seen as a density parameter on the graph, our condition $(1)$ allows the graph to become more and more sparse as $n$ becomes large.
 In fact, we have in mind the situation $\lim_{n \rightarrow \infty} \beta_n = 0$. 

Before stating the results, we start by putting the $n$-agent problem and the graphon problem in the same probabilistic setting. 
\begin{remark}\label{rebranding} Let us re-brand the sequence of $d$-dimensional Brownian motions $(W^i)_{i\in\{ 1,\dots, n\}}$ from \Cref{sec: n-agent} by $(W^{\frac{i}{n}})_{i \in\{ 1,\dots, n\}}$, so that the completion of the natural filtration generated by $(W^{\frac{i}{n}})_{i\in\{ 1,\dots, n\}}$ and $W^*$ is now a subset of $\F$. Consequently, all indices $i \in \N$ that appeared in \Cref{sec: n-agent} should be interpreted as $\frac{i}{n}$. The coefficients for the price evolution in the $n$-agent game, namely, $(\sigma^{i},\sigma^{*i}, \theta^{i})_{i\in\{1, \dots, n\}}$ which are now $(\sigma^{\frac{i}{n}},\sigma^{*\frac{i}{n}},\theta^{\frac{i}{n}})_{i\in\{1,\dots, n\}}$ after this re-branding, should obey the same conditions imposed upon $(\sigma^u, \sigma^{u*}, \theta^u)_{u \in I}$, as stated in \Cref{sec: graphon  game}. To avoid unnecessarily complicated notations, we will keep the original indexing in the following sections. 
This re--branding will come up again in the proofs of the main convergence theorem.
\end{remark}

The following is the main contribution of this work.
It provides convergence of the heterogeneous $n$--player game to the graphon game.
{\color{black}
\begin{theorem}
\label{thm:main.limit}
    Let
    \Cref{cdn:G} 
    be satisfied, assume that that $\E[\e^{\frac{2\rho}{\eta^i\beta_n}|\xi^{i}| }]<\infty$ for all $(i,n)$ and $\xi^u \in L^2(\mu\otimes \P)$.
    Further assume that 
    one of the following two conditions is satisfied:
    \begin{itemize}
        \item[$(i)$] $A^u = \mathbb{R}^d$ for all $u \in I$ and $\delta$ satisfies $\delta < \e^{-(2\|\theta\|_\infty + \frac12)T}(2\|\Sigma\|_\infty\vee\|\theta\|_\infty)^{-1}$.
        \item[$(ii)$] $\sigma^{*u} =0$ for all $u \in I$.
    \end{itemize}
    If the $n$--agent problem \eqref{eq: n-agent obj} admits a Nash equilibrium $(\opt^{i,n})_{i \in \{1,\dots,n\}}$, then for each $i$, the control $\opt^{i,n}$ converges to $\tilde \pi^u$ for some $u$ and a graphon Nash equilibrium $(\tilde\pi^{u})_{u\in I}$ in the sense that, up to a subsequence,
    \begin{align}
    \label{eq:conv.statement}
         \|\widetilde{\pi}_t^{i,n} - \widetilde{\pi}_t^{\frac{i}{n}}\|^2  \xrightarrow[n\to\infty]{} 0 \quad \text{and}\quad \Big|V_0^{i,n}((\opt^{j,n})_{j\neq i}) - V_0^{\frac{i}{n},G}((\tilde{\pi}^v)_{v\neq \frac{i}{n}}) \Big| \xrightarrow[n\to\infty]{} 0 \quad \P\otimes \diff t\text{ a.s.}
    \end{align}
\end{theorem}
}
This result will follow as a consequence of a general propagation of chaos result for (quadratic) FBSDEs in non--homogeneous interaction.
These propagation of chaos results seem to be first of the kind, we devote \Cref{sec:gct} to these results.

{\color{black}
Before going any further, let us present an example where the above result becomes easy in that propagation of chaos is not needed, at least granted our characterization results to come in \Cref{rem:char.non.common} and \Cref{cor:graphon-bsde}.
This example deals with the case of a market with constant coefficients, and it will further motivate the analysis of random coefficients case done in this paper.
\begin{proposition}
\label{pro:constant.coef}
    Assume that for all $u\in I$, $A^u = \R^d$, $\sigma^{*u} = 0$ and $\sigma^u, \mu^u$ are deterministic measurable functions of time. 
    Let us consider a slight modification of the utility maximization problem \eqref{eq: n-agent obj}: $\lambda_{ii} \neq 0$, i.e., agent $i$ takes into account a weighted average of \textit{all} agents' terminal wealth as their benchmark. 
    Under this modification, the utility maximization problem for agent $i$ now reads
\begin{align}\begin{split}
\label{eq:n-agent-obj-modified}
 V_0^{i,n}
    &:=V_0^{i,n}((\pi^j)_{j\neq i})  \\
    &:= \sup_{\pi \in \R^d} \E \bigg[ -\exp\bigg\{-\frac{1}{\etai}\bigg(X_T^{i,\pi} - \frac{\rho}{n\beta_n} \sumall \lambda_{ij} X_T^{j,\pi^j}\bigg)\bigg\} \bigg].
    \end{split}
\end{align}
Then, for all $n\in\mathbb{N}$ there is an $n$--Nash equilibrium $(\widetilde \pi^{i,n})_{i \in \{1,\dots,n\}}$ satisfying
\begin{equation*}
    \sigma^i_t\widetilde \pi^{i,n}_t = \frac{n\beta_n}{n\beta_n - \rho\lambda_{ii}}\eta^i_t\theta^i_t \quad \forall (n,i)\in \N^\ast\times \{1,\dots,n\}\text{ and a.s. }  t,
\end{equation*}
Furthermore, there is a graphon Nash equilibrium $(\widetilde \pi^u)_{u \in I}$ satisfying
\begin{equation*}
    \sigma^u_t\widetilde \pi^u_t = \eta^u\theta^u_t\quad \text{a.e } (u,t)\in I\times[0,T].
\end{equation*}
In particular, $\tilde\pi^{i,n}$ and $\tilde\pi^u$ are deterministic and it holds
\begin{equation}
\label{eq:conv.statement.unconstrained.no.common}
    \|\sigma_t^{i}\widetilde\pi^{i, n}_t - \sigma_t^{\frac in}\widetilde\pi^{\frac in}_t\|  \le \frac{\rho\lambda_{ii}}{n\beta_n - \rho\lambda_{ii}}\|\eta^i\theta^i\|_\infty \quad \forall (n,i)\in \N^\ast\times \{1,\dots,n\}\text{ and a.s. }  t. 
\end{equation}
\end{proposition}
In addition to providing an easy way to prove convergence result, \Cref{pro:constant.coef} is interesting in that it shows that in the present heterogeneous game, when the coefficients are constant, the Nash equilibrium (both in the finite and the graphon games) are constant as well, at least up to the randomness of the graph.
This is in line with the homogeneous case studied by \citeauthor*{lacker2019mean} \cite{lacker2019mean} using PDE techniques and \citeauthor*{espinosa_touzi_2013} \cite{espinosa_touzi_2013} using BSDE techniques.
}

\subsubsection{Competition--indifferent capital}
To conclude this section on the presentation of our main result, we use the rich literature on exponential utility maximization to assess the effect of competition on an individual investor.
As said repeatedly, our results build on characterizations of the Nash and graphon equilibriums by system of (F)BSDEs.
However, in order to numerically simulate equilibria one still needs to simulate the solutions (notably the control process) of a high dimensional system of (F)BSDEs, or of McKean--Vlasov type equations.
And as is well--known in the numerical simulation literature, efficient simulation of the control process is much harder than that of the value process.
One might then wonder whether appropriately choosing the initial capital could make the investor indifferent between being concerned with the relative performance of their peers or not.
That is, denoting\footnote{In the definition of $J$, when $F=0$ we take the elements of $\Ac^i$ to be $\F^i$--progressive, since in this case the agent is not concerned with the performance (and thus investments) of other market participants.}
\begin{equation*}
    J^{i,n}(\xi^i,F): = \sup_{\pi \in \cA^i}\E\Big[-\exp\Big(-\frac{1}{\eta^i}(X^{i,\pi}_T -F) \Big) \Big] \quad \text{where}\quad X^{i,\pi}_0 = \xi^i,
\end{equation*}
we would like to compute $p^{i,n}$ such that
\begin{equation}
\label{eq:indif.price}
    J^{i,n}(\xi^i - p^{i,n},0) = J^{i,n}\Big(\xi^i, \rho \sumj \lambda^n_{ij}X^{j,\opt^{j,n}}_T\Big)
\end{equation}
where $(\opt^{i,n})_{i\in\{1,\dots,n\}}$ is a Nash equilibrium.
This is precisely the (spirit of the) utility indifference pricing of \citeauthor*{hodges1989optimal} \cite{hodges1989optimal}.
In the infinite population game, this indifference capital takes the form
\begin{equation}
\label{eq:indif.price.graphon}
    J^{u}(\xi^u - p^{u},0) = J^{u}\Big(\xi^u, \rho\E\Big[\int_IX^{v,\opt^v}_TG(u,v)\d v \Big|\cF^*_T\Big] \Big)
\end{equation}
with
\begin{equation*}
    J^u(\xi^u,F) := \sup_{\pi \in \cA^u}\E\Big[-\exp\Big(-\frac{1}{\eta^u}(X^{u,\pi}_T -F) \Big) \Big] \quad \text{where}\quad X^{u,\pi}_0 = \xi^u.
\end{equation*}
We thus have the following corollary:
\begin{corollary}
\label{cor:indifference}
    Under the conditions of \Cref{thm: n-bsde}, the competition--indifferent capital $p^{i,n}$ is given by
    \begin{equation*}
        p^{i,n} = \eta^i\log\Big(\frac{\gamma^{i,n}_0}{\gamma_0} \Big)
    \end{equation*}
    where $\gamma^{i,n}$ is the value process of the system \eqref{eq:intro.fbsde} and $(\gamma,\zeta, \zeta^*)$ solves the BSDE
    \begin{align*}
        \gamma_t &=  \int_t^T \left(- \begin{pmatrix} \zeta_s \\ \zeta_s^{*} \end{pmatrix} \cdot \theta_s^i - \frac{\eta^i}{2}|\theta_s^i|^2 + \frac{1}{2 \eta^i} \left|\left(I - P_s^i \right)\left( \begin{pmatrix} \zeta_s \\ \zeta_s^{*} \end{pmatrix} + \eta^i \theta_s^i  \right) \right|^2 \right)\d s - \int_t^T \zeta_s \cdot \d W_s^i - \int_t^T \zeta_s^{*} \d W_s^*
    \end{align*}

    Moreover, if the conditions of \Cref{thm:main.limit} are satisfied, then we have
    \begin{equation*}
        |p^{i,n} - p^{\frac{i}{n}}| \xrightarrow[n\to\infty]{} 0 
    \end{equation*}
    where $p^{u}$ is the competition--indifferent capital of player $u$ in the graphon game.
\end{corollary}
The gist here is that $p^{i,n}$ is given in terms of the value process of a system of BSDEs, so that an investor starting with capital $\xi^i-p^{i,n}$ (only) needs to simulate the control process of a one--dimensional BSDE in order to compute the optimal trading strategy.
\begin{proof}[Proof of \Cref{cor:indifference}]
    The proof starts with the general duality result of \citeauthor*{Delbaen-expo-uti} \cite{Delbaen-expo-uti} which asserts that
    \begin{equation*}
        \sup_{\pi \in \cA^i}\E\Big[-\exp\Big(-\frac{1}{\eta^i}(X^{i,\pi}_T - \rho \sumj \lambda^n_{ij}X^{j,\opt^{j,n}}_T) \Big) \Big] = - \exp\bigg(\frac{1}{\eta^i}\sup_{\Q \in \mathcal{Q}}\Big(\E_{\Q}\Big[ \rho \sumj \lambda^n_{ij}X^{j,\opt^{j,n}}_T \Big] - \xi^i - \eta^iH(\Q|\P) \Big) \bigg)
    \end{equation*}
    where $H(\Q|\P)$ is the relative entropy given by
    \begin{equation*}
        H(\Q|\P) := \begin{cases}
            \E_{\Q}\Big[\log\Big(\log \frac{\d \Q}{\d \P} \Big)\Big]\quad \text{if}\quad \Q \ll \P\\
            + \infty \quad \text{else}
        \end{cases}
    \end{equation*}
    and $\mathcal{Q}$ is the set of probability measures $\Q$ that are absolutely continuous with respect to $\P$, such that the stock price processes are $\Q$--local martingales and $H(\Q|\P)<\infty$.
    Applying this result to both sides of \Cref{eq:indif.price} yields
    \begin{align*}
        p^{i,n} &= \sup_{\Q\in \mathcal{Q}}\Big(\E_{\Q}\Big[ \rho \sumj \lambda^n_{ij}X^{j,\opt^{j,n}}_T \Big] - \xi^i - \eta^iH(\Q|\P) \Big) - \sup_{\Q \in \mathcal{Q}}\Big(- \xi^i - \eta^iH(\Q|\P) \Big)\\
        &= \eta^i\log\Big(- \sup_{\pi \in \cA^i}\E\Big[-\exp\Big(-\frac{1}{\eta^i}(X^{i,\pi}_T - \rho \sumj \lambda^n_{ij}X^{j,\opt^{j,n}}_T) \Big) \Big]\Big) - \eta^i\log\Big(- \sup_{\pi \in \cA^i}\E\Big[-\exp\Big(-\frac{1}{\eta^i}X^{i,\pi}_T  \Big) \Big]\Big)\\
        &=\eta^i\log\Big(\frac{\gamma^{i,n}_0}{\gamma_0} \Big)
    \end{align*}
    where the latter equality follows by \Cref{thm: n-bsde} and \cite[Theorem 7]{hu2005utility}.
    The above argument also shows that $p^u = \eta^u\log(\gamma^u_0/\gamma_0^\prime)$ where $Y^u$ satisfies \Cref{eqn:BSDE-graphon} and $(\gamma\prime, \zeta\prime, \zeta^{*\prime})$ solves

    \begin{align*}
    \gamma_t^\prime =  \int_t^T \left(-\begin{pmatrix} \zeta_s^\prime \\ \zeta_s^{*\prime} \end{pmatrix} \cdot \theta_s^u - \frac{\eta^u}{2}|\theta_s^u|^2 + \frac{1}{2 \eta^u} \left|\left(I - P_s^u \right)\left( \begin{pmatrix} \zeta_s' \\ \zeta_s^{*'} \end{pmatrix} + \eta^u \theta_s^u  \right) \right|^2 \right)\d s - \int_t^T \zeta_s' \cdot \d W_s^u - \int_t^T \zeta_s^{*\prime} \d W_s^*.
    \end{align*}
    The convergence statement therefore follows from \Cref{thm:main.limit}.
\end{proof}

The rest of the paper is dedicated to the proofs of the convergence and existence results.
\section{Characterizations of the utility maximization games}\label{sec:charac}
This section provides characterizations of the Nash equilibria of the two games presented above in terms of solutions of backward SDEs.
These characterizations will play a key role in the proofs of our main results.
\subsection{{\color{black}F}BSDE characterization of the \texorpdfstring{$n$}--agent problem}
\label{sec:charac.bsde}
The following theorem provides a {\color{black}F}BSDE characterization for the $n$-agent utility maximization problem \eqref{eq: n-agent obj}.
In particular, it expresses the Nash equilibrium and the associated utilities as functions of solutions to a system of (quadratic) FBSDEs.
This is an extension of the main result of \citeauthor*{espinosa_touzi_2013} \cite{espinosa_touzi_2013} to the case where both common noise and idiosyncratic noise is considered.
In the statement below and throughout the paper, we denote by $P_t^i(\zeta)$ the projection of a vector $\zeta$ onto the constraint set $\Sigma^i_tA^i$.
Also recall the notation $\overline{X}_t^i := \sum_{j \neq i}\lambda_{ij}^nX_t^{j,\tilde{\pi}^j}$.
\begin{theorem}
\label{thm: n-bsde}
    Assume that $\E[\e^{\frac{2\rho}{\eta^i\beta_n}|\xi^{i}| }]<\infty$.
    If the $n$--player game admits a Nash equilibrirum $(\opt^{i,n})_{i\in \{1,\dots,n\}}$,
    then it holds
    \begin{align}
        \opt_t^{i,n} &= \left(\Sigma_t^i {\Sigma_t^i}^{\top} \right)^{-1}\Sigma_t^i P_t^i\left( \begin{pmatrix} \zeta_t^{ii} \\ \zeta_t^{*i} \end{pmatrix} + \eta^i \theta_t^i  \right) \diff t\otimes \P\text{--a.s.} \quad \text{and}\quad  V^{i,n}_0((\opt^{j,n})_{j\neq i}) = -\e^{-\frac{1}{\eta^i}(\xi^i - \rho\overline{\xi}^i - \gamma_0^i)}\;\; \forall i \in \{1,\dots,n\}
    \end{align} 
    with $(X^i, \gamma^i, \zeta^{ij},\zeta^{*i}) \in \S^1(\R, \F^n)\times \S^1(\R, \F^n)\times  \H^2_{loc}(\R^{d},\F^n) \times \H^2_{loc}(\R,\F^n)$ for all $(i,j)\in \{1,\dots,n\}^2$ solving the FBSDE
\begin{align}
\notag
    \d \gamma_t^i 
    &= \bigg( \begin{pmatrix} \zeta_t^{ii} \\ \zeta_t^{*i} \end{pmatrix} \cdot \theta_t^i + \frac{\eta^i}{2}|\theta_t^i|^2   -\frac{1}{2\eta^i}\sumj|\zeta_t^{ij}|^2  -\frac{1}{2\eta^i}\Big|(I-P_t^i)\left( \begin{pmatrix} \zeta_t^{ii} \\ \zeta_t^{*i} \end{pmatrix} + \eta^i \theta_t^i  \right)\Big|^2 \bigg) \d t + \sumall \zeta_t^{ij} \cdot \d W_t^j + \zeta_t^{*i} \d W_t^*, \quad \P\text{--a.s.} \\\label{eq:intro.fbsde}
    \gamma_T^i &= \rho(\bar{X}_T^i - \bar{\xi}^i)\\\notag
    \d X^{i}_t &= \opt^{i,n}_t \cdot \left( \Sigma_t^i \theta^i_t\d t + \sigma_t^i \d W^i_t + \sigma^{*i}_t\d W^*_t \right), \quad X^{i}_0 = \xi^i.
\end{align}
\end{theorem}
The reader might wonder why our characterizing equation is a multidimensional coupled FBSDE in contrast to BSDEs usually derived in the literature, see for instance \cite{espinosa_touzi_2013,Frei-Reis11}.
We can achieve a characterization by a BSDE by "shifting" the value process $\gamma^i$ and through introducing a function $\uppsi_t:\mathbb{R}^n\to \mathbb{R}^n$ allowing to decouple the FBSDEs \eqref{eq:intro.fbsde} into the BSDE \eqref{eqn: n-bsde} given in the next corollary.


\begin{corollary}
\label{corollary.n.bsde}
    Assume that
    \begin{equation}
    \label{cd:lambda}
        \sum_{j \neq i}\lambda_{ij}^n \in [0,1].
    \end{equation}
    If the $n$--player game admits a Nash equilibrirum $(\opt^{i,n})_{i\in \{1,\dots,n\}}$,
    then it holds
\begin{align*}
    {\Sigma_t^i}^{\top}\opt^{i,n}_t = P_t^i\left(\begin{pmatrix}Z_t^{ii} \\ \uppsi_t(Z_t^*)^i  \end{pmatrix} + \eta_i\theta_t^i \right) := f_t^i(Z_t^{ii}, \uppsi(Z_t^*)^i)\quad\text{and}\quad V_0^{i,n}((\opt^{j,n})_{j\neq i}) = -e^{-\frac{1}{\eta^i}(\xi^i-\rho\overline{\xi}^i -Y_0^i)}
\end{align*}
with $(Y^i,Z^{ij},Z^{*i}) \in \S^1(\R,\F^n) \times \H^2_{loc}(\R^{d},\F^n) \times \H^2_{loc}(\R,\F^n)$ for all $(i,j)\in \{1,\dots,n\}^2$ solving the following $n$-dimensional BSDEs:
    \begin{align}\begin{split}
    \label{eqn: n-bsde}
        Y_t^i &= \int_t^T \bigg(-\frac{\eta^i}{2}\abs{\theta_s^i}^2 - \begin{pmatrix} Z_s^{ii} \\ \uppsi(Z_t^*)^i \end{pmatrix}\cdot \theta_s^i + \frac{1}{2\eta^i} \Big|(I-P_t^i)\left(\begin{pmatrix} Z_s^{ii} \\ \uppsi(Z_s^*)^i \end{pmatrix}+ \eta^i \theta_s^i \right)\Big|^2  \\
        & \quad + \frac{1}{2\eta^i}\sumj\abs{Z_s^{ij}+\rho\lambda_{ij}^n \sigma_s^j f_s^j(Z_s^{jj},\uppsi_s(Z_s^*)^j)}^2 + \sumj \rho\lambda_{ij}^n \paran{f_s^j(Z_s^{jj},\uppsi_s(Z_s^*)^j)\cdot \Sigma_s^j\theta_s^j} \Bigg) \d s \\
        & \quad - \sumall \int_t^T Z_s^{ij} \cdot \d W_s^j - \int_t^T Z_s^{*i} \d W_s^*, \quad \P\text{-a.s., } t \in [0,T]. 
    \end{split}
\end{align}
    where for every fixed $t \in [0,T]$, $\uppsi_t \equiv \uppsi(\zeta,\cdot)$ is the inverse of the mapping $\upphi(\zeta,\cdot):\R^n \to \R^n$ given by
    \begin{align*}
        \upphi^i_t(\zeta, \zeta^{*}) = \zeta^{i,*} - \sumj \rho\lambda_{ij}^n \sigma_t^{j*} \cdot (\Sigma_t^j {\Sigma_t^j}^{\top})^{-1} \Sigma_t^j P_t^j\left(\begin{pmatrix}\zeta^{jj} \\ \zeta^{j*} \end{pmatrix} + \eta^j\theta_t^j \right) \; \text{for all} \; (\zeta,\zeta^{*}) \in \R^{nd}\times \R^n,
    \end{align*}
    where with abuse of notation, $\upphi_t^i$ maps from $\R^n$ to $\R^n$ up to fixing a single trajectory of $(\Sigma_t^j)_{j\in \{ 1,\dots, n\}}$ and $(\theta_t^j)_{j\in \{ 1, \dots, n\}}$.
    Furthermore, for $n\ge3$, $\uppsi_t$ is Lipschitz--continuous with a constant depending on $n$. 
\end{corollary}
Observe that the dimension of the domain of the function $\uppsi$ depends on $n$.
Thus, $\uppsi_t$ will undoubtedly present a major obstacle when studying the limit of the game as $n \rightarrow \infty$. 
For instance, in the infinite population game this decoupling procedure does not seem to work. 
Furthermore, the condition \ref{cd:lambda} will also present an obstacle to the fact that we would like to consider the limit of the game on a relatively sparse graph. To avoid the above difficulties while studying the limit, we will rather work with the FBSDE \eqref{eq:intro.fbsde}.
\begin{remark}
\label{rem:char.non.common}
    In the absence of the common noise $W^*$ (i.e. when $\sigma^{*u}=0$ for all $u\in I$), the complications associated with $\uppsi_t$ discussed above vanish.
    In fact, the system of BSDEs in \Cref{corollary.n.bsde} takes the much simpler form
    \begin{align}
    \notag
        Y_t^i &= \int_t^T \bigg(-\frac{\eta^i}{2}\abs{\theta_s^i}^2 - Z_s^{ii}\cdot \theta_s^i + \frac{1}{2\eta^i} \abs{\paran{I - P_s^i}(Z_s^{ii} + \eta^i\theta_s^i)}^2 + \frac{1}{2\eta^i}\sumj\abs{Z_s^{ij}+ \rho\lambda_{ij}^n P_s^j\left(Z^{jj}_s + \eta^j \theta_s^j \right)}^2  \\\label{eqn:bsde-no-common}
        & \quad + \sumj \rho\lambda_{ij}^n  P_s^j\left(Z^{jj}_s + \eta^j \theta_s^j \right) \cdot \theta_s^j  \bigg) \d s - \sumall \int_t^T Z_s^{ij} \cdot \d W_s^j, \quad t \in [0,T]
\end{align}
and the equilibrium strategy now takes the form
\begin{align}
\label{eq:optim.pi.no.common}
    \opt^{i,n}_t &= (\sigma_t^{i})^{-1}P_t^i\paran{Z_t^{ii}+\eta^i\theta_t^i}, \quad \P\otimes \mathrm{d}t\text{--a.s.}
\end{align}
\end{remark}

\subsection{FBSDE characterization of the graphon problem}
\label{sec:charac.graph}
Similar to the $n$--player game just discussed, we will also derive (F)BSDE characterizations of the graphon game. 
This time, the characterization obtained is with respect to a system of (infinitely many) McKean--Vlasov (F)BSDEs. 
We will call these equations graphon (F)BSDEs to stress the fact that the dependence between the equations occurs through the graphon $G$.
As above, we use the notation $P^u_t(\zeta)$ for the projection of a vector $\zeta$ onto the constrain set $\Sigma^u_tA^u$.
\begin{proposition}\label{prop:graphon-bsde}
    Assume that $\xi^u \in L^2(\mu\otimes \P)$, that the following graphon FBSDE admits a solution $(X^u,Y^u,Z^u,Z^{*u}) \in  \S^2(\R, \F^u)\times \S^2(\R,\F^u)\times \H^2(\R^d,\F^u)\times \H^2(\R,\F^u)$ such that $(u,t,\omega)\mapsto X^u_t$ is measurable:
    \begin{align}\label{eqn:BSDE-graphon}
    \begin{cases}
        \d X_t^u = \tilde\pi_t^u \cdot \{\Sigma_t^u \theta_t^u \d t +  \sigma_t^u \d W_t^u  + \sigma_t^{*u} \d W_t^*\}\\
        \d Y_t^u =  \left(\begin{pmatrix} Z_t^u \\ Z_t^{*u} \end{pmatrix} \cdot \theta_t^u + \frac{\eta^u}{2}|\theta_t^u|^2- \frac{1}{2 \eta^u} \left|\left(I - P_t^u \right)\left( \begin{pmatrix} Z_t^u \\ Z_t^{*u} \end{pmatrix} + \eta^u \theta_t^u  \right) \right|^2 \right)\d t + Z_t^u \cdot \d W_t^u + Z_t^{*u} \d W_t^*  \quad \mu\otimes\P\text{--a.s.}\\
        Y_T^u = \E \Big[ \int_I \rho (X_T^{v}-\xi^v)G(u,v) \d v \big| \cF_T^* \Big],\quad X^u_0 = \xi^u,\quad 
    \opt_t^{u} = \left(\Sigma_t^u {\Sigma_t^u}^{\top} \right)^{-1}\Sigma_t^u P_t^u\left( \begin{pmatrix} Z_t^u \\ Z_t^{*u} \end{pmatrix} + \eta^u \theta_t^u  \right)\,\, \diff t\otimes\mu\otimes\P\text{--a.s.}
    \end{cases}
    \end{align} 
    Then, the graphon game described in \eqref{eq:graphon-obj} admits a graphon Nash equilibrium $(\tilde \pi^u)_{u\in I}$ such that for almost every $u\in I$ it holds  
    \begin{align}\label{eqn:graphon-opt}
        V_0^{u,G} = -\exp\bigg(- \frac{1}{\eta^u} \bigg(\xi^u - \int_I \rho\E[\xi^v]G(u,v) \d v - Y_0^u \bigg) \bigg)
    \end{align}
    and
    \begin{align}\label{eqn:graphon-strat}
        \opt_t^{u} = \left(\Sigma_t^u {\Sigma_t^u}^{\top} \right)^{-1}\Sigma_t^u P_t^u\left( \begin{pmatrix} Z_t^u \\ Z_t^{*u} \end{pmatrix} + \eta^u \theta_t^u  \right)\,\, \diff t\otimes\mu\boxtimes\P\text{--a.s.}
    \end{align}
\end{proposition}
The above result characterizes  the  graphon game with common noise in the sense that solvability of the game reduces to solvability of the system \eqref{eqn:BSDE-graphon}.
Moreover, the value function as well as the equilibrium strategies in the infinite population game are given explicitly in terms of solutions of \eqref{eqn:BSDE-graphon}.
In the case where there is no common noise, i.e. $\sigma^{*u}=0$ for almost all $u\in I$, the above result simplifies as follows:

\begin{corollary}
\label{cor:graphon-bsde}
    Assume that the graphon BSDE
    \begin{align}
    \notag
        \d Y_t^u &=  \bigg(\frac{\eta^u}{2}\abs{\theta_t^u}^2 + Z_t^{u}\cdot \theta_t^u - \frac{1}{2\eta^u} \abs{\paran{I - P_t^u}(Z_t^{u} + \eta^u\theta_t^u)}^2 
        - \E\Big[\int_I \rho P_t^v\paran{Z_t^{v}+\eta^v\theta_t^v}\cdot \theta^v_tG(u,v)\d v\Big] \bigg) \d t
    \\\label{eqn:BSDE-graphon.decoupled}
        &\quad +  Z_t^{u} \cdot \d W_t^u, \quad \mu\otimes \P\text{--a.s.,} \quad t \in [0,T],\quad\text{with}\quad Y_T^u = 0
    \end{align}
    admits a solution $(Y^u,Z^u)_{u\in I}$ such that $(u,t,\omega)\mapsto Z^u_t$ is measurable and $(Y^u,Z^u)\in \S^2(\R,\F^u)\times \H^2(\R^d,\F^u)$ for almost every $u\in I$.
    Then the graphon game described in \Cref{eq:graphon-obj} admits a graphon Nash equilibrium $(\tilde \pi^u)_{u\in I}$ such that for almost every $u\in I$ it holds  
    \begin{align}
    \label{eq:optim.sol.no.common}
        \opt^u_t = (\sigma_t^{u})^{-1}P_t^u\paran{Z_t^{u}+\eta^u\theta_t^u}\,\, \mathrm{d}t\otimes\mu\boxtimes\P\text{--a.s.} \text{ and}\quad V_0^{u,G} = -\exp\Big(- \frac{1}{\eta^u} \Big(\xi^u - \int_I \E[\rho \xi^v]G(u,v) \d v - Y_0^u \Big) \Big).
    \end{align}
\end{corollary}

\section{Proofs of existence and characterization results}
\label{sec:proofs}
The proof of \Cref{thm:main.limit} will be based on general propagation of chaos results that will be given in \Cref{sec:gct}, and the existence \Cref{thm:game.existence} is a consequence of existence of graphon BSDEs discussed in the final section of the paper where we present existence results for graphon (F)BSDEs.

\subsection{Proof of the existence \texorpdfstring{\Cref{thm:game.existence}}{\ref{thm:game.existence}}}
We will distinguish two cases:
The case with common noise and the case without.

$(i)$ Case with common noise: In this case, when $A^u=\R^d$ for all $u$, the FBSDE \eqref{eqn:BSDE-graphon} becomes
   \begin{align}\label{eqn:BSDE-graphon.proof}
    \begin{cases}
        \d X_t^u = b^u(t, Z^u_t, Z^{*u}_t) \d t + h_1^u(t, Z^u_t, Z^{*u}_t) \d W_t^u  + h_2^u(t, Z^u_t, Z^{*u}_t) \d W_t^*\\
        \d Y_t^u = -g^u(t, Z^u_t, Z^{*u}_t)\d t + Z_t^u \cdot \d W_t^u + Z_t^{*u} \d W_t^*  \quad \mu\otimes\P\text{--a.s.}\\
        Y_T^u = \E \Big[ \int_I \rho (X_T^{v} - \xi^v)G(u,v) \d v \big| \cF_T^* \Big],\quad X^u_0 = \xi^u, 
    \end{cases}
    \end{align} 
    with
    \begin{equation*}
        b^u(t,z,z^*) = \left(\Sigma_t^u {\Sigma_t^u}^{\top} \right)^{-1}\Sigma_t^u \left( \begin{pmatrix} z \\ z^{*} \end{pmatrix} + \eta^u \theta_t^u  \right)\Sigma^u_t\theta^u_t,\quad h^u_1(t,z,z^*) = \left(\Sigma_t^u {\Sigma_t^u}^{\top} \right)^{-1}\Sigma_t^u \left( \begin{pmatrix} z \\ z^{*} \end{pmatrix} + \eta^u \theta_t^u  \right)\cdot \sigma^u_t,
    \end{equation*}
    \begin{equation*}
        h^u_2(t,z,z^*) = \left(\Sigma_t^u {\Sigma_t^u}^{\top} \right)^{-1}\Sigma_t^u \left( \begin{pmatrix} z \\ z^{*} \end{pmatrix} + \eta^u \theta_t^u  \right) \cdot \sigma^{*u}_t\quad \text{and}\quad g^u(t,z,z^*) =\begin{pmatrix} z \\ z^{*} \end{pmatrix}\cdot\theta^u_t + \frac{\eta^u}{2}|\theta^u_t|^2.
    \end{equation*}
    In particular, given that the processes $\Sigma^u, \mu^u$ are bounded, the coefficients of this equation satisfy the conditions of \Cref{prop:exist.graphon.FBSDE}.
    Thus, it follows that \Cref{eqn:BSDE-graphon} admits a unique square integrable solution.
    Therefore, the result follows from \Cref{prop:graphon-bsde}.

\medskip

    $(ii)$ Case without common noise: When $\sigma^*=0$, the proof is similar.
In fact, it follows by \Cref{prop:exist.graphon.BSDE} that the graphon BSDE \eqref{eqn:BSDE-graphon.decoupled} admits a unique solution such that $(Y^u, Z^u)\in \mathbb{S}^\infty(\mathbb{F}^u, \mathbb{R}^d) \times \mathbb{H}_{\mathrm{BMO}}(\mathbb{F}^u, \mathbb{R}^d)$ for almost every $u\in I$ with $(u,t,\omega)\mapsto Z^u_t$ measurable and $\sup_u\|Z^u\|_{\H^2(\R^d,\F^u)}<\infty$.
Then, the result follows by Corollary \ref{cor:graphon-bsde}.

\subsubsection{Proof of \texorpdfstring{\Cref{pro:constant.coef}}{\ref{pro:constant.coef}}}

{\color{black}
Under the same assumptions given, the systems of FBSDEs \eqref{eq:intro.fbsde} characterizing the $n$-agent optimization problem simplify to the following
\begin{align}
\notag
    \d \gamma_t^i 
    &= \bigg( \zeta_t^{ii} \cdot \theta_t^i + \frac{\eta^i}{2}|\theta_t^i|^2   -\frac{1}{2\eta^i}\sumj|\zeta_t^{ij}|^2  \bigg) \d t 
    + \sumall \zeta_t^{ij} \cdot \d W_t^j, \qquad \P\text{--a.s.,}\,\, t \in [0,T] \\\label{eqn: n-bsde unconstrained.no.common}
    \gamma_T^i &= \rho(\bar{X}^i_T - \bar{\xi}^i) = \rho\sumall \lambda_{ij}^n \int_0^T \opt^{j,n}_t \cdot \sigma^j_t (\theta^j_t\d t + \d W^j_t), \\ \notag
    \d X_t^i &= \opt_t^{i,n} \sigma_t^i\{\theta_t^i \d t + \d W_t^i\}, \quad X_0^i = \xi^i,
\end{align}
with the equilibirum strategies given by
\begin{align*}
\sigma_t^{i}\opt^{i,n}_t = \zeta_t^{ii}+\eta^i\theta_t^i , \quad \P\otimes \mathrm{d}t\text{--a.s.}
\end{align*}
Let $Y_t^i = \gamma_t^i - \rho\sumall \lambda_{ij}^n \int_0^t \opt_s^{j,n} \sigma_s^j (\theta_s^j \d s + \d W_s^j)$ (recall that $\lambda^n_{ij} = \lambda_{ij}/n\beta_n$). Then we have $Y_T^i = 0$ and we can re-write the FBSDEs \eqref{eqn: n-bsde unconstrained.no.common} as 
\begin{align*}
Y_t^i = \int_t^T \Big( - \zeta_s^{ii}\cdot \theta_s^i - \frac{\eta^i}{2}|\theta_s^i|^2 + \frac{1}{2\eta^i}\sumj |\zeta_s^{ij}|^2 + \rho \sumall \lambda_{ij}^n(\zeta_s^{jj} + \eta^j\theta_s^j)\theta_s^j \Big) \d s - \sumall\int_t^T  \Big(\zeta_s^{ij} - \rho\lambda_{ij}^n (\zeta_s^{jj} + \eta^j\theta_s^j) \Big) \d W_s^j.
\end{align*}
Observe that choosing 
$$
\zeta_t^{ii} = \frac{\rho\lambda_{ii}^n}{1- \rho\lambda_{ii}^n}\eta^i\theta_t^i\quad \text{and}
\quad \zeta_t^{ij} = \frac{\rho\lambda_{ij}^n}{1- \rho\lambda_{jj}^n}\eta^j\theta_t^j
$$
make the stochastic integral in the above BSDE vanish, leaving $Y_t^i$ a deterministic process. 
Thus
\begin{align*}
\begin{cases}
    Y_t^i = \int_t^T \Big( - \zeta_s^{ii}\cdot \theta_s^i - \frac{\eta^i}{2}|\theta_s^i|^2 + \frac{1}{2\eta^i}\sumj |\zeta_s^{ij}|^2 + \sumall \lambda_{ij}^n(\zeta_s^{jj} + \eta_j\theta_s^j)\theta_s^j \Big) \d s, \\
    \zeta_t^{ii} = \frac{\rho\lambda_{ii}^n}{1- \rho\lambda_{ii}^n}\eta^i\theta_t^i, \quad \zeta_t^{ij} = \frac{\rho\lambda_{ij}^n}{1- \rho\lambda_{jj}^n}\eta^j\theta_t^j \quad \text{for} \quad i \neq j
\end{cases}
\end{align*}
is a solution to the above BSDE.

Similarly, the BSDE \eqref{eqn:BSDE-graphon.decoupled} characterizing the graphon game simplifies to 
\begin{align}\label{eqn: graphon-bsde unconstrained.no.common}
 Y_t^u =  \int_t^T \bigg(- \frac{\eta_u}{2}\abs{\theta_s^u}^2 -Z_s^{u}\cdot \theta_s^u 
        + \E\Big[\rho\int_I (Z_s^v + \eta^v\theta_s^v) \cdot \theta^v_sG(u,v)\d v\Big] \bigg) \d s - \int_t^T  Z_s^{u} \cdot \d W_s^u,
\end{align}
with the equilibirum strategy given by 
\begin{align*}
\sigma_t^u\opt_t^u = Z_t^u + \eta^u\theta_t^u.
\end{align*}
Using a change of measure argument, we can rewrite \eqref{eqn: graphon-bsde unconstrained.no.common} as 
\begin{align*}
    Y^i_t&=\int_t^T \bigg(- \frac{\eta^u}{2}\abs{\theta_s^u}^2
    + \E\Big[\rho\int_I (Z_s^v + \eta_v\theta_s^v) \cdot \theta^v_sG(u,v)\d v\Big] \bigg) \d s - \int_t^T  Z_s^{u} \cdot \big(\d W_s^u + \theta_s^u \d s \big)\\
    &= \int_t^T \bigg(- \frac{\eta^u}{2}\abs{\theta_s^u}^2 + \E\Big[\rho\int_I (Z_s^v + \eta^v\theta_s^v) \cdot \theta^v_sG(u,v)\d v\Big] \bigg) \d s - \int_t^T  Z_s^{u} \cdot \d W_s^{u, \Q},
\end{align*}
where $W_s^{u, \Q}$ is a Standard Brownian motion under a new measure $\Q$ such that $\frac{\d \Q}{\d \P} = \e^{\int_t^T - \theta_s^u \cdot \d W_s^u  - \frac12 \int_t^T |\theta_s^u|^2 \d u}$. 
Noting that $Z^u \in \H^2(\R^d, \F^u)$, taking conditional expectation with respect to $\Fc_t^u$ on both sides, we can now conclude that 
\begin{align*}
\begin{cases}
Y_t^u = \int_t^T \Big(-\frac{\eta^u}{2}|\theta_s^u|^2 + \E\big[\rho\int_I  \eta^v|\theta_s^v|^2 G(u,v) \d v \big] \Big) \d s, \\
Z_t^u = 0, \quad \sigma_t^u\opt_t^u = \eta^u\theta_t^u
\end{cases}
\end{align*}
is a solution to the BSDE \eqref{eqn: graphon-bsde unconstrained.no.common}. The convergence results \eqref{eq:conv.statement.unconstrained.no.common} thus follows from the boundedness of $\eta$ and $\theta_t$.

}

\subsection{Proofs for \texorpdfstring{\Cref{sec:charac.bsde}}{\ref{sec:charac.bsde}}}
We now present the proof of the characterization result for the $n$-player game. This section consists of 
the proof for \Cref{thm: n-bsde} and two auxiliary lemmas: \Cref{lem:H} and \Cref{lem:psi}.
\begin{proof}[Proof of \Cref{thm: n-bsde}]
    Assume that $(\opt^{i,n})_{i\in \{1,\dots,n\}}$ is a Nash equilibrium of the problem \eqref{eq: n-agent obj}. 
    First note that our assumptions on $\sigma_t^i$, $\sigma_t^{*i}$, and $\mu^i_t$ imply that $\overline{X}_t^i \in \L^2(\R,\cF_t^n)$. 
    Let $\Tc$ be the set of all $\F^n$--stopping times in $[0,T]$. Define the following family of random variables: 
    \[\Jc^{i,\pi}(\tau) := \E\Big[- \e ^{- \frac{1}{\eta^i}\left(\int_{\tau}^T\pi_s \cdot (\Sigma_s^i\theta_s^i \d s + \sigma_s^i \d W_s^i + \sigma_s^{*i} \d W_s^* ) - \rho(\overline{X}_T^i - \overline{\xi}^i)\right)} \big| \mathcal{F}_\tau^n \Big], \]
    and let 
    \[\Vc^i(\tau) := \esssup_{\pi \in \cA^i}\Jc^{i,\pi}(\tau)\text{ for all } \tau \in \Tc, \text{ so that } \Vc^i(0) = \e^{\frac{1}{\eta^i}(\xi^i - \rho\overline{\xi}^i)}V_0^{i,n}((\opt^j)_{j\neq i}). 
    \]
    Now let 
    \[\upbeta_{\tau}^{i,\pi} := \e^{\frac{1}{\eta^i}\int_0^\tau \pi_u \cdot (\Sigma_u^i\theta_u^i \d u + \sigma_u^i \d W_u^i + \sigma_u^{*i} \d W_u^* )}. \]
    Then it can be checked as in the proof of \cite[Lemma 4.13]{espinosa_touzi_2013} that for all $\pi \in \cA_i$,
    \begin{equation}
    \label{eq:supmart.prop}
        \upbeta_{\tau_1}^{i,\pi}\Vc^i(\tau_1) \geq \E\big[\upbeta_{\tau_2}^{i,\pi}\Vc^i(\tau_2)  \big| \Fc_{\tau_2}^n\big] \text{ for all stopping times } \tau_1 \leq \tau_2,
    \end{equation}
    and by \cite[Proposition I.3.14]{karatzas1991brownian}, the process $\Vc^i$ has a c\`adl\`ag modification again denoted $(\Vc^i_t)_{t\in[0,T]}$.
    Moreover, this process also satisfies \eqref{eq:supmart.prop}, so that for any $\pi \in \cA_i$, the process $\beta^{i,\pi}\Vc^i$ is a $\P$-supermartingale. 
    Now the definition of Nash equilibrium implies that $\opt^{i,n}$ is optimal for agent $i$. In other words, 
    \begin{equation*}
        \Vc_0^i = \sup_{\pi \in \cA_i} \E\big[-\e^{-\frac{1}{\eta^i}(X_T^{i,\pi} - \xi^i - \rho(\overline{X}_T^i - \overline{\xi}^i))} \big] = \E\big[-\e^{-\frac{1}{\eta^i}(X_T^{i,\opt^{i,n}} - \xi^i - \rho(\overline{X}_T^i - \overline{\xi}^i))} \big].
    \end{equation*}
    The above implies that $\upbeta^{i,\opt}\Vc^i$ is a $\P$-martingale, where we write $\upbeta^{i,\opt}$ as a shorthand notation for $\upbeta^{i,\opt^{i,n}}$.
    Denote  $\tilde{M}^i:=\upbeta^{i,\tilde{\pi}}\Vc^i$. 
    We now proceed to show that the adapted and continuous process 
    \begin{equation}
    \label{eqn:gamma}
        \gamma_t^i = X_t^{\opt^{i,n}} - \xi^i + \eta^i \ln(-\tilde{M}_t^i), \quad t \in [0,T]
    \end{equation}
    solves a BSDE. 
    Note already that by definition of $\Vc^i_t$ and $\tilde{M}^i_t$, we have 
    \[
        V_0^{i,n}((\opt^{j,n})_{j\neq i}) = -\e^{-\frac{1}{\eta^i}(\xi^i - \rho\overline{\xi}^i - \gamma_0^i)}.
    \]
    This proves the representation of $V^{i,n}_0((\opt^{j,n})_{j\neq i})$. 

\medskip

    We first need to check that $\gamma^i$ is indeed in $\S^1(\R,\F^n)$. 
    {\color{black}
    On the one hand, using Jensen's inequality, we have
    \begin{align} \label{bound:gamma} 
        \frac{1}{\eta^i}\E\big[X_T^{i,\opt^{i,n}} - \xi^i - \rho(\overline{X}_T^i - \overline{\xi}^i) \big| \Fc_t^n \big] \leq \ln(-\tilde{M}_t^i).
    \end{align}
    On the other hand, by definition of $\mathcal{V}^i$, we have $-\tilde{M}_t^i = -\upbeta^{i,\opt}_t\mathcal{V}^i_t \le \upbeta^{i,\opt}_t\E[\e^{\frac{\rho}{\eta^i}(\overline{X}_T^i - \overline{\xi}^i)}\mid \mathcal{F}_{t}^n]$.
    Thus, using the inequality $\ln(x)\le x$ we have
    \begin{align}
    \notag
        \ln(-\tilde M^i_t) &\le \ln(\upbeta^{i,\opt}_t) + \ln\Big( \E[\e^{\frac{\rho}{\eta^i}(\overline{X}_T^i - \overline{\xi}^i)}\mid \mathcal{F}_{t}^n]\Big)\\
        \label{eq:log.rep.M}
        &\le \frac{1}{\eta^i}\int_0^t \opt^{i,n}_u \cdot (\Sigma_u^i\theta_u^i \d u + \sigma_u^i \d W_u^i + \sigma_u^{*i} \d W_u^* ) + \E\big[\e^{\frac{\rho}{\eta^i}(\overline{X}_T^i - \overline{\xi}^i)}\mid \mathcal{F}_{t}^n\big].
    \end{align}
    {\color{black}
    Now, combining this with \Cref{bound:gamma} and the definition of $\upbeta_t^{i, \tilde{\pi}}$, we obtain
    \begin{align*}
         \E\Big[\sup_{t\in [0,T]}|\ln(-\tilde M^i_t)|\Big]
         &\leq \E\Big[ \sup_{t\in [0,T]} \frac{1}{\eta_i} \E \big[|X_t^{\tilde{\pi}^{i,n}} - \xi^i - \rho (\overline{X}_T^i - \overline{\xi}^i) | \big| \Fc_t^n \big] \Big] \\
         &\quad + \E\big[\sup_{t\in [0,T]} \frac{1}{\etai}| X_t^{\tilde{\pi}^{i,n}} - \xi^i | \big] + \E\Big[\sup_{t\in [0,T]} \E\big[\e^{\frac{\rho}{\eta^i}(\overline{X}_T^i - \overline{\xi}^i)}\mid \mathcal{F}_{t}^n\big] \Big] \\ 
         &\leq \frac{1}{\eta_i}\E\big[\sup_{t\in [0,T]}  \E[ |X_T^{ \tilde{\pi}^{i,n}}| \big| \Fc_t^n] \big] + \E\big[\sup_{t\in [0,T]} |X_t^{ \tilde{\pi}^{i,n}}| \big] + 2 \E[\xi^i]\\
         &\quad +  \E\Big[ \sup_{t\in [0,T]} \frac{1}{\eta_i} \E \big[|\rho (\overline{X}_T^i - \overline{\xi}^i) | \big| \Fc_t^n \big] \Big] + \E\Big[\sup_{t\in [0,T]} \E\big[\e^{\frac{\rho}{\eta^i}(\overline{X}_T^i - \overline{\xi}^i)}\mid \mathcal{F}_{t}^n\big] \Big].
    \end{align*}
    It is then sufficient to bound the last term. By Jensen's and H\"older's inequalities we have
    \begin{align*}
    \E\Big[\sup_{t\in [0,T]} \E\big[\e^{\frac{\rho}{\eta^i}(\overline{X}_T^i - \overline{\xi}^i)}\mid \mathcal{F}_{t}^n\big] \Big] \leq \frac{1}{n-1}\sumj \E\big[\e^{\frac{2\rho}{\etai}\frac{\lambda_{ij}}{\beta_n} \xi^j} \big]^{\frac12}\E\Big[ \sup_{t\in [0,T]} \E\big[\e^{\frac{2\rho}{\etai}\frac{\lambda_{ij}}{\beta_n}X_T^{\opt^j}}  \mid \Fc_t^n \big]^{\frac12} \Big].
    \end{align*} }
    By the admissibility condition on $\pi$, it follows that $\ln(-\tilde M^i) \in \mathbb{S}^1(\R, \F^n)$.}
    It thus follows that $\gamma^i \in \S^1(\R,\F^n)$ for every $i\in \{1,\dots,n\}$.
    For an arbitrary $\pi \in \Ac_i$, define
    \[
    M_t^{i,\pi} := \e^{-\frac{1}{\eta^i}(X_t^{\pi} - \xi^i - \gamma_t^i)} = \tilde{M}_t^i\e^{-\frac{1}{\eta^i}(X_t^{\pi}- X_t^{\opt^{i,n}})}.\]
    It follows from the same argument as in \cite[Theorem 4.7 2(b)]{espinosa_touzi_2013} that $M^{i,\pi}$ is a supermartingale. 
Now by \Cref{eqn:gamma}, Doob--Meyer decomposition and It\^o's formula, there is $( \zeta^i,\zeta^{*i})\in \H^2_{loc}(\R^{nd},\F) \times \H^2_{loc}(\R^d,\F)$ such that
\begin{align*}
    \d \gamma_t^i = - b_t^i \d t + \sumall \zeta_t^{ij} \cdot \d W_t^j + \zeta^{*i}_t \d W_t^*.
\end{align*}
We will proceed by first computing $b^i$,$\gamma^i$ and $\opt^i$, and next deriving the BSDEs satisfied by $(\gamma^i, \zeta^i, \zeta^{*i})$.

By It\^{o}'s formula, we have
\begin{align}\nonumber \label{eqn:ito-decomp}
    - \d \e^{-\frac{1}{\eta^i}(X_t^{\pi} - \xi^i - \gamma_t^i)} 
      &= \e^{-\frac{1}{\eta^i}(X_t^{\pi} - \xi^i - \gamma_t^i)} \cdot \Bigg\{ \frac{1}{\eta^i} \bigg(\left(\sigma_t^i \pi_t^i \right) \cdot \d W_t^i +\sumall \zeta_t^{ij} \cdot \d W_t^j + \left( \sigma_t^{*i} \cdot  \pi_t^i + \zeta_t^{*i} \right) \d W_t^* \bigg) \\ \nonumber
    &\qquad +  \frac{1}{\eta^i}\left(b_t^i + \pi_t^i \cdot \Sigma_t^i\theta_t^i \right) \d t + \frac{1}{(\eta^i)^2}\bigg(\sigma_t^{i}\pi_t^i \cdot \zeta_t^{ii} + \sigma_t^{*i}  \pi_t^i  \zeta^{*i}_t \bigg) \d t \\
    &\qquad-\frac{1}{2(\eta^i)^2}\bigg(|\sigma_t^{i}\pi_t|^2 + |\sigma_t^{*i} \pi_t|^2 
    + \sumall |\zeta_t^{ij}|^2 + |\zeta_t^{*i}|^2 \bigg)  \d t\Bigg\}.
\end{align}
Using the supermartingale property of $M^{i,\pi}$, the martingale property of $\tilde{M}^i$, together with \eqref{eqn:ito-decomp}, keeping in mind that $\Sigma_t^i:= (\sigma_t^i, \sigma_t^{*i})$, and writing $\zeta^i_t := (\zeta^{ii}_t, \zeta^{*i}_t)$, we get 
\begin{align}
\label{eq:const.b}
    b_t^i \leq \frac{1}{2\etai}\abs{{\Sigma_t^i}^{\top}\pi_t^i - (\zeta_t^i  + \eta^i\theta_t^i)}^2 + \frac{1}{2\eta^i}\sumj |\zeta_t^{ij}|^2 - \frac{\etai}{2}|\theta_t^i|^2 - \zeta_t^{i} \cdot\theta_t^i,   
\end{align}
    and 
\begin{align*}
    b_t^i =\frac{1}{2\eta^i}\abs{{\Sigma_t^i}^{\top} \tilde{\pi}_t^i - (\zeta_t^i  + \eta^i\theta_t^i)}^2 + \frac{1}{2\eta^i}\sumj |\zeta_t^{ij}|^2 - \frac{\eta^i}{2}|\theta_t^i|^2 - \zeta_t^{i} \cdot\theta_t^i.   
\end{align*}
Thus, $\opt^{i,n}_t$ minimizes the function (in $\pi^i$) on the right hand side of \eqref{eq:const.b}.
Therefore, 
we can express $\opt_t^{i,n}$ and $b_t^i$ as follow:
\begin{align}
    \opt_t^{i} &= \left( {\Sigma_t^i \Sigma_t^i}^{\top}\right)^{-1} {\Sigma_t^i} P_t^i\big( \zeta_t^i  + \eta^i\theta_t^i \big), \\ \nonumber
    b_t^i &= \frac{1}{2\eta^i} \mathrm{dist}\Big(\zeta_t^i + \eta_i\theta_t^i, \Sigma_t^i A^i \Big)^2 + \frac{1}{2\eta^i}\sumj |\zeta_t^{ij}|^2 - \frac{\eta^i}{2}|\theta_t^i|^2 - \zeta_t^{i} \cdot \theta_t^i.
\end{align}
Therefore, $(\gamma^i, \zeta^i,\zeta^{*i})\in \S^1(\R, \F)\times \H^2_{loc}(\R^{nd},\F) \times \H^2_{loc}(\R^d,\F)$ solves the BSDE
\begin{align}\begin{split}
\label{eq:def.b.gamma.bsde}
    \d \gamma_t^i 
    &= \bigg( \zeta_t^{i} \cdot \theta_t^i + \frac{\eta^i}{2}|\theta_t^i|^2   -\frac{1}{2\eta^i}\sumj|\zeta_t^{ij}|^2  -\frac{1}{2\eta^i}\Big|(I-P_t^i)\left(\zeta_t^i + \eta^i \theta_t^i \right)\Big|^2 \bigg) \d t\\
    &\quad + \sumall \zeta_t^{ij} \cdot \d W_t^j + \zeta_t^{*i} \d W_t^*, \\
    \gamma_T^i &= \rho(\overline{X}_T^i - \overline{\xi}^i) = \rho\sum_{j \neq i}{ \lambda_{ij}^n\int_{0}^{T} \opt^j_s \cdot \big\{  \Sigma_s^j \theta^j_s \d s + \sigma_s^{j} \d W_s^j  + \sigma_s^{j*} \d W_s^* \big\}
    }.
    \end{split}
    \end{align}
\end{proof}

\begin{proof}(of Corollary \ref{corollary.n.bsde})
    The proof of this corollary builds upon that of Theorem \ref{thm: n-bsde}, with exactly the same notation.
    Define the process
    $$Y_t^i := \gamma_t^i -\sum_{j \neq i}{ \rho\lambda_{ij}^n \int_{0}^{t} \opt^j_s \cdot \big\{  \Sigma_s^j \theta^j_s \d s + \sigma_s^{j} \d W_s^j  + \sigma_s^{j*} \d W_s^* \big\}} $$
    as well as
    \begin{align}
    \label{eq:def.upphi}
        Z_t^{ij} := \zeta_t^{ij}-\rho \lambda_{ij}^n \sigma_t^j \opt_t^j,\quad \text{and}\quad Z_t^{*i} := \upphi_t^i(\zeta_t^*) = \zeta_t^{*i} -\sumj \rho \lambda_{ij}^n \sigma_t^{j*} \cdot \opt_t^j.
    \end{align}
Here, $\upphi_t$ is a mapping from $\R^d$ to $\R^d$ defined component-wise as above. Moreover, notice that $Z_t^{ii} = \zeta_t^{ii}$ since $\lambda_{ii}^n = 0$, and that $\gamma_0^i = Y_0^i$.

The processes $(Y^i, Z^{ij}, Z^{*i})$ thus satisfies 
\begin{align*}
    Y_t^i &= \int_t^T \bigg(-\frac{\eta^i}{2}|\theta_s^i|^2 - \zeta_s^{i} \cdot \theta_s^i + \frac{1}{2 \eta^i} \Big|(I-P_t^i)\left(\zeta_t^i + \eta^i \theta_t^i \right)\Big|^2 + \frac{1}{2\etai}\sumj |\zeta_s^{ij}|^2 \\
    & \quad + \sumj \rho\lambda_{ij}^n(\opt_s^j \cdot \Sigma_s^j\theta_s^j) \bigg) \d s - \sumall \int_t^T Z^{ij}_s \cdot \d W_s^j - \int_t^T Z^{*i}_s  \d W_s^*.
\end{align*}
 By \Cref{lem:psi}, $\upphi_t$ has an inverse $\uppsi_t$, so that $\zeta^{*i}_t = \uppsi_t(Z^*_t)^i$.
 We can thus express the equilibrium strategy for player $i$ as 
 \begin{equation}
     {\Sigma_t^i}^{\top}\opt_t^i = P_t^i\left(\begin{pmatrix}Z_t^{ii} \\ \uppsi_t(Z_t^*)^i  \end{pmatrix} + \eta_i\theta_t^i \right) : = f_t^i(Z_t^{ii}, \uppsi(Z_t^*)^i), \quad t \in [0,T]
    \end{equation}
    and
    \[
        V_0^{i,n}((\opt^j)_{j\neq i}) = -\e^{\frac{1}{\eta^i}(\xi^i - \rho\overline{\xi}^i - \gamma_0^i)}.
    \]
    By construction, $(Y^i, Z^i,Z^{*i})\in \S^1(\R,\F^n) \times \H^2_{loc}(\R^{nd},\F^n) \times \H^2_{loc}(\R,\F^n)$ solves the BSDE \eqref{eqn: n-bsde}.
 \end{proof}

\begin{lemma}\label{lem:H}
    For any $t \in [0,T]$, fixed $\alpha \in \R^d$ and $\beta \in \R^{d + 1}$, the map 
    \begin{align*}
    H_{\alpha, \beta}(x) = x +\frac{1}{n-1}{\sigma_t^{*i}}^{\top}\left(\Sigma_t^i{\Sigma_t^i}^{\top}\right)^{-1}\Sigma_t^i \cdot P_t^i\left(\begin{pmatrix}\alpha \\ x \end{pmatrix} + \beta \right)
    \end{align*} 
    is a bijection on $\mathbb{R}$ for every $i$. Furthermore, its inverse is a contraction.
\end{lemma}

\begin{proof}
    Fix $t \in [0,T]$ and $i$. $H_{\alpha}(\cdot)$ is a bijection if and only if the map 
    \begin{align*}
    M^y(x) = y - \frac{1}{n-1}{\sigma_t^{*i}}^{\top}\left(\Sigma_t^i{\Sigma_t^i}^{\top}\right)^{-1}\Sigma_t^i \cdot P_t^i\left(\begin{pmatrix}\alpha \\ x \end{pmatrix} + \beta \right)
    \end{align*} 
    has a unique fixed point. 
    Notice that since the projection operator is $1$--Lipschitz,
    \begin{align}
        \abs{M^y(x)- M^y(x')} \leq \frac{1}{n-1}\Big|{{\sigma_t^{*i}}^{\top}\left(\Sigma_t^i{\Sigma_t^i}^{\top}\right)^{-1}\Sigma_t^i} \Big| \abs{x-x'}.
    \end{align}
    It is thus sufficient to show that $\Big|{{\sigma_t^{*i}}^{\top}\left(\Sigma_t^i{\Sigma_t^i}^{\top}\right)^{-1}\Sigma_t^i} \Big| < 1$. For notational convenience, let us omit all $t$ subscripts. First notice that $\Sigma^i{\Sigma^i}^{\top} = \sigma^{i}\sigma^{i} + \sigma^{*i} {\sigma^{*i}}^{\top}$. Using the Sherman-Morrison formula, we have that 
    \begin{align*}
    \left(\Sigma^i{\Sigma^i}^{\top}\right)^{-1} = \sigma^{-i}\sigma^{-i} - \frac{\sigma^{-i}\sigma^{-i}\sigma^{*i}{\sigma^{*i}}^{\top}\sigma^{-i}\sigma^{-i}}{1 + {\sigma^{*i}}^{\top}\sigma^{-i}\sigma^{-i}\sigma^{*i}},
    \end{align*}
    and 
    \begin{align*}
    {{\sigma^{*i}}^{\top}\left(\Sigma^i{\Sigma^i}^{\top}\right)^{-1}\Sigma^i} 
    &= \begin{bmatrix}{\sigma^{*i}}^{\top}\left(\Sigma^i{\Sigma^i}^{\top}\right)^{-1}\sigma^i & {\sigma^{*i}}^{\top}\left(\Sigma^i{\Sigma^i}^{\top}\right)^{-1}\sigma^{*i} \end{bmatrix} \\
    &=\begin{bmatrix}\left(1 - \frac{{\sigma^{*i}}^{\top}\sigma^{-i}\sigma^{-i}\sigma^{*i}}{1+ {\sigma^{*i}}^{\top}\sigma^{-i}\sigma^{-i}\sigma^{*i}} \right){\sigma^{*i}}^{\top}\sigma^{-i} &  \frac{{\sigma^{*i}}^{\top}\sigma^{-i}\sigma^{-i}\sigma^{*i}}{1+ {\sigma^{*i}}^{\top}\sigma^{-i}\sigma^{-i}\sigma^{*i}} \end{bmatrix}.
    \end{align*}
Thus
    \begin{align*}
    \Big|{{\sigma^{*i}}^{\top}\left(\Sigma^i{\Sigma^i}^{\top}\right)^{-1}\Sigma^i} \Big|
     = \frac{{\sigma^{*i}}^{\top}\sigma^{-i}\sigma^{-i}\sigma^{*i}}{1 + {\sigma^{*i}}^{\top}\sigma^{-i}\sigma^{-i}\sigma^{*i}} < 1,
    \end{align*}
where the last line follows from the fact that $\sigma^i$ is uniformly elliptic for every $i$.
We now proceed to show that the inverse of $H_{\alpha}$, which we denote by $H_{\alpha}^{-1}$, is a contraction. For $x \neq x'$, we have
\begin{align}\nonumber
    &\Big|x-x'+\frac{1}{n-1}{\sigma^{*i}}^{\top}\left(\Sigma^i{\Sigma^i}^{\top}\right)^{-1}\Sigma^i \cdot \left( P^i\left(\begin{pmatrix}\alpha \\ x \end{pmatrix} + \beta \right) - P^i\left(\begin{pmatrix}\alpha \\ x' \end{pmatrix} + \beta \right) \right)\Big|^2 \ \\ \nonumber
    & = \abs{x-x'}^2+\frac{2}{n-1}(x-x') \cdot {\sigma^{*i}}^{\top}\left(\Sigma^i{\Sigma^i}^{\top}\right)^{-1}\Sigma^i \cdot \left( P^i\left(\begin{pmatrix}\alpha \\ x \end{pmatrix} + \beta \right) - P^i\left(\begin{pmatrix}\alpha \\ x' \end{pmatrix} + \beta \right) \right) \\ \label{eqn:lipschitz-projection}
    & \qquad \qquad \quad\: + \frac{1}{(n-1)^2}\Big|{\sigma^{*i}}^{\top}\left(\Sigma^i{\Sigma^i}^{\top}\right)^{-1}\Sigma^i \cdot \left( P^i\left(\begin{pmatrix}\alpha \\ x \end{pmatrix} + \beta \right) - P^i\left(\begin{pmatrix}\alpha \\ x' \end{pmatrix} + \beta \right) \right)\Big|^2 \\ \nonumber
    & \geq \abs{x - x'}^2,
\end{align}
since the property of projection onto closed convex sets implies that the second term is nonnegative.
\end{proof}

\begin{lemma} \label{lem:psi}
    Consider the map $\upphi_t: \R^d \rightarrow \R^n$ introduced in the statement of  \Cref{thm: n-bsde} and defined component-wise below as
    \begin{align}\label{phi}
        \upphi_t^i(\zeta_t^{*}) = \zeta_t^{i,*} - \sumj \lambda_{ij}^n \sigma_t^{j*} \cdot (\Sigma_t^j {\Sigma_t^j}^{\top})^{-1} \Sigma_t^j P_t^j\left(\begin{pmatrix}Z_t^{jj} \\ \zeta_t^{j*} \end{pmatrix} + \eta^j\theta_t^j \right).   
    \end{align}
    Under the assumption that $\sum_{j \neq i}\lambda_{ij}^n \in [0,1]$, for $t \in [0,T]$,  $\upphi_t$ is a bijection on $\mathbb{R}^n$ and has an inverse that we denote by $\uppsi_t$. Furthermore, $\uppsi_t$ is measurable and Lipschitz--continuous with a constant that depends only on $n$ when $n \geq 3$.
\end{lemma}

\begin{proof}
    Omit all $t$ subscripts for notational convenience. 
    Let $Z^*$ and $\zeta^*$ denote the column vectors $(Z^{1,*},\dots,Z^{n,*})^\top$ and $(\zeta^{1,*},\dots,\zeta^{n,*})^\top$ respectively. 
    Further, let $\left[\sigma^{j*} \cdot (\Sigma^j {\Sigma^j}^{\top})^{-1} \Sigma^j P^j\left(\begin{pmatrix}Z^{jj} \\ \zeta^{j*} \end{pmatrix} + \eta^j\theta^j \right) \right]$ denote the column vector with the $j$-th component equal to  $\sigma^{j*} \cdot (\Sigma^j {\Sigma^j}^{\top})^{-1} \Sigma^j P^j\left(\begin{pmatrix}Z^{jj} \\ \zeta^{j*} \end{pmatrix} + \eta^j\theta^j \right)$. 
    By \Cref{eq:def.upphi}, we have
    \begin{align*}
        Z^* =  \zeta^* -\frac{1}{n-1}\Lambda  \left[\sigma^{j*} \cdot (\Sigma^j {\Sigma^j}^{\top})^{-1} \Sigma^j P^j\left(\begin{pmatrix}Z^{jj} \\ \zeta^{j*} \end{pmatrix} + \eta^j\theta^j \right)\right],
\end{align*}
where $\Lambda$ is the matrix $(\lambda_{ij})_{0\le i,j\le n}$ and 
\begin{align*}
    \frac{1}{n-1}\bar{1}^i(\Lambda+I)\left[\sigma^{j*} \cdot (\Sigma^j {\Sigma^j}^{\top})^{-1} \Sigma^j P^j\left(\begin{pmatrix}Z^{jj} \\ \zeta^{j*} \end{pmatrix} + \eta_j\theta^j \right)\right] + Z^{i,*} = H_{Z^{ii}, \eta_i\theta^i}(\zeta^{i,*})
\end{align*}
where $1^i$ denote the $n$-dimensional vector with $1$ at the $i$-th position and $0$'s in all other positions.  By \Cref{lem:H}, $H_{Z^{jj}, \eta_j\theta^j}(\zeta^{i,*})$ is invertible.  
Using \Cref{eq:def.upphi} again we have

\begin{align}\begin{split}\label{N}
    \zeta^{i,*} &= Z^{i,*} + \sumj \lambda_{ij}^n\sigma^{j*} \cdot (\Sigma^j {\Sigma^j}^{\top})^{-1} \Sigma^j \\
    &\qquad \qquad  P^j\left(\begin{pmatrix}Z^{jj} \\ H_{Z^{jj}, \eta_j\theta^j}^{-1}\left(\frac{1}{n-1}\bar{1}^j(\Lambda+I)\left[\sigma^{k*} \cdot (\Sigma^k {\Sigma^k}^{\top})^{-1} \Sigma^k P^k\left(\begin{pmatrix}Z^{kk} \\ \zeta^{k*} \end{pmatrix} + \eta_k\theta^k \right)\right] + Z^{i,*} \right) \end{pmatrix} + \eta_j\theta^j \right) \\&:= N^{i,Z^*}(\zeta^{*}).
\end{split}
\end{align}
We then proceed to showing that $N^{i, Z^*}(\zeta^{*})$ has a unique fixed point. Notice that for $x \neq y$, following the inequality in \eqref{eqn:lipschitz-projection}, 
\begin{align*}
    \abs{H_{Z^{jj}, \eta^j\theta^j}(x)-H_{Z^{jj}, \eta^j\theta^j}(y)}^2 \geq \paran{1 + \frac{1}{n-1}}^2 \Big|\left( P^j\left(\begin{pmatrix}Z^{jj} \\ x \end{pmatrix} + \eta^j\theta^j \right) - P^j\left(\begin{pmatrix}Z^{jj} \\ y \end{pmatrix} + \eta^j\theta^j \right) \right) \Big|^2,
\end{align*}
Thus for fixed $Z^{jj}$ and $\eta^j\theta^j$, the map $P^j\left(\begin{pmatrix}Z^{jj} \\ H^{-1}_{Z^{jj}, \eta^j\theta^j}(\cdot) \end{pmatrix} + \eta^j\theta^j\right)$ is $1/(1+ \frac{1}{n-1})$-Lipschitz. 
For $X,Y \in \mathbb{R}^d$ and $X \neq Y$,

\begin{align*}
    \abs{N^{i,Z^*}(X)-N^{i,Z^*}(Y)} &\leq \sumj \lambda_{ij}^n\frac{\abs{\tilde{\sigma}^{j*}}\sum_{k \neq j}\lambda_{jk}^n\abs{(X-Y)_k}} {n} + \sumj \lambda_{ij}^n\frac{\abs{\tilde{\sigma}^{j*}}\abs{X-Y}}{n} \\
    &\leq \frac{1}{n-1}\big|X-Y\big|.
\end{align*}
where the last inequality follows since $\sum_{j\neq i} \lambda_{ij}^n \in [0,1]$ for all $i$, and $\abs{\tilde{\sigma}^{j*}} < 1$ for all $j$ (see proof of  \cref{lem:H}). We can now conclude that for $n \geq 3$, $N^{i, Z^*}$ admits a unique fixed point which we denote by $\uppsi(Z^*)^i$, and that $\boldsymbol\zeta^* = \uppsi(\boldsymbol Z^*)$ is the unique solution to \Cref{phi}.

Finally we proceed to prove that $\uppsi$ is Lipschitz with a constant that depends only on $n$ when $n \geq 3$. From \eqref{N}, we have that for all $i$ and $n \geq 3$,

\begin{align*}
 &\abs{\uppsi(Z_1^*)^i - \uppsi(Z_2^*)^i} \leq \abs{(Z_1^*)^i - (Z_2^*)^i} +  \frac{2\abs{\uppsi(Z_1^*)^i - \uppsi(Z_2^*)^i}}{n(n-1)}+
 \frac{1}{n}\sup_{1 \leq j \leq n}\abs{(Z_1^*)^j-(Z_2^*)^j}.
\end{align*}
Then we have $\sup_{1 \leq j \leq n} \abs{\uppsi(Z_1^*)^j - \uppsi(Z_2^*)^j} \leq \frac{n-1}{n-2} \sup_{1 \leq i \leq n}\abs{(Z_1^*)^j-(Z_2^*)^j}$. 
Therefore, the function $\uppsi_t$ is Borel measurable. 
\end{proof}

\subsection{Proofs for \texorpdfstring{\Cref{sec:charac.graph}}{\ref{sec:charac.graph}}}
We now prove results pertaining to the characterization of the infinite population game.
These are direct consequences of the work of \citeauthor{hu2005utility} \cite{hu2005utility}. 
\begin{proof}[Proof of \Cref{prop:graphon-bsde}]
    Let $(X^u, Y^u, Z^u, Z^{*u})_{u\in I}$ solve \Cref{eqn:BSDE-graphon} with $(X^u,Y^u,Z^u,Z^{*u}) \in  \S^2(\R, \F^u)\times \S^2(\R,\F^u)\times \H^2(\R^d,\F^u)\times \H^2(\R,\F^u)$.
    Then, for almost every $u \in I$ the processes $(Y^u, Z^u, Z^{*u})$ solves the BSDE 
    \begin{align}
        \begin{cases}
            \d Y_t^u =  Z_t^u \cdot \d W_t^u + Z_t^{*u} \d W_t^*  + \left(\frac{\eta^u}{2}|\theta_t^u|^2 +\begin{pmatrix}Z_t^u \\ Z_t^{*u} \end{pmatrix} \cdot \theta_t^u - \frac{1}{2\eta^u} \bigg| \left(I - P_t^u \right)\left(\begin{pmatrix}Z_t^u \\ Z_t^{*u} \end{pmatrix} + \eta^u\theta_t^u \right) \bigg|^2 \right)\d t.\\
            Y_T^u = F
        \end{cases}
    \end{align}
    with $F:=\E \left[\rho \int_I  (X_T^{v,\tilde{\pi}^v}-\xi^v)G(u,v) \d v | \cF_T^* \right]$.
    Thus, it follows from \cite[Theorem 7]{hu2005utility} that $\opt^u$ given by \eqref{eqn:graphon-strat} is an optimal strategy for the utility maximization problem \eqref{eq:graphon-obj} while the value function satisfies \eqref{eqn:graphon-opt}.
    By linear growth of the projection operator, it follows that $\opt^u\in \H^2(A^u,\F^u)$ for almost every $u\in I$.
    And by measurability of $Z^u$, we have that $\widetilde\pi^u$ is measurable.
\end{proof}
\begin{remark}
    \cite[Theorem 7]{hu2005utility} assumes bounded terminal condition $F$, but examining the proof reveals that the boundedness assumption is needed only to guarantee existence of the BSDE and BMO property of $Z^u\cdot\diff W^u + Z^{*u}\diff W^*$.
\end{remark}
\begin{proof}[Proof of Corollary \ref{cor:graphon-bsde}]
    Let $(Y^u, Z^u)_{u\in I}$ solve \Cref{eqn:BSDE-graphon.decoupled} and introduce the processes
    \begin{equation*}
        \gamma_t^u := Y_t^u + \int_0^t\E\Big[\rho\int_I  P_s^v\paran{Z_s^{v}+\eta^v\theta_s^v} \cdot \theta^v_sG(u,v)\d v \Big] \d s.
    \end{equation*}
    Then, $(\gamma^u, Z^u)_u$ satisfies
    \begin{align}
    \notag
        \d \gamma_t^u &=  \bigg(-\frac{\eta^u}{2}\abs{\theta_t^u}^2 - Z_t^{u}\cdot \theta_t^u + \frac{1}{2\eta^u} \abs{\paran{I - P_t^u}(Z_t^{u} + \eta^u\theta_t^u)}^2 
         \bigg) \d t -  Z_t^{u} \cdot \d W_t^u, \quad \mu\otimes \P\text{--a.s.,} \quad t \in [0,T],
    \end{align}
    and it follows by Fubini theorem and the martingale property that
    \begin{equation*}
        \gamma_t^u = Y_t^u + \E\Big[\int_I\rho (X^v_t - \xi^v)G(u,v)\d v\Big].
    \end{equation*}
    In particular, $\gamma_T^u =  \E\Big[\int_I\rho (X^v_T - \xi^v)G(u,v)\d v\Big]$.
    Thus, by \cite[Theorem 7]{hu2005utility}, the value function of the utility maximization problem \eqref{eq:graphon-obj} (when $\sigma^*=0$) satisfies \eqref{eq:optim.sol.no.common} and the process $\opt^u$ given by \eqref{eq:optim.sol.no.common} is an optimal strategy that is square--integrable.
    In the present case, we even have that
    $$\bigg\{ \exp\Big(-\frac{1}{\eta^u}X^{\tilde\pi^u}_\tau\Big), \tau\,\,\F^u\text{--stopping times}\bigg\}$$ 
    is uniformly integrable.
    In particular, $(\tilde\pi^u)_{u\in I}$ is admissible.
\end{proof}

\section{General backward propagation of chaos theorem: proof of \Cref{thm:main.limit}}\label{sec:gct}
In this section we present  backward propagation of chaos results that are central in the proof of our main convergence result.
We will start by proving the case with common noise and then we will come back to the case without common noise.
The two proofs are similar, but the case with common noise is slightly more involved because the representing backward particle system if fully coupled with a forward process.

\subsection{Proof of \Cref{thm:main.limit}: The common noise case}
Consider an interacting particle system $(X^{i,n}, Y^{i,n}, Z^{ij,n}, Z^{*i,n})$ with the processes $(Y^{1,n}, Y^{2,n},\dots, Y^{n,n})$ evolving backward in time, and $(X^{1,n}, X^{2,n},\dots, X^{n,n})$ evolving forward in time and characterizing the Nash equilibrium, i.e. such that
    \begin{align}
        \opt_t^{i,n} &= \left(\Sigma_t^i {\Sigma_t^i}^{\top} \right)^{-1}\Sigma_t^i P_t^i\left( \begin{pmatrix} Z_t^{ii,n} \\ Z_t^{*i,n} \end{pmatrix} + \eta^i \theta_t^i  \right) \diff t\otimes \P\text{--a.s.} \quad \text{and}\quad  V^{i,n}_0((\opt^{j,n})_{j\neq i}) = -\e^{-\frac{1}{\eta^i}(\xi^i - \rho\overline{\xi}^i - Y_0^{i,n})}\;\; \forall i \in \{1,\dots,n\},
    \end{align}
    see \Cref{thm: n-bsde}.
We can find functions $h$ and $g$ such that they satisfy the following FBSDEs:
\begin{equation}
\label{eqn:N.BSDE}
	\begin{cases}
	    \d X^{i,n}_t = h^i(t,Z^{ii,n}_t, Z^{*i,n}_t)\Big\{\theta_t^i\d t +\sigma_t^i \d W^i_t + \sigma_t^{*i}\d W^*_t \Big\},  \quad X^{i,n}_0 =\xi^i.\\
		\d Y^{i,n}_t = -\Big\{g^i(t,Z^{ii,n}_t, Z^{*i,n}_t) + \frac{1}{2\eta^i} \sumj|Z^{ij,n}_t|^2 \Big\}\d t + \sum_{j=1}^n Z^{ij,n}_t \d W^{j}_t + Z_t^{*i,n}\d W^*_t\\
		Y^{i,n}_T = \rho\sumj \lambda_{ij}^n (X^{j,n}_T - X_0^{j,n}).
	\end{cases}
\end{equation}
Observe that due to the graph $(\lambda_{ij})_{1\le i,j\le n}$, the particles in the above system are not indistinguishable as in the homogeneous case considered by \citeauthor*{lauriere2022backward} \cite{lauriere2022backward,lauriere2022convergence} and \citeauthor*{possamai2021non} \cite{possamai2021non}.
Our goal here is to show that as the number of particles in the system approaches infinity, the above particle system converges to the infinite particle system $(X^u,Y^u,Z^u, Z^{*u})_{0\le u\le 1}$ given by
\begin{equation}
\label{eqn:MkV.BSDE}
\begin{cases}
	\d X^u_t = h^u(t,Z^u_t, Z^{*u}_t)\Big\{\theta^u_t\d t +\sigma_t^u \d W^u_t + \sigma_t^{*u}\d W^*_t \Big\},\quad X^u_0 = \xi^u.\\
	\d Y^u_t = -g^u(t,Z^u_t, Z^{*u}_t)\d t + Z_t^u \d W^u_t + Z^{*u}_t\d W^*_t\\
	Y_T^u = \E[ \rho\int_I(X_T^{v})G(u,v)\d v\mid \Fc_T^*] - \int_I \rho X_0^v G(u,v) \d v.
\end{cases}
\end{equation}
As above, this system is understood in the sense that the mapping $(u,t,\omega)\mapsto  (X^u_t, Y^u_t, Z^u_t, Z^{*u}_t)$ is measurable and for almost every $u\in I$, we have $(X^u, Y^u, Z^u, Z^{*u}) \in \S^2(\R, \F^u)\times \S^{2}(\R, \F^u)\times \H^{2}(\R^d, \F^u)\times \H^{2}(\R, \F^u)$.
In particular, if we consider a specific particle $u = \frac{i}{n}$ in the continuum, we will show that $(Y_t^{i,n}, Z_t^{ii,n}, Z_t^{*i,n})$ and $(Y_t^{\frac{i}{n}}, Z_t^{\frac{i}{n}}, Z_t^{\frac{i}{n}*})$ are ``close'' when $n \rightarrow \infty$. 
We will consider the following assumption on the coefficients of the FBSDEs:

\begin{condition}\label{cdn:generator}
    $h^u:[0,T]\times \Omega\times\R^d\times \R\to \R^d$ and $g^u:[0,T]\times \Omega\times\R^d\times \R\to \R$ are two functions such that there exist nonnegative constants $\ell_g$ and $\ell_h$, so that for almost every $u \in I$, it holds
\begin{align*}
	|h^u(t,z,z^*) - h^u(t,z',z^{*'})| \le {\color{black}\ell_h}\big( \|z-z'\| + \|z^* - z^{*'}\|\big) \text{ and } \|h^u(t, x, z^*)\|_\infty\le \ell_h(1 + \|z\| + |z^*|)
\end{align*}
and
\begin{align*}
    |g^u(t,z,z^*) - g^u(t,z',z^{*'})| \le \ell_g(\|z-z'\| + |z^* - z^{*'}|)
\end{align*}
for all $(t, z,z', z^*,z^{*'}) \in [0,T]\times (\R^d)^2\times \R^2$.
\end{condition} 

{\color{black}
\begin{remark}
    Recall that we use the same probability setting as described in \Cref{rebranding}. In other words, the indices in \eqref{eqn:N.BSDE} should be considered as $\frac{i}{n}$.
    Further recall the link betwee $\lambda_{ij}$ and $\beta_n>0$ and the graphon $G$ is made in \Cref{cdn:G}.
\end{remark}
}
\Cref{thm:main.limit}.$(i)$ is then a direct corollary of the following theorem:
\begin{theorem}\label{thm:main.conv}
    Assume that \Cref{cdn:G,cdn:generator} are satisfied.
    Further assume that the FBSDE \eqref{eqn:N.BSDE} and \eqref{eqn:MkV.BSDE} admit respective solutions $(X^{i,n},Y^{i,n}, Z^{ij,n}, Z^{*i,n})_{(i,j)\in \{1,\dots,n\}^2} $ and $(X^u, Y^u, Z^u, Z^{*u})_{u \in I}$ such that $ (X^{i,n},Y^{i,n}, Z^{ij,n}, Z^{*i,n}) \in \S^2(\R, \F^n)\times \S^{2}(\R, \F^n)\times \H^{2}(\R^d, \F^n)\times \H^{2}(\R, \F^n)$ for every $i,j$ and $(X^u, Y^u, Z^u, Z^{*u}) \in \S^2(\R, \F^u)\times \S^{2}(\R, \F^u)\times \H^{2}(\R^d, \F^u)\times \H^{2}(\R, \F^u)$ for almost every $u\in I$.
    Then for every $i\in \N^*$, it holds
    \begin{equation}
    \label{eq:statement.main.y}
        |Y^{i,n}_0 - Y^{\frac in}_0| \xrightarrow[n\to\infty]{} 0.
	\end{equation}
    Moreover, up to a subsequence, it holds
    \begin{equation}
    \label{eq:statement.main.z}
        \E\big[\|Z^{ii,n}_t - Z^{\frac{i}{n}}_t\| + |Z^{*i,n}_t - Z^{*\frac{i}{n}}_t|\big] \xrightarrow[n\to\infty]{} 0\quad \text{for almost every } t\in [0,T].
    \end{equation}
\end{theorem}
\begin{proof}
    Using \Cref{cdn:generator}, in light of \Cref{rmk: fubini property} and the definition of $\cF_T^*$, we have that for almost all $(t,v) \in [0,T] \times I$, 
    \begin{align}
    \notag
       \E\Big[\int_IX_t^vG(u,v)\d v \Big|\Fc^*_T\Big] &= \int_I \E[X_0^v]G(u,v)\d v + \int_0^t\E\Big[\int_Ih^v(s,Z^v_s, Z^{*v}_s)\theta_s^vG(u,v) \d v \big| \Fc^*_T\Big]\d s \\\notag
        &\quad + \E\Big[\int_0^t \int_I h^v(s,Z^v_s, Z^{*v}_s)\sigma_s^{v}G(u,v) \d v \d W^v_s \big| \Fc^*_T \Big]\\\notag
        &\quad + \E\Big[\int_0^t\int_I h^v(s,Z^v_s, Z^{*v}_s)\sigma_s^{*v}G(u,v)\d v \big| \Fc^*_T \Big]\d W^*_s\\\notag
       &= \int_I \E[X_0^v]G(u,v)\d v + \int_0^t\E\Big[\int_Ih^v(s,Z^v_s, Z^{*v}_s)\theta_s^vG(u,v) \d v \big| \Fc^*_T\Big]\d s \\
        \label{eq:formula.x.common}
       &\quad + \int_0^t\E\Big[\int_I h^v(s,Z^v_s, Z^{*v}_s)\sigma_s^{*v}G(u,v)\d v \big| \Fc^*_T \Big]\d W^*_s
    \end{align}
    where the first equality uses the fact that $X^u_0$ is independent of $W^*$, and the second equality follows from \cite[Lemma B.1]{lacker2022superposition}.
    Let us now introduce the ``shifted'' processes
    \begin{align*}
       \Zc^{*u}_t &:= Z^{*u}_t - \E\Big[\rho\int_Ih^v(t,Z_t^{v}, Z_t^{*v})\sigma^{*v}G(u,v)\d v \big|\Fc^*_T\Big], \qquad\qquad \Zc_t^u:= Z^u_t, \\
    \text{and} \quad \Yc_t^u &:= Y_t^u -\rho\bigg(\E\Big[\int_IX_t^vG(u,v)\d v\big|\Fc_T^*\Big] -  \int_I\E[X_0^v]G(u,v)\d v \bigg)
\end{align*}
so that using \Cref{eqn:MkV.BSDE}, the processes $(\Yc^u, \Zc^u, \Zc^{*u})$ satisfy
    \begin{align*}
        \Yc^u_t = \int_t^T g^u(s,Z^u_s, Z^{*u}_s) + \E\Big[\rho\int_Ih^v(Z^v_s, Z^{*v}_s)\theta_s^vG(u,v)\d v\big|\Fc_T^*\Big] \d s -\int_t^T\Zc^u_s\d W^u_s - \int_t^T\Zc^{*u}_s \d W^*_s.
    \end{align*}
Observe that the drift term is not written with respect to the newly defined $(\Yc^u, \Zc^u, \Zc^{*u})$, but rather with respect to the original $(Y^u, Z^u, Z^{*u})$. 
Similarly, for the prelimits, consider
\begin{align*}
        \Zc_t^{*i,n}&:= Z_t^{*i,n} - \rho \sumj \lambda_{ij}^n\sigma_t^{*j}h^j(t,Z_t^{jj,n}, Z_t^{*j,n}),\qquad \Zc_t^{ij,n}:= Z_t^{ij,n} - \rho \lambda_{ij}^n\sigma_t^jh^j(t,Z_t^{jj,n}, Z_t^{*j,n}), \\
        \text{and} \quad 
        \Yc_t^{i,n} &:= Y_t^{i,n} - \rho \sumj\lambda_{ij}^n(X_t^{j,n} - X_0^{j,n}),
    \end{align*}
    so that using \Cref{eqn:N.BSDE}, the processes $(\Yc^{i,n}, \Zc^{ij,n}, \Zc^{*i,n})$ satisfy
    \begin{align*}
         \Yc^{i,n}_t& = \int_t^Tg^i(s,Z^{ii,n}_s, Z^{*i,n}_s) + \frac{1}{2\etai}\sumj\|Z_s^{ij,n}\|^2 +\rho \sumj \lambda_{ij}^n h^j(s,Z^{jj,n}_s, Z^{*j,n}_s)\theta^j_s \d s \\
        &\qquad - \sum_{j=1}^n \int_t^T\Zc^{ij,n}_s\d W^j_s - \int_t^T\Zc^{*i,n}_s\d W^*_s.
    \end{align*}
    To further simplify the notation, let us put\footnote{Here, $\delta_{\{i = j\}}$ is an indicator function for $\{i = j\}$.} 
    \begin{equation*}
        \begin{cases}
            \Delta \Yc^{i,n}_t : = \Yc_t^{i,n} - \Yc_t^{\frac{i}{n}}\\
             \Delta \Zc_t^{*i,n} := \Zc_t^{*i,n} - \Zc_t^{*\frac{i}{n}}\\
            \Delta \Zc_t^{ij,n} = \Zc_t^{ij,n} - \delta_{\{i=j\}} \Zc_t^{\frac{i}{n}}
        \end{cases}
        \quad \text{and}\qquad
        \begin{cases}
            \Delta Y_t^{i,n} = Y_t^{i,n} - Y_t^{\frac{i}{n}}\\
            \Delta Z_t^{*i,n} := Z_t^{*i,n} - Z_t^{*\frac{i}{n}}\\
            \Delta Z_t^{ij,n} = Z_t^{ij,n} - \delta_{\{i=j\}} Z_t^{\frac{i}{n}}.
        \end{cases}   
    \end{equation*}
    Let $t\in [0,T]$ be fixed.
    We now define the sequence of stopping times $\tau_k$ such that for every positive $k$,
    \begin{equation*}
       \tau_k := \inf\bigg\{s\ge t: \sup_{r \in [t,s]} |\Delta  \Yc_r^{i,n}|^2 + \int_t^s \sum_{j=1}^n\big(\|\Delta Z^{ij,n}_r\|^2 + \|\Delta Z^{jj,n}_r\| + \|\Delta Z^{*j,n}_r\|^2 + \|h^j(r, Z^{jj,n}_r, Z^{*j,n}_r)\|^2  \big) \d r \ge k \bigg\}\wedge T.
    \end{equation*}
    Observe that $(\tau_k)_k$ depends on $i$ and $n$, but this dependence will be omitted to simplify notation.
    Since $ (X^{i,n},Y^{i,n}, Z^{ij,n}, Z^{*i,n}) \in \S^2(\R, \F^n)\times \S^{2}(\R, \F^n)\times \H^{2}(\R^d, \F^n)\times \H^{2}(\R, \F^n)$ and $(X^u, Y^u, Z^u, Z^{*u}) \in \S^2(\R, \F^u)\times \S^{2}(\R, \F^u)\times \H^{2}(\R^d, \F^u)\times \H^{2}(\R, \F^u)$
    it follows that for each $n$ and $i$, $\tau_k$ converges to $T$ $\P$-a.s as $k \rightarrow \infty$.
   Furthermore, put 
    \begin{equation}
    \label{def:gamma} 
        \Gamma^{i,n}_s:= \rho\sumj \lambda_{ij}^n h^{\frac jn}(s,Z^{\frac jn}_s, Z^{* \frac jn}_s)\cdot\theta^{\frac jn}_s - \rho\E\Big[\int_Ih^v(s,Z^v_s, Z^{*v}_s)\cdot\theta^v_sG(\frac in,v)\d v\big|\Fc^*_T\Big],
    \end{equation}
    and
    \[\Gamma^{*i ,n}_s:= \rho \sumj \lambda_{ij}^n h^{\frac jn}(s,Z^{\frac jn}_s, Z^{* \frac jn}_s)\cdot\sigma^{* \frac jn}_s - \rho\E\Big[\int_Ih^v(s,Z^v_s, Z^{*v}_s)\cdot\sigma^{*v}_sG(\frac{i}{n},v)\d v\big|\Fc^*_T\Big]. \]
    Now, applying It\^o's formula to $ |\Delta\Yc_t^{i,n}|^2$, we get 
    \begin{align}
    \notag
        &|\Delta \Yc_t^{i,n}|^2 + \int_t^{\tau_k} \Big( \sumall\| \Delta \Zc^{ij,n}_s\|^2 + \|\Delta \Zc^{*i,n}_s\|^2 \Big) \diff s \\ \notag
        &= |\Delta \Yc_{\tau_k}^{i,n}|^2 + \int_t^{\tau_k} 2\Delta \Yc_s^{i,n}\bigg(g^i(s,Z^{ii,n}_s, Z^{*i,n}_s) - g^{\frac{i}{n}}(s,Z^{\frac{i}{n}}_s, Z^{*\frac{i}{n}}_s) + \sumj \|Z_s^{ij,n}\|^2  \bigg) \d s\\\notag
        &\quad + \int_t^{\tau_k} 2\Delta \Yc^{i,n}_s\rho \sum_{j \neq i}\lambda_{ij}^n\theta^{j}_s\Big(h^j(s,Z^{jj,n}_s, Z^{*j,n}_s) - h^{\frac jn}(s,Z^{\frac jn}_s, Z^{* \frac jn}_s) \Big)\diff s  \\\label{Ito:Yc}
        &\quad + \int_t^{\tau_k}2\Delta \Yc_s^{i,n}\Gamma_s^{i,n} \d s -  \sum_{j=1}^n \int_t^{\tau_k}2\Delta\Yc_s^{i,n}\Delta \Zc^{ij,n}_s\cdot\d W^j_s - \int_t^{\tau_k}2\Delta\Yc_s^{i,n}\Delta \Zc^{*i,n}_s\d W^*_s.
    \end{align}
    Recall that 
    $\Delta \Zc^{ij,n} = \Delta Z^{ij,n} - \frac{\rho}{n\beta_n}\lambda_{ij}\sigma^jh^j(Z^{jj,n}, Z^{*j,n})$ for $i \neq j$ and
    $\Delta \Zc^{ii,n} = \Delta Z^{ii,n}$.
    \Cref{Ito:Yc} now takes the form
    \begin{align}
    \notag
        &|\Delta \Yc_t^{i,n}|^2 + \int_t^{\tau_k} \Big( \sumall\| \Delta \Zc^{ij,n}_s\|^2 + \|\Delta \Zc^{*i,n}_s\|^2 \Big) \diff s \\ \notag
        &= |\Delta \Yc_{\tau_k}^{i,n}|^2 + \int_t^{\tau_k} 2\Delta \Yc_s^{i,n}\left(g^i(s,Z^{ii,n}_s, Z^{*i}_s) - g^{\frac{i}{n}}(s,Z^{\frac{i}{n}}_s, Z^{*\frac{i}{n}}_s)  \right) \d s  \\ \notag
        &\quad + \int_t^{\tau_k} 2\Delta \Yc^{i,n}_s\rho \sum_{j \neq i}^n\lambda_{ij}^n\theta^j_s\Big(h^j(s,Z^{jj,n}_s, Z^{*j,n}_s) - h^{\frac jn}(s,Z^{\frac jn}_s, Z^{* \frac jn}_s) \Big)\diff s\\\notag
    &\quad +  \int_t^{\tau_k}2\Delta \Yc_s^{i,n}\Gamma_s^{i,n} \d s + \int_t^{\tau_k}2\Delta \Yc_s^{i,n}\Gamma_s^{*i,n} \d s + \sumj \rho \lambda_{ij}^n \int_t^{\tau_k}2 \Delta\Yc_s^{i,n}  h^j(s,Z_s^{jj,n}, Z_s^{*j,n}) \sigma_s^j \cdot \Delta Z_s^{ij,n} \d s \\\label{Ito:Yc:no g}
    & \quad - \sumj \int_t^{\tau_k}2\Delta\Yc_s^{i,n}\Delta \Zc^{ij,n}_s \cdot \Big(\d W^j_s - \Delta Z_s^{ij,n} \d s \Big) - \int_t^{\tau_k}2\Delta\Yc_s^{i,n}\Delta Z^{ii,n}_s\cdot \d W_s^i   
     - \int_t^{\tau_k}2\Delta\Yc_s^{i,n}\Delta \Zc^{*i,n}_s \d W^*_s.
\end{align}
Let $\QQ$ be the probability  measure with  density
\begin{align}\nonumber
    \frac{\d \QQ}{\d \PP} = &\exp \Bigg( \sumj \int_t^{\tau_k} \Delta Z_s^{ij,n} \cdot \d W_s^j  - \frac{1}{2}  \sumj \int_t^{\tau_k}  \|\Delta Z_s^{ij,n} \|^2  \d s  \Bigg).
\end{align}
The probability measure $\Q$ depends on $i$ and $n$, but its density has second moment bounded by a constant $C_k$ depending on $k$, but not on $i$ and $n$.
Taking conditional expectation under $\QQ$ with respect to $\cF_t^n$ in \eqref{Ito:Yc:no g}, we obtain the following:
\begin{align}\notag\begin{split}
    &|\Delta \Yc_t^{i,n}|^2 + \E^\Q\bigg[\int_t^{\tau_k} \Big( \sumall\| \Delta \Zc^{ij,n}_s\|^2 + \|\Delta \Zc^{*i,n}_s\|^2 \Big) \diff s \Big| \Fc^n_t\bigg]\\
     &\leq  \E^{\QQ}\left[|\Delta\Yc_{\tau_k}^{i,n}|^2 | \Fc_t^n \right] + \E^\Q\bigg[\int_t^{\tau_k}\frac{2\ell_g^2}{\varepsilon}|\Delta \Yc^{i,n}_s|^2 + \varepsilon|\Delta \Zc^{ii,n}_s|^2 + \varepsilon|\Delta \Zc^{*i,n}_s|^2 \Big| \Fc^n_t\bigg]\\
    &\quad  + \E^{\QQ}\Big[\int_t^{\tau_k} 2 \Delta \Yc_s^{i,n} \Gamma_s^{i,n}  \d s \big| \cF_t^n \Big] + \E^{\QQ}\Big[\int_t^{\tau_k} 2 \Delta \Yc_s^{i,n} \Gamma_s^{*i,n}  \d s \big| \cF_t^n \Big]\\ 
    &\quad+ C\E^{\QQ}\bigg[\int_t^{\tau_k} \left( |\Delta \Yc^{i,n}_s|^2 + \frac{\rho}{(n-1)^2\beta_n^2} \Big(\sum_{j=1}^n\lambda_{ij}^2\Big)\sum_{j \neq i}^n|\theta^j_s|^2 \Big\|h^j(s,Z^{jj,n}_s, Z^{*j,n}_s) - h^{\frac jn}(s,Z^{\frac jn}_s, Z^{* \frac jn}_s) \Big\|^2  \right)  \d s \big| \cF_t^n \bigg] \\
     &\quad + C\E^{\QQ}\bigg[\sumj \int_t^{\tau_k} 2 \rho\lambda_{ij}^n |\Delta\Yc_s^{i,n}|  \|h^j(s,Z_s^{jj,n}, Z_s^{*j,n}) \sigma_s^j \cdot \Delta Z_s^{ij,n}\| \d s \big| \Fc_t^n \bigg]. 
\end{split}\end{align}
Using $\E^{\mathfrak{P}}[\lambda^2_{ij}] \le \beta_n$ and by definition of the stopping time $\tau_k$, this estimate can be simplifed to 
\begin{align}\label{Ito:Yc:Q}\begin{split}
    &|\Delta \Yc_t^{i,n}|^2 + (1-\varepsilon)\E^\Q\bigg[\int_t^{\tau_k} \Big( \sumall\| \Delta \Zc^{ij,n}_s\|^2 + \|\Delta \Zc^{*i,n}_s\|^2 \Big) \diff s \big|\Fc^n_t\bigg]\\
     &\leq  \E^{\QQ}\left[|\Delta\Yc_{\tau_k}^{i,n}|^2 | \Fc_t^n \right] + \E^\Q\bigg[\int_t^{\tau_k}\Big( 1 + \frac{2\ell_g^2}{\varepsilon}\Big)|\Delta \Yc^{i,n}_s|^2\diff s \big|\Fc^n_t\bigg] \\
    &\quad  + \E^{\QQ}\bigg[\int_t^{\tau_k} 2 |\Delta \Yc_s^{i,n} |\Big(|\Gamma_s^{i,n}| + |\Gamma_s^{*i,n}|\Big)  \d s \big| \cF_t^n \bigg] + \frac{C_{\rho, \theta, h,k}}{(n - 1) \beta_n} \\ 
     &\quad + C\E^{\QQ}\bigg[\sumj \int_t^{\tau_k} 2 \rho \lambda_{ij}^n|\Delta\Yc_s^{i,n}| \|h^j(Z_s^{jj,n}, Z_s^{*j,n}) \sigma_s^j \cdot \Delta Z_s^{ij,n}\| \d s \big| \Fc_t^n \bigg].
\end{split}\end{align}
Applying Young's inequality and recalling the definition of $\tau_k$, the last term above can be estimated as
\begin{align}  \nonumber
    \E^{\QQ}\Bigg[\sumj &\int_t^{\tau_k} 2 \rho \lambda_{ij}^n |\Delta\Yc_s^{i,n}| \|h^j(Z_s^{jj,n}, Z_s^{*j,n}) \sigma_s^j \cdot \Delta Z_s^{ij,n}\| \d s \big| \Fc_t^n \Bigg] 
    \\ \nonumber
    &\le \frac{\rho\|\sigma\|_\infty}{(n - 1) \beta_n}\E^\Q\bigg[\sup_{t\le s\le \tau_k}|\Delta \Yc^{i,n}_s|\bigg( \int_t^{\tau_k}\sum_{j \neq i}^n\|h^j(s, Z^{jj,n}_s, Z^{*j,n}_s)\|^2\diff s +  \sum_{j \neq i}^n\int_t^{\tau_k}\|\Delta Z^{ij,n}_s\|^2\diff s \bigg) \big| \Fc_t^n \bigg]\\
    &\leq \frac{C_{\rho,\sigma,h, T,k}}{(n - 1) \beta_n} .
\end{align}
Thus, choosing $\varepsilon<1$ and subsequently using in \eqref{Ito:Yc:Q} Gronwall's inequality, taking expectation with respect to $\P$, Cauchy--Schwarz inequality and Doob's inequality, we are left with
\begin{align}
    \E\Big[|\Delta \Yc_t^{i,n}|^2 \Big] 
     &\leq  \E\bigg[\E^\Q\Big[|\Delta\Yc_{\tau_k}^{i,n}|^2 \big| \Fc_t^n \Big]\bigg]+ C_{k,T} \E\Big[\Big(\frac{\d \QQ}{\d \PP}\Big)^4 \Big]^{\frac{1}{4}}\E\bigg[\int_0^T |\Gamma_s^{i,n}|^2 + |\Gamma^{*i,n}_s|^2 \d s \bigg]^{\frac{1}{2}}
     + \frac{C_{\rho, \sigma, \theta, h, T,k}}{(n - 1) \beta_n}.
     \label{eqn:E Yc-square}
\end{align}
Observe that using again Cauchy--Schwarz we have
\begin{align}\label{eqn:sup delta yc bound part1}
    \E\bigg[  \E^\Q\Big[|\Delta\Yc_{\tau_k}^{i,n}|^2 \big| \Fc_t^n \Big] \bigg] 
&\leq C\E\Big[\Big(\frac{\d \QQ}{\d \PP}\Big)^{2} \Big]^{\frac{1}{2}} \E\Big[|\Delta \Yc_{\tau_k}|^{4}\Big]^{\frac{1}{2}} .
\end{align}
To proceed, first notice that for every $n$, we have that $(|\Delta\Yc_{\tau_k}^{i,n}|^4)_{k\ge1}$ converges to $0$ in probability as $k \rightarrow \infty$ almost surely, since $\tau_k$ converges to $T$ $\P$-a.s . There thus exists a fast sub-sequence $\Delta \Yc_{\tau_{k,m}}^{i,n}$ such that 
\begin{equation*}
    \P(|\Delta\Yc_{\tau_{k,m}}^{i,n}|^4 > \varepsilon) \leq \frac{\e^{-k^3}}{m}.
\end{equation*}
Therefore, for every $\varepsilon > 0$ we have
\begin{align*}
    \E\big[|\Delta\Yc_{\tau_{k,m}}^{i,n}|^4\big] = \E\Big[|\Delta\Yc_{\tau_{k,m}}^{i,n}|^4 \delta_{\{|\Delta\Yc_{\tau_{k,m}}^{i,n}|^4 \leq \varepsilon\}} \Big] + \E\Big[|\Delta\Yc_{\tau_{k,m}}^{i,n}|^4 \delta_{\{|\Delta\Yc_{\tau_{k,m}}^{i,n}|^4 > \varepsilon\}} \Big] \leq \varepsilon + k^2 \cdot \frac{\e^{-k^3}}{m}.
\end{align*}
Our definition of $\tau_k$ implies that all moments of $\frac{\d \QQ}{\d \PP}$ are upper bounded by $e^{Ck^2}$ for some constant $C>0$ independent of $k$. 
Thus, coming back to \eqref{eqn:E Yc-square} we continue the estimation as
\begin{align}
    \E\Big[|\Delta \Yc_t^{i,n}|^2 \Big]
    \label{eq:estimate.cond.z}
     &\leq  \e^{Ck^2}\Big(\varepsilon  + k^2\frac{\e^{ - k^3}}{m} \Big)^{\frac 12 } + 
    \frac{C_{\rho, \sigma, \theta, h, T,k}}{(n - 1)\beta_n}  + C_{k,T}\E\bigg[\int_0^{T}|\Gamma^{*i,n}_s|^2\d s\bigg]^{\frac12} + C_{k,T}\E\bigg[\int_0^{T}|\Gamma^{i,n}_s|^2\d s\bigg]^{\frac12}.
\end{align}
Applying \Cref{lem:bound:Gamma}, first fix $k$ and let $n \rightarrow \infty$, followed by letting $m \rightarrow \infty$ and $\varepsilon\to0$, we conclude that 
\[\E\Big[|\Delta \Yc_t^{i,n}|^2 \Big]\xrightarrow[n\to\infty]{} 0. \]
In particular, starting with $t=0$, it follows that the sequence
$(\Delta \Yc^{i,n}_0)_{n\ge1}$ converges to zero, and since 
$\Delta \Yc^{i,n}_0 = \Delta Y^{i,n}_0$, we obtain \Cref{eq:statement.main.y}.

\medskip

Let us now turn to the convergence of the control processes.
By  \Cref{Ito:Yc:Q}, \Cref{eq:estimate.cond.z}, Cauchy--Schwarz inequality and the above estimates we have
\begin{align*}
    \E\bigg[\int_t^{\tau_{k,m}}  \| \Delta \Zc^{ii,n}_s\| + \|\Delta \Zc^{*i,n}_s\| \diff s \bigg] &\le T\E\bigg[\Big(\frac{\diff \Q}{\diff \P}\Big)^2\bigg]^{\frac 12}\E^\Q\bigg[\int_t^{\tau_{k,m}}  \| \Delta \Zc^{ii,n}_s\|^2 + \|\Delta \Zc^{*i,n}_s\|^2 \diff s \bigg]^{\frac 12}\\
    &\le T\cdot \e^{Ck^2}\bigg\{\varepsilon + k^2\frac{\e^{- k^3}}{m} + \frac{C_{h,\theta, T,k}}{(n - 1)\beta_n} +  C_{k,T}\E\bigg[\int_0^{T}|\Gamma^{*i,n}_s|^2\d s\bigg]^{\frac12}\\
    &\quad + C_{k,T}\E\bigg[\int_0^{T}|\Gamma^{i,n}_s|^2\d s\bigg]^{\frac12}\bigg\}^{\frac 12}.
\end{align*}
Using 
\Cref{lem:bound:Gamma}, first fix $k$ and let $n \rightarrow \infty$, followed by letting $m \rightarrow \infty$ and $\varepsilon\to0$, we conclude that, up to a subsequence, it holds
\begin{equation*}
   \lim_{m\to\infty}\lim_{n\to \infty} \|\Delta \Zc^{ii,n}_s\|\delta_{\{s\le \tau_{k,m}\}} + \|\Delta \Zc^{*i,n}_s\|\delta_{\{s\le \tau_{k,m}\}} =0\quad \P\text{--a.s.; } \text{ for a.e. } s\in [t,T].
\end{equation*}
Since $\P(\tau_{k,m}\ge t) = 1$, this shows that
\begin{equation*}
   \|\Delta \Zc^{ii,n}_t\| + \|\Delta \Zc^{*i,n}_t \| \xrightarrow[n\to\infty]{} 0\quad\P \text{--a.s.}
\end{equation*}
By the identity
 $\Delta \Zc^{ii,n} = \Delta Z^{ii,n}$, we have thus obtained that $(\Delta Z^{ii,n}_t)_{n\ge1}$ converges to zero.
For the convergence of $\Delta Z^{*i,n}_t$, observe that
\begin{align*}
    |\Delta Z^{*i,n}_s| &\le |\Delta \Zc^{*i,n}_s| + |\Gamma^{*i,n}_s| + \frac{\rho}{n\beta_n}\sum_{j \neq i}^n\lambda_{ij}|\sigma^{* \frac jn}_s|\Big|h^j(s, Z^{jj,n}_s, Z^{*j,n}_s) - h^j(s, Z^{\frac jn}_s, Z^{* \frac jn}_s)\Big|\\
    &\le |\Delta \Zc^{*i,n}_s| + |\Gamma^{*i,n}_s| + \rho \|\sigma^{*i}\|_\infty\ell_h\frac{1}{(n - 1)\beta_n}\sum_{j \neq i}^n\lambda_{ij}\Big( \|\Delta Z^{jj,n}_s\|+ |\Delta Z^{*j,n}_s|\Big).
\end{align*}
Thus,
\begin{align*}
    \E\bigg[\int_0^{\tau_{k,m}}|\Delta Z^{*i,n}_s|\diff s \bigg] &\le \E\bigg[\int_0^{\tau_{k,m}}|\Delta \Zc^{*i,n}_s|^2 + |\Gamma^{*i,n}_s|^2\diff s\bigg] + \frac{C_{\rho, \sigma^*, h, k}}{(n - 1) \beta_n}. 
\end{align*}
Therefore, arguing as above and using again \Cref{lem:bound:Gamma} we have that, up to a subsequence,
\begin{equation*}
    \|\Delta Z^{*i,n}_t \| \xrightarrow[n\to\infty]{} 0\quad \P\text{--a.s.} 
\end{equation*}
Therefore, \Cref{eq:statement.main.z} follows by dominated convergence.
This concludes the proof.
\end{proof}
\begin{lemma}\label{lem:bound:Gamma}
    Under the conditions of \Cref{thm:main.conv}, 
    it holds
    \[
        \E\bigg[\int_0^T |\Gamma_s^{*i,n}|^2 \d s \bigg] + \E\bigg[\int_0^T |\Gamma_s^{i,n}|^2 \d s \bigg] \xrightarrow[n\to\infty]{} 0\quad \text{for every } i \in \N^*.
    \] 
\end{lemma} 
\begin{proof}
     We will consider only the term $\Gamma^{i,n}$; the term $\Gamma^{*i,n}$ is dealt with similarly. Using \Cref{cdn:G}, especially that $\lambda_{ij}$ are i.i.d. and independent of $(\Omega, \Fc, \P)$ and $(W^1,\dots, W^n,W^*)$, we have
%
\begin{align*}
   \E |\Gamma^{i,n}_s|^2 &= \E\bigg|\frac{1}{(n - 1) \beta_n}\sumj \lambda_{ij}h^j(s,Z^{\frac{j}{n}}_s, Z^{*\frac{j}{n}}_s)\cdot\theta^{\frac{j}{n}}_s - \E\Big[\int_Ih^v(s,Z^v_s, Z^{*v}_s)\cdot\theta^v_sG(\frac{i}{n},v)\d v\big|\Fc^*_T\Big]\bigg|^2 \\
    &\leq 2\E \bigg|\frac{1}{n-1}\sumj \left( \frac{\lambda_{ij}}{\beta_n}h^j(s,Z^{\frac jn}_s, Z^{*\frac jn}_s) \theta^{\frac jn}_s -  h^{\frac jn}(s,Z^{\frac jn}_s, Z^{* \frac jn}_s) \theta^{\frac jn}_sG_n(\frac in, \frac jn) \right)\bigg|^2\\
    &\quad+2\E\bigg| \frac{1}{n - 1}\sumj h^{\frac jn}(s,Z^{\frac jn}_s, Z^{*\frac jn}_s) \theta^{\frac jn}_sG_n(\frac in, \frac jn) -  \E\Big[\int_Ih^v(s,Z^v_s, Z^{*v}_s)\cdot\theta^v_sG(\frac{i}{n},v)\d v\big|\Fc^*_T\Big] \bigg|^2 \\
    &\le \frac{C_{\theta}}{(n - 1)^2\beta_n^2}\mathrm{Var}(\lambda_{ij}) \E \sumj  \big| h^j(s,Z^{\frac jn}_s, Z^{*\frac jn}_s) \big| ^2\\
    &\quad +4\E\bigg| \frac1n\sumj h^{\frac jn}(s,Z^{\frac jn}_s, Z^{*\frac jn}_s) \theta^{\frac jn}_sG_n(\frac in, \frac jn) -  \E\Big[\frac1n\sum_{j=1}^nh^{\frac jn}(s,Z^{\frac jn}_s, Z^{*\frac jn}_s) \theta^{\frac jn}_sG_n(\frac in, \frac jn)\big|\Fc^*_T\Big] \bigg|^2 \\
    &\quad+ \E\bigg|\E\Big[\frac1n\sum_{j=1}^nh^{\frac jn}(s,Z^{\frac jn}_s, Z^{*\frac jn}_s) \theta^{\frac jn}_sG_n(\frac in, \frac jn)\big|\Fc^*_T\Big] -  \E\Big[\int_Ih^v(s, Z^v_s, Z^{*v}_s)\cdot\theta^v_sG(\frac{i}{n},v)\d v\big|\Fc^*_T\Big] \bigg|^2.
    \end{align*}
    Using that the step function $F^n_s(u)$ given by
    \begin{equation*}
        F^n_s(u) := \sum_{j=1}^nh^{\frac jn}(s,Z^{\frac jn}_s, Z^{*\frac jn}_s) \theta^{\frac jn}_s\delta_{\{u\in (\frac jn, \frac{j+1}{n}]\}}
    \end{equation*}
    approximates the function $F_s:v \mapsto h^v(s,Z^v_s, Z^{*v}_s)\cdot\theta^v_s$ in $L^2(I,\Bc(I),\mu)$,
    we have
    \begin{align}
    \notag
    \E |\Gamma^{i,n}_s|^2    
      &\le \frac{C_{\theta}}{(n - 1)^2\beta_n^2}\mathrm{Var}(\lambda_{ij}) \E \sumj  \big| h^j(s,Z^{\frac jn}_s, Z^{*\frac jn}_s)\big| ^2\\ \notag
      &\quad +\frac{4}{(n-1)^2}\sumj\E\bigg|  \Big(h^{\frac jn}(s,Z^{\frac jn}_s, Z^{*\frac jn}_s) \theta^{\frac jn}_s -  \E\big[h^{\frac jn}(s,Z^{\frac jn}_s, Z^{*\frac jn}_s) \theta^{\frac jn}_s\big|\Fc^*_T\big]\bigg|^2 \\
      \notag
    &\quad+ \E\bigg[\Big|\int_IF^{n}_s(v)G_n(\frac{i}{n},v)\diff v - \int_Ih^v(s, Z^v_s, Z^{*v}_s)\cdot\theta^v_sG_n(\frac{i}{n},v)\d v\Big|^2\Big]\\\label{eq:boxnorm.conve}
    &\quad + \E\Big[\Big|\int_Ih^v(s, Z^v_s, Z^{*v}_s)\cdot\theta^v_sG_n(\frac{i}{n},v)\d v - \int_Ih^v(s, Z^v_s, Z^{*v}_s)\cdot\theta^v_sG(\frac{i}{n},v)\d v\Big|^2\Big]\\\notag
    &\le \frac{C_{\theta}}{(n - 1)^2\beta_n^2}\E \Big| \sumj (\lambda_{ij} - \E[\lambda_{ij}]) h^j(s,Z^{\frac jn}_s, Z^{*\frac jn}_s) \Big|^2  + \frac{C_{\theta}}{(n-1)^2}\sumj \E\Big|\Big(h^{\frac jn}(s,Z^{\frac jn}_s, Z^{*\frac jn}_s)\Big|^2 \\\notag
    &\quad + \|F^n_s - F_s\|_{L^2(I,\Bc(I),\mu)} +C_\theta\E\bigg[\int_I\|h^v(s,Z^v_s, Z^{*v}_s)\|^2\Big(G_n(\frac in, v) - G(\frac in, v)\Big)^2\diff v \bigg].
\end{align}
Because the Lipschitz constants of $h^u$ and $g^u$ do not depend on $u$, standard FBSDE estimates show that $\sup_{u\in I}\|(Z^u,Z^{*u})\|_{\H^2(\R^{d+1}, \F^u)}<\infty$.
Hence, integrating on both sides above and using \Cref{cdn:generator}, we have
\begin{align}
\notag
        \frac1n\sum_{i=1}^n\E\bigg[\int_0^T |\Gamma^{i,n}_s|^2 \diff s\bigg] &\le 
        \big( \frac{C_{h,\theta,T}}{(n - 1)\beta_n^2} + \frac{C_{\theta}}{(n-1)}\big) \big(\sup_{u\in I}\|(Z^u,Z^{*u})\|_{\H^2(\R^{d+1}, \F^u)} + 1\big) + C_T\|F^n_s - F_s\|_{L^2(I,\Bc(I),\mu)}\\
        \notag
        &\quad + \ell_h^2C_\theta\big(\sup_{u\in I}\|(Z^u,Z^{*u})\|_{\H^2(\R^{d+1}, \F^u)}+1\big)\frac1n\sum_{i=1}^n\int_I\Big(G_n(\frac in, v) - G(\frac in, v)\Big)^2\diff v\\\notag
        &\le \frac{C_{h,\theta,T,Z}}{(n - 1)\beta_n^2} + \frac{C_{\theta, Z}}{(n-1)}  + C_T\|F^n_s - F_s\|_{L^2(I,\Bc(I),\mu)}\\\notag
        &\quad + \ell_h^2C_{\theta,Z}\int_I\int_I\Big(G_n(\frac {\lfloor nu\rfloor}{n}, v) - G(\frac {\lfloor nu\rfloor}{n}, v)\Big)^2\diff v\diff u\\
        \label{e{}q:operator.vs.boxnorm}
        &\le  \frac{C_{h,\theta,T,Z}}{(n - 1)\beta_n^2} + \frac{C_{\theta, Z}}{(n-1)} + C_T\|F^n_s - F_s\|_{L^2(I,\Bc(I),\mu)} + \ell_h^2C_{\theta,Z}\|G_n - G\|_2^2 . 
\end{align}
Therefore, since $n\|G_n-G\|^2_2\to0$, it follows that for each $i$, we have
\[
    \E\bigg[\int_0^T |\Gamma_s^{i,n}|^2 \d s \bigg] \xrightarrow[n\to\infty]{} 0.
    \] 
\end{proof}
{\color{black}
\begin{remark}
    If the function $h$ is bounded (which is the case when the graphon equilibrium $(\widetilde \pi^u)_{u\in I}$ is bounded), it is enough the require that $n\|G_n - G\|_\Box\to0$, which is weaker that $\L^2$--convergence.
    This is due to the fact that the last term in \Cref{eq:boxnorm.conve} can be estimated as
    \begin{align*}
        \E\bigg[\bigg(\int_Ih^v(s,Z^v_s, Z^{*v}_s)\cdot\theta^u_s\Big(G_n(\frac in, v) - G(\frac in, v)\Big)\diff v\bigg)^2 \bigg] &\le C_{h, \theta}\bigg| \int_I\Big(G_n(\frac in, v) - G(\frac in, v)\Big)\diff v\bigg|^2.
    \end{align*}
    Taking the average, we obtain the following estimation
    \begin{align}
    \bigg(\frac1n \sum_{i = 1}^n \Big|\int_IG_n(\frac{i}{n},v)- G(\frac{i}{n},v)\d v \Big| \bigg)^2 = \bigg(\int_I \Big| \int_I \Big(G_n(\frac{\lceil nu\rceil}{n}, v) - G(\frac{\lceil nu\rceil}{n}, v)\Big) \d v \Big| \d u\bigg)^2 \leq 4\|G_n - G\|^2,
    \end{align}
     where $\|G\|$ is the so--called operator norm given by
    \begin{equation*}
        \|G\| :=\sup_{\|h\|_\infty\le 1}\int_I\bigg|\int_Ih(u)G(u,v)\diff v\bigg|\diff u.
    \end{equation*}
    It follows from \citeauthor*{Lovasz} \cite[Lemma 8.11]{Lovasz} that the $\|G\|_\Box$ and $\|G\|$ are equivalent norms. Therefore the last term in \eqref{eq:operator.vs.boxnorm} can be replaced by $\ell_h^2C_{\theta,Z}\|G_n - G\|_{\Box}^2$.
\end{remark}}
\subsection{Proof of \Cref{thm:main.limit}: The non--common noise case}
Let us not present the proof of \Cref{thm:main.limit}.$(ii)$.
Throughout this subsection, we assume $\sigma^{*u}=0$ for all $u \in I$.
By \Cref{thm: n-bsde} and \Cref{rem:char.non.common}, the Nash equilibrium $(\widetilde \pi^{i,n})_{i \in \{1,\dots,n\}}$ is characterized by the BSDE \ref{eqn:bsde-no-common}.
That is, it holds
\begin{align*}
    \opt^{i,n}_t = (\sigma_t^{i})^{-1}P_t^i\paran{Z_t^{ii}+\eta^i\theta_t^i} \quad\text{and}\quad V_0^{i,n}((\opt^{j,n})_{j\neq i}) = -e^{-\frac{1}{\etai}(\xi^i-\rho\overline{\xi}^i -Y_0^i)} \quad \P\otimes \mathrm{d}t\text{--a.s.}
\end{align*}
with $(Y^{i,n}, Z^{ij,n})_{(i,j)\in \{1,\dots,n\}^2}$ solving BSDE \ref{eqn:bsde-no-common}.
Moreover, by \Cref{cor:graphon-bsde} and \Cref{prop:exist.graphon.BSDE}, there is a graphon equilibrium $(\widetilde \pi^u)_{u\in I}$ such that
\begin{align*}
        \opt^u_t = (\sigma_t^{u})^{-1}P_t^u\paran{Z_t^{u}+\eta^u\theta_t^u}\,\, \mathrm{d}t\otimes\mu\boxtimes\P\text{--a.s.} \text{ and}\quad V_0^{u,G} = -\exp\Big(- \frac{1}{\eta^u} \Big(\xi^u - \int_I \E[\rho \xi^v]G(u,v) \d v - Y_0^u \Big) \Big)
\end{align*}
with $(Y^u, Z^u)_{u\in I}$ solving \Cref{eqn:BSDE-graphon.decoupled}.
It thus suffices to show that
\begin{equation*}
    \big|Y^{i,n}_0 - Y^{\frac in}_0\big|^2 + \big| Z^{ii,n}_t - Z^{\frac in}_t\big|^2   \xrightarrow[n\to\infty]{} 0 \quad \diff t\otimes \P\text{--a.s.}
\end{equation*}

\medskip

Let us put $\Delta Y^{i,n}:= Y^{i,n} - Y^{\frac in} $ and $\Delta Z^{ij,n} := Z^{ij,n} - Z^{\frac in}\delta_{\{i=j\}}$.
Let $t\in [0,T]$ be fixed and consider the stopping time
\begin{equation*}
    \tau_k:= \inf\bigg\{s\ge t: \sup_{t\le r\le  s}|\Delta Y^{i,n}_r|^2 + \int_t^s\sum_{j=1}^n\|P^j_r(Z^{jj,n}_r + \eta^j\theta^j_r)\|^2 + \|\Delta Z^{jj,n}_r\|^2\diff s\ge k\bigg\}\wedge T.
\end{equation*}
Observe that for each $i,n$ the sequence $(\tau_k)_{k\ge1}$ converges to $T$ $\P$--a.s.
Applying It\^o's formula to $\e^{\kappa t}(\Delta Y^{i,n})^2$ for some $\kappa>0$ to be chosen, we have
\begin{align*}
    &\e^{\kappa t}(\Delta Y^{i,n}_t)^2\\
     & = \e^{\kappa \tau_k}(\Delta Y^{i,n}_{\tau_k})^2 + \int_t^{\tau_k}2\e^{\kappa t}\Delta Y^{i,n}_s\Big\{ \theta^i_s\cdot \Delta Z^{ii,n}_s + \frac{1}{2\eta^i}\Big( \big|(I - P^i_s)(Z^{ii,n}_s + \eta^i\theta^i_s) \big|^2 - \big|(I - P^{\frac in}_t)(Z^{\frac in}_s + \eta^{\frac in}\theta^{\frac in}_s) \big|^2 \Big) \Big\}\\
    &\quad + \int_t^{\tau_k}2\e^{\kappa s}\Delta Y^{i,n}_s\sumj\big|Z^{ij,n}_s + \sigma^j\lambda^n_{ij}\rho P^j_s(Z^{jj,n}_s + \eta^i\theta^i_s) \big|^2\diff s\\
    &\quad + \int_t^{\tau_k}2\e^{\kappa s}\Delta Y^{i,n}_s\rho\Big\{\sumj \lambda^n_{ij}P^j_s(Z^{jj,n}_s + \eta^j\theta^j_s)\theta^j_s - \E\Big[\int_IP^v_s(Z^v_s + \eta^v\theta^v_s)\theta^v_sG(u,v)\diff v \Big] \Big\}\diff s\\
    &\quad - \int_t^{\tau_k}\kappa\e^{\kappa s}(\Delta Y^{i,n}_s)^2 \diff s - \sum_{j=1}^n\int_t^{\tau_k}\e^{\kappa s}\|\Delta Z^{ij,n}_s\|^2\diff s - \sum_{j=1}^n\int_t^{\tau_k}2\e^{\kappa s}\Delta Y^{i,n}_s\Delta Z^{ij,n}_s\diff W^j_s.
\end{align*}
Let us introduce the measure $\Q$ with density
\begin{equation*}
    \frac{\diff \Q}{\diff \P} = \Ec\bigg(\int_t^{\tau_k}\big(\theta^i_s + \frac{1}{2\eta^i}\gamma_s(Z^{ii,n}_s, Z^{\frac in}_s)\big)\diff W^i_s + \sumj \int_t^{\tau_k}\Big(Z^{ij,n}_s + 2Z^{ij,n}_s\sigma^j_s\lambda^n_{ij}\rho P^j_s(Z^{jj,n}_s + \eta^j\theta^j_s) \Big)\diff W^j_s \bigg),
\end{equation*}
where $\gamma_s$ is the (linearly growing) function such that
\begin{equation*}
    \big|(I - P^i_t)(Z^{ii,n}_s + \eta^i\theta^i_s) \big|^2 - \big|(I - P^{\frac in}_t)(Z^{\frac in}_s + \eta^{\frac in}\theta^{\frac in}_s) \big|^2  = \gamma_s(Z^{ii,n}_s, Z^{\frac in}_s)\Delta Z^{ii,n}_s.
\end{equation*}
This follows by Lipschitz--continuity of the projection operator since $A^j$ is convex (also recall the rebranding $i\equiv \frac in$). 
Thus, by Girsanov's theorem, the $\mathrm{BMO}$ martingale property of $Z^{\frac in}$, and square integrability of $Z^{ij,n}$ we have
\begin{align*}
    \e^{\kappa t}(\Delta Y^{i,n}_t)^2 &=\E^\Q\bigg[\e^{\kappa \tau_k}(\Delta Y^{i,n}_{\tau_k})^2 + \int_t^{\tau_k}2e^{\kappa s}\Delta Y^{i,n}_s\sumj\rho^2\frac{\lambda_{ij}^2}{n^2\beta_n^2}\|\sigma^j_s\|^2\|P^j_s(Z^{jj,n}_s + \theta^j_s\eta^j)\|^2 \diff s  \Big| \Fc_t^n \bigg]\\
    &\quad + \E^\Q\bigg[\int_t^{\tau_k}2\e^{\kappa s}\Delta Y^{i,n}_s\rho\Big\{\frac{1}{n\beta_n}\sumj \lambda_{ij}P^{ j}_t(Z^{ jj,n}_s + \eta^{ j}\theta^{j}_s)\theta^{j}_s - \frac{1}{n\beta_n}\sumj\lambda_{ij}P^{\frac jn}_s(Z^{\frac jn}_s + \eta^{\frac jn}\theta^{\frac jn}_s) \theta^{\frac jn}_s \Big\}\diff s\\
    &\qquad\qquad + \int_t^{\tau_k}2\e^{\kappa s}\Delta Y^{i,n}_s\rho\Gamma^{i,n}_s - \kappa e^{\kappa s}(\Delta Y^{i,n}_s)^2 - \sum_{j=1}^n\e^{\kappa s} \|\Delta Z^{ij,n}_s\|^2 \diff s\Big| \Fc_t^n\bigg]
\end{align*}
where $\Gamma^{i,n}_s$ is the process given by
\begin{equation*}
    \Gamma^{i,n}_s := \frac{1}{n\beta_n}\sumj\lambda_{ij}P^{\frac jn}_s(Z^{\frac jn}_s + \eta^{\frac jn}\theta^{\frac jn}_s) \theta^{\frac jn}_s - \E\Big[\int_IP^v_s(Z^v_s + \eta^v\theta^v_s)\theta^v_sG(\frac in,v)\diff v \Big].
\end{equation*}
Using Lipschitz--continuity of the projection operator and boundedness of $\Sigma$, we continue the estimation as
\begin{align*}
    \e^{\kappa t}(\Delta Y^{i,n}_t)^2 + \E^\Q\bigg[ \sum_{j=1}^n\int_t^{\tau_k}\e^{\kappa s} \|\Delta Z^{ij,n}_s\|^2 \diff s\Big| \Fc_t^n\bigg] &=\E^\Q\bigg[\e^{\kappa \tau_k}(\Delta Y^{i,n}_{\tau_k})^2 + \frac{C_{k,\rho,\sigma}}{n^2\beta^2_n} +\int_t^{\tau_k}e^{\kappa s}\Big(\frac{C_{\theta,\sigma,\rho}}{\varepsilon} - \kappa \Big)|\Delta Y^{i,n}_s|^2 \diff s \Big| \Fc_t^n \bigg]\\
    &\quad + \varepsilon\E^\Q\bigg[\int_t^{\tau_k}\e^{\kappa s}\Big\{\frac{1}{n\beta_n}\sumj \lambda_{ij}\|\Delta Z^{ jj,n}_s\| \Big\}^2 + \e^{\kappa s}|\Gamma^{i,n}_s|^2\diff s\Big| \Fc_t^n\bigg],
\end{align*}
where we also used Young's inequality with some $\varepsilon>0$.
Choosing $\kappa>0$ large enough, and using Cauchy--Schwarz inequality, it follows that
\begin{align*}
    \E^\Q\bigg[\e^{\kappa t}(\Delta Y^{i,n}_t)^2 +  \sum_{j=1}^n\int_t^{\tau_k}\e^{\kappa s} \|\Delta Z^{ij,n}_s\|^2 \diff s\bigg] &\le\E^\Q\Big[\e^{\kappa \tau_k}(\Delta Y^{i,n}_{\tau_k})^2   \Big]+ \frac{C_{k,\rho}}{n^2\beta^2_n}\\
    &\quad + \varepsilon\E^\Q\bigg[\int_t^{\tau_k}\e^{\kappa s}\frac{1}{n\beta_n^2}\Big(\sumj\lambda^2_{ij}\Big)\frac{1}{n}\sumj \|\Delta Z^{ jj,n}_s\|^2 + \e^{\kappa s}|\Gamma^{i,n}_s|^2\diff s\bigg]\\
    &\le  \E^\Q\Big[\e^{\kappa \tau_k}(\Delta Y^{i,n}_{\tau_k})^2   \Big]+ \frac{C_{k,\rho}}{n^2\beta^2_n} + \frac{C_{k,\kappa}}{n\beta_n} + \varepsilon\E^\Q\bigg[\int_t^{\tau_k} \e^{\kappa s}|\Gamma^{i,n}_s|^2\diff s\bigg],
\end{align*}
where we used the fact that $\lambda_{ij}$ is independent of $W^1,\dots,W^n$ and $\E^{\mathfrak{P}}[\lambda^2_{ij}]\le \beta_n$, and definition of the stopping time $\tau_k$.
Because $(\Delta Y^{i,n}_{\tau_k})_{k\ge1}$ converges to $0$ in $\P$--probability and thus in $\Q$--probability for each $n$,  we can find a fast sub--sequence $\Delta Y^{i,n}_{\tau_{k,m}}$ such that
\begin{equation*}
    \Q\big(|\Delta Y^{i,n}_{\tau_{k,m}}| \ge \varepsilon \big) \le \frac{\e^{-k^2}}{m}.
\end{equation*}
Thus, for every $\varepsilon>0$, we have
\begin{equation*}
    \E^\Q[|\Delta Y^{i,n}_{\tau_{k,m}}|^2] \le \varepsilon + k\frac{\e^{-k^2}}{m}.
\end{equation*}
Hence, using again definition of $\tau_k$,
\begin{align*}
    \E^\Q\bigg[\e^{\kappa t}(\Delta Y^{i,n}_t)^2 +  \sum_{j=1}^n\int_t^{\tau_{k,m}}\e^{\kappa s} \|\Delta Z^{ij,n}_s\|^2 \diff s\bigg] &\le \varepsilon + k\frac{\e^{-k^2}}{m} + \frac{C_{k,\rho,\sigma}}{n\beta_n} + \varepsilon\E^\Q\bigg[\int_t^{\tau_k} \e^{\kappa s}|\Gamma^{i,n}_s|^2\diff s\bigg].
\end{align*}
Using Cauchy--Schwarz inequality, we further have
\begin{align*}
    \E\bigg[\e^{\kappa t}|\Delta Y^{i,n}_t| +  \int_t^{\tau_{k,m}}\e^{\kappa s} \|\Delta Z^{ii,n}_s\| \diff s\bigg] &\le 2T\E\Big[\Big(\frac{\diff \Q}{\diff \P}\Big)^2 \Big]^{1/2}\E^\Q\bigg[\e^{\kappa t}(\Delta Y^{i,n}_t)^2 + \int_t^{\tau_{k,m}}\e^{\kappa s} \|\Delta Z^{ii,n}_s\|^2 \diff s \bigg]^{1/2}\\
    &\le  C_k\bigg( \varepsilon + k\frac{\e^{-k^2}}{m} + \frac{C_{k,\rho,\sigma}}{n\beta_n} + \varepsilon\E^\Q\bigg[\int_t^{\tau_{k,m}} \e^{\kappa s}|\Gamma^{i,n}_s|^2\diff s\bigg]\bigg)^{1/2}.
\end{align*}

We will show below that for each $k$ fixed
\begin{equation}
\label{eq:conv.gamma.proof.ncn}
    \E^\Q\bigg[\int_t^{\tau_{k,m}} |\Gamma^{i,n}_s|^2\diff s\bigg]   \xrightarrow[n\to\infty]{} 0.
\end{equation}
Thus, first taking the limit in $n$, then in $m$ and and then letting $\varepsilon\to0$, it follows that 
\begin{equation*}
   \E\bigg[\e^{\kappa t}|\Delta Y^{i,n}_t| +  \int_t^{\tau_{k,m}}\e^{\kappa s} \|\Delta Z^{ii,n}_s\| \diff s\bigg]   \xrightarrow[m,n\to\infty]{} 0.
\end{equation*}
We thus obtain that $\Delta Y^{i,n}_0 \to 0$ as $n\to\infty$ and, up to a subsequence, 
\begin{equation*}
    \|\Delta Z^{ii,n}_s\|\delta_{\{s\le\tau_{k,m}\}}   \xrightarrow[m,n\to\infty]{} 0 \quad \text{for almost every } s\in [t,T]\quad \P\text{--a.s. for a.e. } s\in [t, T].
\end{equation*}
In particular, because $\P(\tau_{k,m}\ge1) = 1$, $\Delta Z^{ii,n}_t \to 0$ $\P$-a.s. as $n\to \infty$.

\medskip

Let us now come back to \eqref{eq:conv.gamma.proof.ncn}.
Since the random variables $(Z^{u})_{u\in I}$ are e.p.i., it follows by the exact law of large numbers, see \citeauthor*{sun2006exact} \cite[Corollary 3.10]{sun2006exact} that it holds
\begin{equation*}
   \| \Gamma^{i,n}_s\| \le \Big\| \frac{1}{n\beta_n}\sumj\lambda_{ij}P^{\frac jn}_s(Z^{\frac jn}_s + \eta^{\frac jn}\theta^{\frac jn}_s) \theta^{\frac jn}_s - \int_IP^v_s(Z^v_s + \eta^v\theta^v_s)\theta^v_sG(\frac in,v)\diff v\Big\|.
\end{equation*}
Therefore, using triangular inequality and the fact that 
\begin{equation*}
    \int_IF^n_s(v)G_n(\frac in,v)(v)\diff v = \frac{1}{n}\sum_{j=1}^n P^{\frac jn}_s(Z^{\frac jn}_s + \eta^{\frac jn}\theta^{\frac jn}_s) \theta^{\frac jn}_sG_n(\frac in,\frac jn)
\end{equation*}
with 
\begin{equation*}
    F^n_s(u) := \sum_{j=1}^n P^{\frac jn}_s(Z^{\frac jn}_s + \eta^{\frac jn}\theta^{\frac jn}_s) \theta^{\frac jn}_s \delta_{\{u\in (\frac jn, \frac{j+1}{n}]\}},
\end{equation*}
it follows that
\begin{align*}
    \| \Gamma^{i,n}_s\| &\le \Big\|\frac{1}{n\beta_n}\sum_{j=1}^n\lambda_{ij}P^{\frac jn}_s(Z^{\frac jn}_s + \eta^{\frac jn}\theta^{\frac jn}_s) \theta^{\frac jn}_s - \frac{1}{n}\sum_{j=1}^n P^{\frac jn}_s(Z^{\frac jn}_s + \eta^{\frac jn}\theta^{\frac jn}_s) \theta^{\frac jn}_sG_n(\frac in,\frac jn)  \Big\|\\
    &\quad + \Big\|\int_IF^n_s(v)G_n(\frac in,v)(v)\diff v - \int_IF^n_s(v)G(\frac in,v)(v)\diff v\Big\|\\
    &\quad + \Big\|\int_IF^n_s(v)G(\frac in,v)\diff v - \int_IP^v_s(Z^v_s + \eta^v\theta^v_s)\theta^v_sG(\frac in,v)\diff v\Big\|.
\end{align*}
Proceeding as in the proof of \Cref{lem:bound:Gamma}, we have
\begin{align*}
    \E^\Q\bigg[\int_0^{\tau_k}\| \Gamma^{i,n}_s\|^2\diff s\bigg] &\le \frac{\mathrm{var}(\lambda_{ij})}{n\beta_n^2}\|\theta\|_\infty\Big(\|Z^{\frac{i}{n}}\cdot W^{\frac in}\|_{\mathrm{BMO}} + C_{\theta,\eta}\Big) + \E^\Q\bigg[\int_I \int_0^{\tau_k}\|F^n_s(v)\|^2\diff s\Big(G_n(\frac in, v) - G(\frac in, v)\Big)^2\diff v\bigg]\\
    &\quad + \E^{\Q}\bigg[\int_0^{\tau_k}\int_I\|F^n_s(v) - P^v_s(Z^v_s + \eta^v\theta^v_s)\theta^v_s\|^2\diff v \diff s \bigg]\\
    &\le \frac{C_{\theta,\eta}}{n\beta^2_n} + \int_I\Big(G_n(\frac in, v) - G(\frac in, v)\Big)^2\E^\Q\bigg[\int_0^T
    \|F^n_s(v)\|^2\bigg]\diff v\\
    &\quad + \E\Big[\Big(\frac{\diff \Q}{\diff \P}\Big)^2\Big]^{1/2}\E\bigg[\int_0^T\bigg(\int_I\|F^n_s(v) - P^v_s(Z^v_s + \eta^v\theta^v_s)\theta^v_s\|^2\diff v\bigg)^2\bigg]^{1/2}.
\end{align*}
Since the intervals $(\frac jn, \frac{j+1}{n}]$ form a partition of $I$, and using linear growth of the projection operator, it follows that
\begin{equation*}
    \|F^n_s(u)\|^2\le \sum_{j=1}^n\|Z^{\frac jn}_s\|^2\delta_{u\in (\frac jn, \frac{j+1}{n}]} + C_{\theta,A}
\end{equation*}
Thus, using the inequality $\|\cdot\|_{\H^2(\R^d,\F^u)}\le \|\cdot\|_{\H^2_{\mathrm{BMO}}(\R^d,\F^u)}$, that the BMO norm does not depend on the underlying measure and the fact that $\sup_{u\in I}\|Z^{u}\|_{\H^2_{\mathrm{BMO}}(\R^d,\F^u)}<\infty$, we have
\begin{align*}
    \E^{\Q}\bigg[\int_0^T\|F^n_s(u)\|^2\diff s\bigg]&\le \sum_{j=1}^n\|Z^{\frac in}\|_{\H^2_{\mathrm{BMO}}(\R^d,\F^u)}\delta_{u\in (\frac jn, \frac{j+1}{n}]} + C_{\theta,A}\\
    &\le C.
\end{align*}
Hence, we have
\begin{align*}
   \E^\Q\bigg[\int_0^{\tau_k}\| \Gamma^{i,n}_s\|^2\diff s\bigg]
    &\le \frac{C_{\theta,\eta}}{n\beta^2_n} + C\int_I\Big(G_n(\frac in, v) - G(\frac in, v)\Big)^2\diff v + C_k\E\bigg[\int_0^T\|F^n_s - F_s\|_{L^2(I,\mu)}^2\diff s\bigg]^{1/2}
\end{align*}
where $F_s(v):= P^v_s(Z^v_s + \eta^v\theta^v_s)\theta^v_s$. 
Since the sequence of step functions $F^n_s$ converges to $F$ in $L^2(I,\mu)$, it follows by dominated convergence that, fixing $k$, we have \eqref{eq:conv.gamma.proof.ncn}.
The convergence to zero of the term $\int_I\Big(G_n(\frac in, v) - G(\frac in, v)\Big)^2\diff v$ is proved as in at the end of the proof of \Cref{lem:bound:Gamma}.
This concludes the proof.

\section{Wellposedness of graphon McKean--Vlasov BSDEs and FBSDEs} \label{sec:wellposedness}
We conclude the article with two existence results for graphon McKean--Vlasov (F)BSDEs used in the proof of existence of graphon equilibria.
In the ensuing statements and proofs, we will use the space $\S^p(\F,\R^d,I)$ defined as the space of families of processes $(Y^u)_{u\in I}$ such that $(u,t,\omega)\mapsto Y^u$ is $\Ic\otimes \F$--measurable and for almost every $u$, it holds $Y^u\in \S^p(\F^u,\R^d)$.
This space is equipped with the norm 
$$\|Y\|_{\S^p(\F,\R^d,I)}:= \int_I\|Y^u\|_{\S^p(\F^u,\R^d)}\d u$$
which makes it a Banach space.
We similarly define $\H^p(\F,\R^d,I)$.
We further denote by $\mathbb{H}_{\mathrm{BMO}}(\F^u, \mathbb{R}^{d})$ the space of  $\F^u$--predictable processes $Z$ with values in $\R^d$ such that the process $\int Z\d W^u$ is a $(\P,\F^u)$--BMO martingale. 
The space $\mathbb{H}_{\mathrm{BMO}}(\F, \mathbb{R}^{d}, I)$ is defined analogously to $\S^p(\F,\R,I)$ with the norm
$$\|Z\|_{\H_{\mathrm{BMO}}(\F,\R^d,I)}:= \int_I\|Z^u\|_{\H_{\mathrm{BMO}}(\F^u,\R^d)}\d u.$$

\subsection{Graphon McKean--Vlasov FBSDEs}
We start by proving existence of the graphon McKean--Vlasov FBSDEs with Lipschitz coefficients.
Observe that this is a system involving a continuum of coupled equation, where the coupling is due to the graphon term.
\begin{proposition}
\label{prop:exist.graphon.FBSDE}
    Assume that the functions $g:I\times [0,T]\times \R^d\times \R\to \R$; $b,h_2:I\times [0,T]\times\Omega\times\R^{d+1}\to \R$ and $h_1:I\times[0,T]\times\Omega\times\R^{d+1}\to \R^d$
    are Borel measurable and Lipschitz--continuous in the sense that
    \begin{align*}
        |g^u(t,z,z^*) - g^u(t,\bar z,\bar z^*)| &\le \ell_g(\|z - \bar z\| + |z^* - \bar z^*|)\\
        |b^u(t,z,z^*) - b^u(t,\bar z,\bar z^*)|&+\|h^u_1(t,z,z^*) -  h^u_1(t,\bar z,\bar z^*)\| + |h^u_2(t,z,z^*) -  h^u_2(t,\bar z,\bar z^*)|  \le \ell_h(\|z - \bar z\| + |z^* - \bar z^*|)
    \end{align*}
    for some $\ell_g,\ell_h>0$ and $(t, z,\bar z, z^*,\bar z^*)  \in [0,T]\times (\R^d)^2\times \R^2$, and $\int_0^T|g^u(t,0,0)|\d t<\infty$.
    Further assume that we are given a family  $(\xi^u)_{u\in I}$ such that $\xi^u \in L^2(\Bc(I)\otimes\mathcal{F}^u_0,\mu\otimes\P)$.
    Then, if {\color{black}$\rho< \frac{1}{2\ell_h}\e^{-(2\ell_g^2 + \frac12)T}$}, the graphon system
    \begin{equation}
    \label{eqn:MkV.FBSDE.exists}
        \begin{cases}
        \d X^u_t = b^u(t,Z^u_t, Z^{*u}_t)\d t +h^u_1(t,Z^u_t,Z^{*u}_t) \d W^u_t + h^u_2(t,Z_t^u, Z^{*u}_t)\d W^*_t ,\quad X^u_0= \xi^u.\\
        \d Y^u_t = -g^u_t(Z^u_t, Z^{*u}_t)\d t + Z_t^u \d W^u_t + Z^{*u}_t\d W^*_t\\
        Y_T^u = \E[ \rho\int_IX_T^{v}G(u,v)\d v\mid \Fc_T^*]
    \end{cases}
    \end{equation}
    admits a unique solution $(X^u,Y^u,Z^u,Z^{*u})_{u \in I} \in \mathbb{S}^2(\F,\mathbb{R},I)\times \mathbb{S}^2(\F, \mathbb{R},I)\times \mathbb{H}^2(\F, \mathbb{R}^{d},I)\times \mathbb{H}^2(\F, \mathbb{R},I)$.
\end{proposition}
\begin{proof}
    Let $(z^u,z^{*u})_{u\in I}\in \mathbb{H}^2(\F, \mathbb{R}^d,I)\times \mathbb{H}^2(\F, \mathbb{R},I)$ be a given family of processes and consider $(X^u,Y^u,Z^u,Z^*_u)_{u\in I}$ given by
    \begin{equation}
    \label{eq:fixed.point.fbsde}
        \begin{cases}
            X^u_t := \xi^u +\int_0^tb(s, z^u_s,z^{*u}_s)\d s + \int_0^th_1(s, z^u_s, z^{*u}_s)\d W^u_s + \int_0^th_2(s, z^u_s, z^{*u}_s)\d W^*_s\\
            Y^u_t = \E\Big[ \rho\int_IX^v_TG(u,v)\d v\Big|\Fc^*_T \Big] + \int_t^Tg^u(s,Z^u_s, Z^{*u}_s)\d s - \int_t^TZ^u_s\d W^u_s - \int_t^TZ^{*u}_s\d W^*_s.
        \end{cases}
    \end{equation}
    It follows by \citeauthor*{Stricker-Yor78} \cite[Section 4]{Stricker-Yor78} that $(u,t,\omega)\mapsto X^u_t$ is measurable, and thus that $\int_IX^v_TG(u,v)\d v$ is well--defined.
    Arguing as in the proof of \cite[Section 4]{Stricker-Yor78} (in particular using Picard iteration), one establishes that $(u, t, \omega) \mapsto (Y^u_t, Z^u_t, Z^{*u}_t)$ is measurable.
    Moreover, since it is square--integrable, it follows by the standard result of \citeauthor*{pardoux1990adapted} \cite{pardoux1990adapted} on Lipschitz BSDEs that $(Y^u, Z^u)$ exists and is unique in $\mathbb{S}^2(\F^u, \mathbb{R})\times \mathbb{H}^2(\F^u, \mathbb{R}^{d})\times \mathbb{H}^2(\F^u, \mathbb{R})$ for almost every $u$.
    Thus, the function
    \begin{equation*}
        \Psi((z^u,z^{*u})_{u\in I}) := (Z^u,Z^{*u})_{u\in I}
    \end{equation*}
    maps the Banach space $\mathbb{H}^2(\F, \mathbb{R}^d,I)\times \mathbb{H}^2(\F, \mathbb{R},I)$ into itself.
    It remains to show that $\Psi$ admits a unique fixed point.

    Let $(z^u,z^{*u})_{u\in I}, (\bar z^u,\bar z^{*u})_{u\in I} \in \mathbb{H}^2(\F, \mathbb{R}^d,I)\times \mathbb{H}^2(\F, \mathbb{R},I)$ be given. Put $\Psi((z^u, z^{*u})_{u\in I}) = (Z^u, Z^{*u})_{u\in I}$ and $\Psi((\bar z^u, \bar z^{*u})_{u\in I}) = (\bar Z^u, \bar Z^{*u})_{u\in I}$ such that $( X^u, Y^u, Z^u,Z^*_u)_{u\in I}$ and $(\bar X^u, \bar Y^u, \bar Z^u,\bar Z^*_u)_{u\in I}$ satisfy \eqref{eq:fixed.point.fbsde}.
    Let us introduce the shorthand notation $\Delta X^u:= X^u - \bar X^u$, $\Delta Y^u:= Y^u - \bar Y^u$, $\Delta Z^u:= Z^u - \bar Z^u$ and $\Delta Z^{*u} := Z^{*u} - \bar Z^{*u}$.
    Given some constant $\kappa>0$, we apply It\^o's formula to $\e^{\kappa t}|\Delta Y^u_t|^2$ to obtain
    \begin{align*}
        \e^{\kappa t}|\Delta Y^u_t|^2 &\le \e^{\kappa T}\rho^2\E\Big[\int_I\|\Delta X^v_T\|^2G(u,v)^2 \d v\Big|\mathcal{F}^*_T \Big] + \int_t^T \e^{\kappa s}\big(2\frac{\ell_g^2}{\varepsilon} - \kappa \big)|\Delta Y^u_s|^2 + (\varepsilon-1)\int_t^T\e^{\kappa s}\big(\|\Delta Z^u_s\|^2 + |\Delta Z^{*u}_s|^2 \big) \d s\\
           &\quad -\int_t^T\e^{\kappa s}\Delta Y^u_s\Delta Z^u_s\d W^u_s - \int_t^T\e^{\kappa s}\Delta Y^u_s\Delta Z^{*u}_s\d W^{*}_s. 
    \end{align*}
    Taking expectation on both sides and choosing $\kappa=2\ell_g^2/\varepsilon$ , we have
    \begin{align*}
        \E\bigg[\e^{\kappa t}|\Delta Y^u_t|^2  + (1-\varepsilon)\int_t^T\e^{\kappa s}\|\Delta Z^u_s\|^2 + \e^{\kappa s}|\Delta Z^{*u}_s|^2\d s \bigg] \le \e^{\kappa T}\rho^2\int_I\E\big[\|\Delta X^v_T\|^2\big]\d v.
    \end{align*}
    On the other hand, applying It\^o's formula to $\e^{\kappa t}\|\Delta X^u_t\|^2$ and using Lipschitz--continuity of $b^u, h^u_1$ and $h^u_2$, we have
    \begin{align*}
        \E\big[\e^{\kappa t}\|\Delta X^u_t\|^2\big] &\le \E\bigg[\int_0^t2\e^{\kappa s}\ell_h(\|z^u_s - \bar z^{u}_s\| + |z^{*u}_s - \bar z^{*u}_s|)\|\Delta X_s^u\| + \ell_h^2\e^{\kappa s}(\|z^u_s - \bar z^{u}_s\|^2 + |z^{*u}_s - \bar z^{*u}_s|^2)\d s \bigg] \\
        &\quad + \kappa \E\bigg[\int_0^t\e^{\kappa s}\|\Delta X^u_s\|^2\diff s\bigg]\\
        &\le \E\bigg[(1 + \kappa)\int_0^t\e^{\kappa s}\|\Delta X_s^u\|^2\d s\bigg]  + 2\ell_h^2\E\bigg[\int_0^t\e^{\kappa s}\big(\|z^u_s - \bar z^{u}_s\|^2 + |z^{*u}_s - \bar z^{*u}_s|^2\big) \d s \bigg],
    \end{align*}
    where the last inequality follows from Young's inequality. Thus, by Gronwall's inequality, we have
    \begin{equation*}
        \E\big[\e^{\kappa t}\|\Delta X^u_t\|^2\big] \le 2\ell_h^2\e^{(\kappa + 1)T}\E\bigg[\int_0^t\e^{\kappa s}\big(\|z^u_s - \bar z^{u}_s\|^2 + |z^{*u}_s - \bar z^{*u}_s|^2\big) \d s \bigg].
    \end{equation*}
    Thus, if $\varepsilon = 1/2$, we have
    \begin{align*}
        \int_I\E\bigg[\int_0^T\e^{\kappa s}\|\Delta Z^u_s\|^2 + \e^{\kappa s}|\Delta Z^{*u}_s|^2\d s \bigg]\d u \le 
        4\ell_h^2\e^{(4\ell_g^2 + 1)T}\rho^2\int_I\E\bigg[\int_0^T\e^{\kappa s}\big( \|z^u_s - \bar z^{u}_s\|^2 + |z^{*u}_s - \bar z^{*u}_s|^2\big) \d s \bigg]\d u.
    \end{align*}
    Thus, by the choice of $\rho$ and the Banach fixed point theorem, the mapping $\Psi$ admits a unique fixed point, implying that the graphon FBSDE \eqref{eqn:MkV.FBSDE.exists} admits a unique solution in $\mathbb{S}^2(\F,\mathbb{R},I)\times \mathbb{S}^2(\F, \mathbb{R},I)\times \mathbb{H}^2(\F, \mathbb{R}^{d},I)\times \mathbb{H}^2(\F, \mathbb{R},I)$.
\end{proof}

\subsection{Graphon McKean--Vlasov BSDE}
Let us now turn to the wellposedness of graphon McKean--Vlasov FBSDEs with Lipschitz--continuous coefficients.
\begin{proposition}
\label{prop:exist.graphon.BSDE}
    Assume that the functions $g:I\times [0,T]\times \R^d\to \R$ and $f:I\times [0,T]\times\R^d\to \R$ are Borel--measurable and satisfy the  locally Lipschitz and Lipschitz--continuity conditions
    \begin{equation}
    \label{eq:growth.cond.g}
        |g^u(t,z) - g^u(t,z^\prime)| \le \ell_g(\|z\|+ \|z^\prime\|)\|z- z^\prime\|\quad \text{and}\quad |g^u(t,z)|\le \ell_g(1 + \|z\|^2)
    \end{equation} 
    and
    \begin{equation*}
       |f^u(t,z) - f^u(t,z^\prime)| \le \ell_f\|z- z^\prime\|\quad \text{and}\quad |f^u(t,z)|\le \ell_f(1 + \|z\|)
    \end{equation*}
    for some constants $\ell_g,\ell_f>0$ and every $(t,z,z^\prime)\in [0,T]\times (\R^d)^2$ and almost all $u \in I$.
    Further assume that we are given $\mathcal{F}^u_T$--measurable random variables $F^u$ such that $(u,\omega)\mapsto F^u$ is measurable and uniformly bounded.
    Then, the graphon system
    \begin{equation*}
        Y^u_t = F^u + \int_t^T \Big(g^u(s,Z^u_s) + \int_I\E[f^v(s,Z^v_s)]G(u,v)\d v \Big) \,ds - \int_t^TZ^u_s\d W^u_s
    \end{equation*}
    admits a unique solution $(Y^u,Z^u)_{u \in I}$ such that we have $(Y^u,Z^u )_{u\in I} \in \mathbb{S}^\infty(\mathbb{F}, \mathbb{R},I) \times \mathbb{H}_{\mathrm{BMO}}(\mathbb{F}, \mathbb{R}^d,I)$\\ and $\sup_{u\in I}\|Z^u\|_{\mathbb{H}_{\mathrm{BMO}}(\mathbb{F}, \mathbb{R}^d)}<\infty$. 
\end{proposition}
\begin{proof}
    Let $(y^u,z^u)_{u\in I} \in \mathbb{S}^\infty(\mathbb{F}, \mathbb{R}^d,I) \times \mathbb{H}_{\mathrm{BMO}}(\mathbb{F}, \mathbb{R}^d,I)$ be given and consider the (decoupled) quadratic BSDEs
    \begin{equation}
    \label{eq:aux.q.bsde.graph}
        Y^u_t = F^u + \int_t^T \Big(g^u(s,Z^u_s) + \int_I\E[f^v(s,z^v_s)]G(u,v)\d v \Big) \,ds - \int_t^TZ^u_s\d W^u_s.
    \end{equation}
    It follows by \citeauthor{hu2005utility}  \cite{hu2005utility} that for almost every $u \in I$, this equation admits a unique solution $(Y^u,Z^u) \in \mathbb{S}^\infty(\F^u, \mathbb{R})\times \mathbb{H}_{\mathrm{BMO}}(\F^u, \mathbb{R}^{d}) $.
    Moreover, it follows by the arguments of \citeauthor*{Stricker-Yor78} that $(u,t,\omega)\mapsto(Y^u_t,Z^u_t)$ is measurable.
    Thus, the function 
    \[
        \Psi((y^u,z^u)_{u \in I}) := (Y^u,Z^u)_{u \in I}
    \]
    is well--defined and maps the Banach space $\mathbb{S}^\infty(\F, \mathbb{R}, I)\times \mathbb{H}_{\mathrm{BMO}}(\F, \mathbb{R}^{d},I)$ into itself.
    It therefore remains to show that this mapping admits a unique fixed point.
    
    \medskip

    Let $(y^u, z^u)_{u\in I}, (\bar y^u,\bar z^u)_{u\in I} \in \mathbb{S}^\infty(\F, \mathbb{R}, I)\times \mathbb{H}_{\mathrm{BMO}}(\F, \mathbb{R}^{d},I)$ be given and put
    $\Psi((y^u,z^u)_{u\in I}) = (Y^u, Z^u)_{u\in I}$ and $\Psi((\bar y^u,\bar z^u)_{u\in I}) = (\bar Y^u,\bar Z^u)_{u\in I}$.
    Let $\kappa>0$ be a constant to be determined and let $\tau$ be an $\F^u$--stopping time.
    Apply It\^o's formula to $\e^{\kappa t}|\Delta Y^u_t|^2:= \e^{\kappa t}|Y^u_t - \bar Y^u_t|^2$ to obtain
    \begin{align*}
        \e^{\kappa \tau}|\Delta Y_\tau^u|^2 & = \int_\tau^T2\e^{\kappa s}\Delta Y^u_s\Big(g^u(s,Z^u_s) - g^u(s,\bar Z^u_s) + \int_I\E[f^v(s,z^v_s) - f^v(s,\bar z^v_s)]G(u,v)\d v\Big)\d s\\
        &\quad - \kappa\int_\tau^T\e^{\kappa s}|\Delta Y^u_s|^2 \d s- \int_\tau^T \e^{\kappa s}\|\Delta Z^u_s\|^2\d s - \int_t^T2\e^{\kappa s}\Delta Y^u_s\Delta Z^u_s\d W^u_s\\ 
        &\le \Big(\frac{1}{\varepsilon}- \kappa\Big)\int_\tau^T\e^{\kappa s}|\Delta Y^u_s|^2 \d s + \varepsilon \ell^2_f\int_\tau^T\e^{\kappa s} \int_I\E[\|\Delta z^u_s\|^2]G(u,v)^2\d s -\int_\tau^T\e^{\kappa s}|\Delta Z^u_s|^2\d s\\
        &\quad - \int_\tau^T2\e^{\kappa s}\Delta Y^u_s\Delta Z^u_s\d W^{u,\Q}_s
    \end{align*}
    where we used the short hand notation $\Delta Z^u:=Z^u -\bar Z^u$ and $\Delta z^u:= z^u-\bar z^u$, and where $W^{u,\Q}$ is a Brownian motion under the probability measure
    \begin{equation*}
        \frac{\d \Q}{\d \P} := \mathcal{E}\Big(\int_0^\cdot\beta^u(s,Z^u_t,\bar Z^u_t)\d W^u_t \Big)
    \end{equation*}
    with $\beta$ being a linearly growing function such that $g^u(s,z) - g^u(s,\bar z) = \beta^u(s,z,\bar z)\cdot(z - \bar z)$.
    Choose $\kappa$ such that $\kappa>\frac{1}{\varepsilon}$. 
    Taking conditional expectation on both sides yields
    \begin{align*}
        \e^{\kappa \tau}|\Delta Y_\tau^u|^2 + \E^\Q\bigg[ \int_\tau^T\e^{\kappa s}|\Delta Z^u_s|^2\d s\Big|\mathcal{F}^u_\tau \bigg] &\le \varepsilon \ell_f^2\int_I\E\bigg[ \int_\tau^T\e^{\kappa s} \|\Delta z^v_s\|^2\d s \Big| \mathcal{F}^u_\tau\bigg]\d v .
    \end{align*}
    Taking the supremum over $\tau$ and integrating on both sides in $u$ therefore gives
    \begin{equation*}
        \|\Delta Y\|^2_{\mathbb{S}^{\infty}(\mathbb{R}, \mathbb{F},I)} + \|\Delta Z\|^2_{\mathbb{H}^2_{\mathrm{BMO}}(\mathbb{R}^d,\mathbb{F},I) } \le \varepsilon \ell_f^2 \|\delta z\|^2_{\mathbb{H}^2_{\mathrm{BMO}}(\mathbb{R}^d,\mathbb{F},I) },
    \end{equation*}
    and where we used the fact that the BMO norm does not depend on the underlying probability measure and $\|\cdot\|_{\mathbb{H}^2(\mathbb{R}^d,\mathbb{F}) }\le \|\cdot\|^2_{\mathbb{H}^2_{\mathrm{BMO}}(\mathbb{R}^d,\mathbb{F}) }$.
    Choosing $\varepsilon>0$ small enough allows to conclude that $\Psi$ is a contraction, and thus it follows by the Banach fixed point theorem that $\Psi$ admits a unique fix point in $\mathbb{S}^{\infty}(\mathbb{R}, \mathbb{F},I)\times \mathbb{H}^2_{\mathrm{BMO}}(\mathbb{R}^d,\mathbb{F},I)$.
\end{proof}

\bibliographystyle{abbrvnat}
\bibliography{ref}

\end{document}